\documentclass[aps,prb,floatfix,superscriptaddress]{revtex4-2}

\usepackage{amsmath,amssymb}
\usepackage{graphicx}
\usepackage[dvipsnames]{xcolor}
\usepackage{dsfont}
\usepackage[retain-unity-mantissa=false]{siunitx}
\usepackage{hyperref}  
\begin{document}

\title{All Optical Excitation of Spin Polarization in d-wave Altermagnets}

\DeclareSIUnit\angstrom{\text {Å}}
\DeclareSIUnit\bar{bar}

\author{Marius Weber}
       \thanks{these authors contributed equally}
       \affiliation{Department of Physics and Research Center OPTIMAS, RPTU University Kaiserslautern-Landau, 67663 Kaiserslautern, Germany}
       \affiliation{Institute of Physics, Johannes Gutenberg University Mainz, 55099 Mainz, Germany}
\author{Stephan Wust}
        \thanks{these authors contributed equally}
        \affiliation{Department of Physics and Research Center OPTIMAS, RPTU University Kaiserslautern-Landau, 67663 Kaiserslautern, Germany}
\author{Luca Felipe Haag}
    \affiliation{Department of Physics and Research Center OPTIMAS, RPTU University Kaiserslautern-Landau, 67663 Kaiserslautern, Germany}
\author{Paul Herrgen}
    \affiliation{Department of Physics and Research Center OPTIMAS, RPTU University Kaiserslautern-Landau, 67663 Kaiserslautern, Germany}
\author{Akashdeep Akashdeep}
    \affiliation{Institute of Physics, Johannes Gutenberg University Mainz, 55099 Mainz, Germany}
\author{Kai Leckron}
    \affiliation{Department of Physics and Research Center OPTIMAS, RPTU University Kaiserslautern-Landau, 67663 Kaiserslautern, Germany}
\author{Christin Schmitt}
    \affiliation{Institute of Physics, Johannes Gutenberg University Mainz, 55099 Mainz, Germany}
\author{Rafael~\surname{Ramos}}
\affiliation{Centro Singular de Investigación en Química Biolóxica e Materiais Moleculares (CIQUS), Universidade de Santiago de Compostela, Santiago de Compostela, Spain}
\affiliation{Departamento de Química-Física, Universidade de Santiago de Compostela, Santiago de Compostela, Spain}
\affiliation{WPI Advanced Institute for Materials Research, Tohoku University, Sendai 980-8577, Japan}
    
\author{Takashi~\surname{Kikkawa}}
    \affiliation{Department of Applied Physics, The University of Tokyo, Tokyo 113-8656, Japan}
\affiliation{Advanced Science Research Center, Japan Atomic Energy Agency, Tokai 319-1195, Japan}

\author{Eiji~\surname{Saitoh}}
    \affiliation{WPI Advanced Institute for Materials Research, Tohoku University, Sendai 980-8577, Japan}
    \affiliation{Department of Applied Physics, The University of Tokyo, Tokyo 113-8656, Japan}
    \affiliation{Institute for AI and Beyond, The University of Tokyo, Tokyo 113-8656, Japan}
    \affiliation{RIKEN Center for Emergent Matter Science (CEMS), Wako 351–0198, Japan}
\author{Mathias Kläui}
       \affiliation{Institute of Physics, Johannes Gutenberg University Mainz, 55099 Mainz, Germany}
\author{Libor \v{S}mejkal}
    \affiliation{Max Planck Institute for the Physics of Complex Systems, 01187 Dresden, Germany }
    \affiliation{Max Planck Institute for Chemical Physics of Solids, 01187 Dresden, Germany}
\author{Jairo Sinova}
       \affiliation{Institute of Physics, Johannes Gutenberg University Mainz, 55099 Mainz, Germany}
    \affiliation{Department of Physics, Texas A\&M University, College Station, Texas 77843-4242, USA}
\author{Martin Aeschlimann}
    \affiliation{Department of Physics and Research Center OPTIMAS, RPTU University Kaiserslautern-Landau, 67663 Kaiserslautern, Germany}
\author{Gerhard Jakob}
       \affiliation{Institute of Physics, Johannes Gutenberg University Mainz, 55099 Mainz, Germany}
\author{Benjamin Stadtmüller}
    \affiliation{Department of Physics and Research Center OPTIMAS, RPTU University Kaiserslautern-Landau, 67663 Kaiserslautern, Germany}
    \affiliation{Experimentalphysik II, Institute of Physics, Augsburg University, 86159 Augsburg, Germany}
\author{Hans Christian Schneider}
    \affiliation{Department of Physics and Research Center OPTIMAS, RPTU University Kaiserslautern-Landau, 67663 Kaiserslautern, Germany}

\date{2026-06-04}

\begin{abstract}
The recently discovered altermagnets exhibit collinear magnetic order with zero net magnetization but with unconventional spin-polarized d/g/i-wave band structures, expanding the known paradigms of ferromagnets and antiferromagnets.
In addition to novel current-driven electronic transport effects, the unconventional time-reversal symmetry breaking in these systems also makes it possible to obtain a spin response to \emph{linearly polarized} fields in the optical frequency domain. We show through ab-initio calculations of the prototypical d-wave altermagnet RuO$_2$, with a symmetry combining twofold spin rotation with fourfold lattice rotation, $[C_2\|C_{4z}]$,
that there is an optical analogue of a spin splitter effect, as the coupling to a linearly polarized exciting laser field makes the d-wave character of the altermagnet directly visible.
By magneto-optical measurements on RuO$_2$ films of different thicknesses ranging from $2$ to $8\,$nanometers, we demonstrate the predicted connection of the linear polarization of an ultrashort pump pulse to the sign and magnitude of the optically excited electronic spin polarization in the ultrathin RuO$_2$ films. The possibility of exciting and controlling an electronic spin polarization by linearly polarized optical pulses in a compensated system is a unique consequence of the altermagnetic material properties. Our experimental results  therefore establish an optical pump-probe based protocol for detection of altermagnetic characteristics in ultrathin RuO$_2$ films, but our all-optical approach should apply more generally to materials in this altermagnetic symmetry class. 
\end{abstract}


\maketitle

\section{Introduction}

The discovery of altermagnetism was based on the realization that certain combinations of real-space rotations and spin-space symmetries may lead to compensated magnetic materials with an unconventional time-reversal symmetry broken electronic structure~\cite{Smejkal_2020_Crystal} and corresponding alternating spin polarization in momentum space forming d, g, or i-wave magnetic quantum phases~\cite{Smejkal2022b}. 

Materials with altermagnetic symmetries promise novel electronic and transport properties that cannot be realized in conventional ferromagnets and antiferromagnets on the one hand, or in non-magnetic semiconductors or metals on the other. Depending on the underlying spin and real-space symmetries, the spin and momentum of electronic Bloch states in altermagnets are intertwined in a characteristic fashion, which can be of d-wave, g-wave or i-wave form~\cite{Smejkal_2020_Crystal,Smejkal2022b,Hayami_2019_Momentum-Dependent, Naka2019,Ahn_2019_Antiferromagnetism, Reimers2024}. In particular the d-wave altermagnets can be expected to show novel exotic current-driven spin functionalities such as unconventional spin torques, spin-orbit torques, and spin-splitter currents~\cite{Bose2022, Bai2022-new, Karube2022-new, GonzalezHernandez2021, Han2024}. Several experimental investigations of these transport effects have detected time-reversal symmetry broken responses that are consistent with or strongly suggestive of altermagnetic properties and functionalities~\cite{naka_2020_anomalous,Yuan_2020_Giant,naka2021,Feng2022,Reichlova2024-new,Wang2023a,Kluczyk2023,Betancourt2023, Samanta2020-new, shao2021, hariki2024xray}. Using photoemission spectroscopy makes it possible to image directly the alternating spin splitting in the band structure. This has been achieved for samples of MnTe and CrSb, thereby proving the existence of altermagnetism in these materials~\cite{Krempasky2024,Lee2024a, Osumi2024, Hajlaoui2024,Reimers2024}.

\begin{figure}[b]
    \centering
    \includegraphics[width=\linewidth]{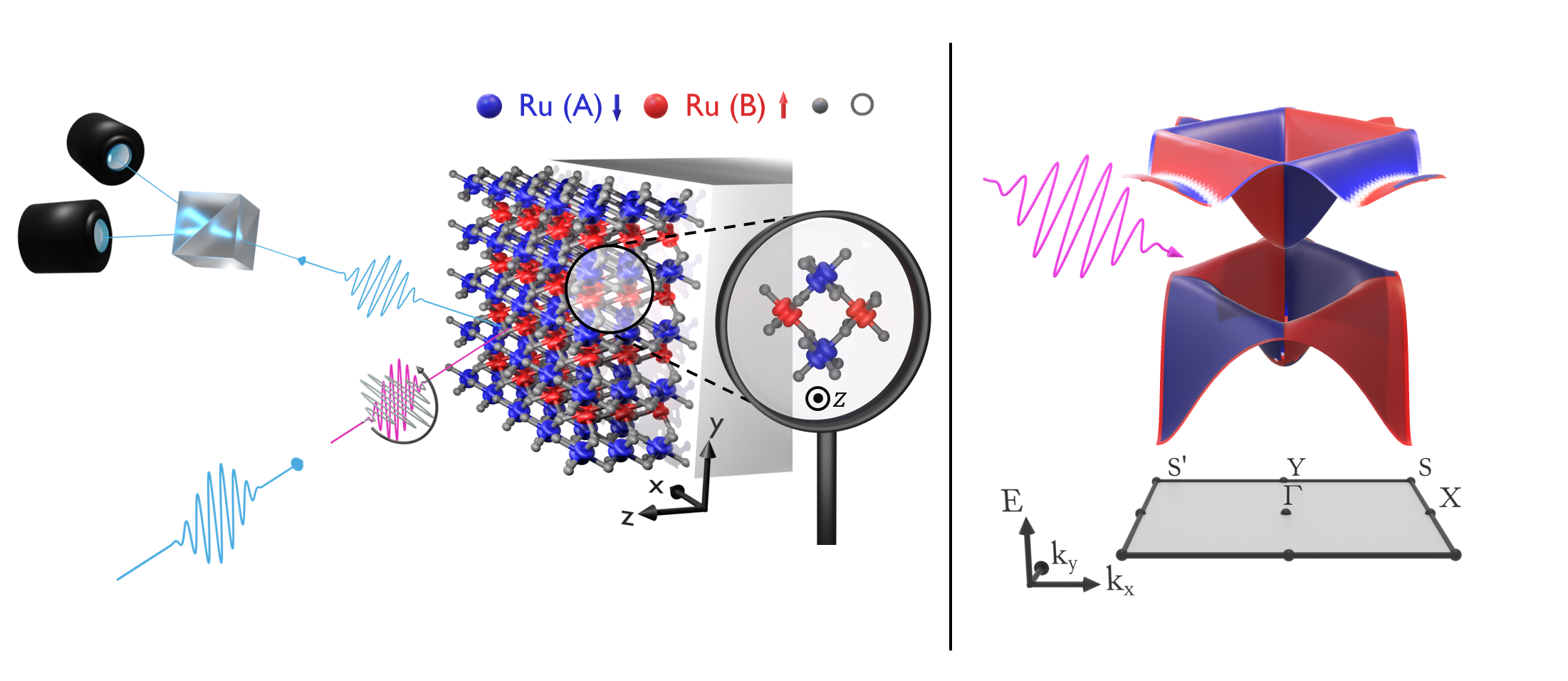}
    \caption{\textbf{Probing the optical response of an altermagnet} Left hand side: Sketch of experimental pump-probe setup. The blue pulses represent the probe laser pulse before and after reflection on the sample consisting of a 3.5\,nm thick layer of RuO$_2$ grown on a TiO$_2$ substrate. The red pulse represents the pump whose linear polarization can be rotated. Note that the polarization plane of the reflected blue pulse is changed due to the magneto-optical Kerr effect. Right hand side: Part of the band structure accessible by the pump pulse together with the geometry of the $k_x-k_y$ 2-dimensional Brillouin zone slice at $k_z=0$, i.e., cut through the $\Gamma$ point.}
    \label{fig:1}
\end{figure}

We focus here on the ultrafast optical response of altermagnets as a purely optical way to reveal possible signatures of altermagnets in their excited states. We study here RuO$_2$ as a prototypical candidate material for an optically active altermagnet with a characteristic altermagnetic spin splitting, which is shown for a slice of the Brillouin zone in Figure~\ref{fig:1}. As this particular planar d-wave Fermi surface geometry supports a spin-splitter effect for electronic currents it has received significant attention in the last few years~\cite{Feng2022, Bai2022-new, Bose2022, Liu2023-new, Fedchenko2024, Zhu2024, Lin2024-arxiv, Smejkal2022GMR, Smejkal2022a, Smejkal2022b}. 

Despite the recently reported experimental and theoretical indications for the absence of magnetic order in clean bulk RuO$_2$ and thick films \cite{Smolyanyuk2024, Liu2024a, Kessler2024, Hiraishi2024a}, there has been mounting experimental evidence for altermagnetic properties in ultrathin RuO$_2$ of a few nanometer thickness which gradually disappear with increasing film thickness~\cite{he2025evidence,jeong2025metallicity,akashdeep2025interface,akashdeep2025surface}. 

In the present manuscript, we first show theoretically for the altermagnetic phase of this material the existence of a spin-splitter effect in the optical domain, in which a linearly polarized optical pump field induces a spin polarization in the d-wave altermagnet whose polarization depends on the direction of the E-field vector of the transverse pump beam. This is substantially different from conventional magnets, where one has to employ a higher fluence in order to change the electronic spin polarization via an optically excited pronounced non-equilibrium in the electron system that triggers energy and angular momentum transfer processes to the lattice and the spin system ~\cite{Beaurepaire.1996,boeglin_distinguishing_2010,tengdin_critical_2018,koopmans_explaining_2010}. Our predictions are then corroborated by experiments on RuO$_2$ films of different thickness. Using the experimental set-up depicted schematically in Figure~\ref{fig:1}, we detect a signal in the ultrathin film of RuO$_2$ that, in accordance with the theoretical predictions, clearly arises from the altermagnetic character of the sample. As this signal vanishes with increasing film thickness, our results also contribute to the ongoing debate regarding the magnetic order of RuO$_2$ by providing evidence of the existence of altermagnetism in ultrathin RuO$_2$ films using a magneto-optical approach specifically adapted to altermagnets. Our experimental finding of optically excited spin polarizations that live up to at least $250\,$fs is in agreement with our recent calculation of spin-dependent electron dynamics in another prototypical planar d-wave material, KRu$_4$O$_8$\cite{weber2025newton}. 

The theoretical and experimental results reported here promise a faster and purely \emph{optical manipulation} of the spin degrees of freedom, which cannot be achieved in ultrafast magneto-optics of conventional ferromagnets because the s-wave character of the Fermi surface prevents such a controlled and selective excitation of spins in different regions of the ferromagnetic band structure. Thus, our study lays the foundation for a new type of ultrafast altermagnetic spintronics that can exploit novel dynamic spin functionalities ranging from ultrafast spin-splitter currents to unconventional torques at interfaces between altermagnets and other spintronic materials.

\section{Results}

\subsection{Model of the optical excitation process}
We begin our exposition with the description of the spin-polarized band structure and single-particle electron states as calculated for the altermagnetic phase of RuO$_2$ by density-functional theory including spin-orbit coupling~\cite{methods}.

\begin{figure}[t]
    \centering
    \includegraphics[width=\linewidth]{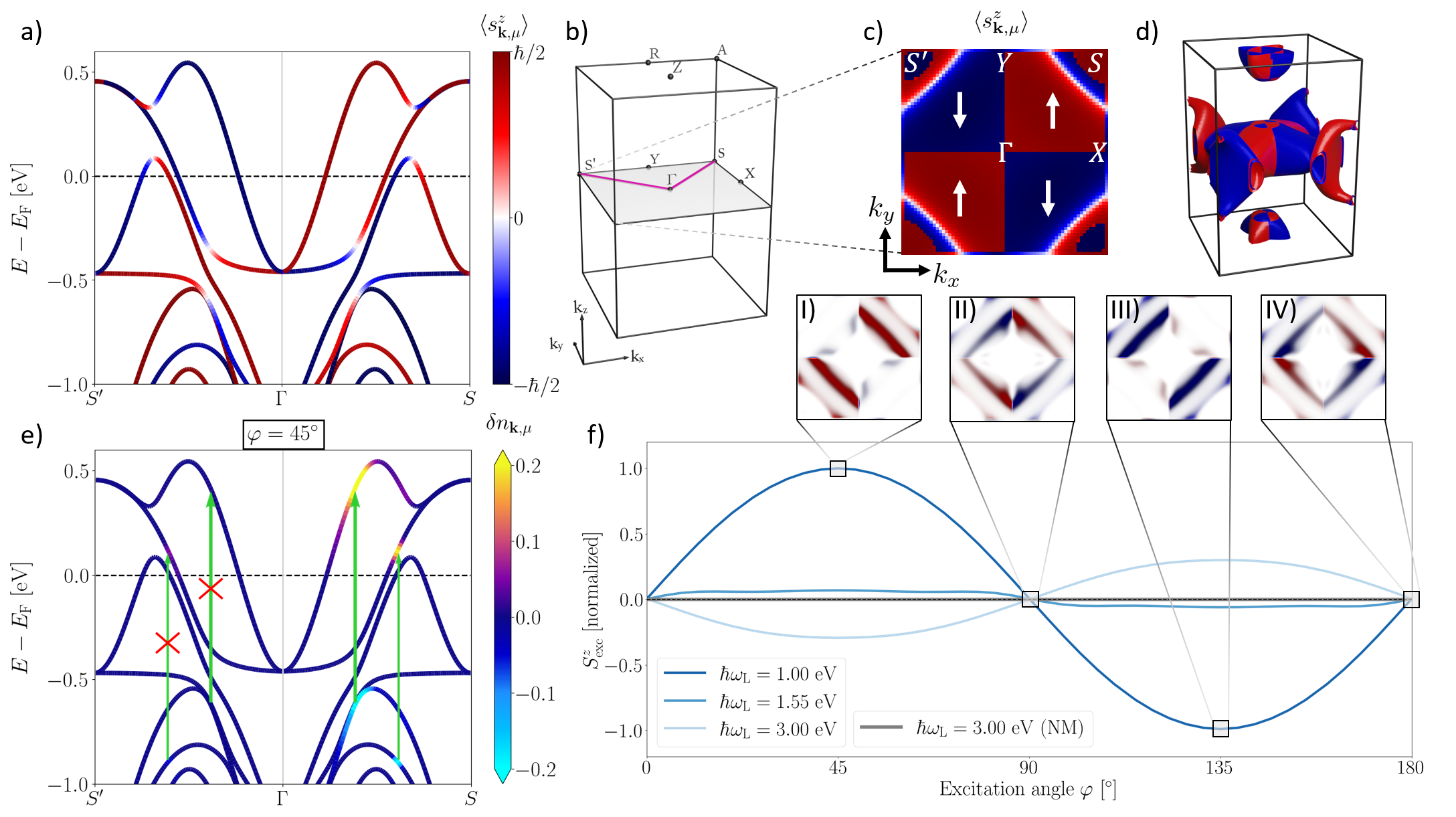}
    \caption{\textbf{Computed electronic transitions and their resulting spin polarization} (A) Band and spin structure of RuO$_2$ along the high symmetry path $S'$-$\Gamma$-$S$, (B) Sketch of the 3D BZ with the high symmetry path used in (A) and (E) indicated by pink lines in the 2-dimensional slice at $k_z = 0$, (C) spin texture of the uppermost altermagnetic band in the greyish shaded 2-dimensional slice in (B), (D) Fermi surface in the 3D BZ, (E) band structure and optical transitions for excitation with angle $45^{\circ}$ at $\hbar\omega_{\mathrm{L}} = 1\,\mathrm{eV}$, (F) excited spin polarization as a function of excitation angle for different pump-photon energies (the blueish curves correspond to results calculated for altermagnetic phase). Note that no spin polarization occurs for the paramagnetic phase (grey curve). The insets above (F) show the spin-resolved change in electronic distribution functions in the 2-dimensional slice of the BZ after excitation with $\hbar\omega = 1\,\mathrm{eV}$ for four different angles. Again, no such spin dependence in the BZ occurs for the paramagnetic system.} 
    \label{fig:2}  
\end{figure}

Figures~\ref{fig:2}A~and~B show the band structure for two high symmetry directions and the 3D Brillouin zone, respectively. The path in Figure~\ref{fig:2}A, along which the band structure is plotted and which corresponds to the pink line in Figure~\ref{fig:2}B, connects the center of the Brillouin zone ($\Gamma$-point) with the $S$ and $S'$ points at the Brillouin zone boundary. The color code of the band structure plot represents the spin expectation value of the Bloch states ranging from red (spin up) to blue (spin down). The spin-resolved band structure already reveals the characteristic alternating  spin polarization in the two orthogonal high-symmetry directions typical of a planar d-wave altermagnet such as  RuO$_2$. Figure~\ref{fig:2}C shows the 2-dimensional spin-texture plot of the unoccupied valence bands in the shaded region of the Brillouin zone, cf.~Figure~\ref{fig:2}B with its typical d-wave symmetry. Finally, Figure~\ref{fig:2}D contains a Fermi surface plot that clearly displays the planar character of the calculated structure for altermagnetic RuO$_2$. Cutting the Fermi surface at $k_z=0$ reveals an interlinked shape with two loops pointing in the directions of the maximum spin splitting, which is often shown schematically for the Fermi surface of planar d-wave altermagnets~\cite{Smejkal2022b}. The classification of altermagnetic symmetry in terms of spin groups does not include spin-orbit coupling, so that the electronic Bloch states in an ideal planar d-wave altermagnet are pure spin states. Spin mixed states due to spin-orbit coupling show up as off-white regions in the spin color-map of the Bloch states in Figure~\ref{fig:2}A. The influence of spin-orbit coupling is seen to be largely limited to avoided crossing points and therefore does not strongly affect the altermagnetic characteristics of the band structure. We note that the spin-orbit coupling is needed for the calculation of MOKE spectra, cf.~section~S8 (Supplementary). 

Based on the calculated ab-initio band structure and Bloch states and using a method based on Fermi's Golden Rule, cf.~Eq.~\eqref{optical}, we determine the optically excited spin density via the momentum-resolved dipole-matrix elements for any combination of photon energy and orientation of the linearly polarized pump E-field vector~\cite{methods}.
Figure~\ref{fig:2}E shows the computed change of electron occupations with respect to the ground state for a pump photon energy of $\hbar\omega_{\mathrm{L}} = 1.00\,\mathrm{eV}$ and an angle of $\varphi = 45^{\circ}$ between the polarization vector of the E-field and the [100] axis of the RuO$_2$ lattice. We will refer to this angle simply as ``excitation angle'' to avoid confusion with the spin polarization vector. The plot of the occupation change~$\delta n_{\boldsymbol{k},\mu}$ uses a color code ranging from turquoise (missing electrons below the Fermi level) via blue (no change of $n_{\boldsymbol{k},\mu}$) to yellow (additional electrons above the Fermi level). The pump photon energy of $1.00\,\mathrm{eV}$ leads to two dominant transitions in the band structure in the $\Gamma$--$S$ direction as indicated by the green arrows. Crucially, similar transitions between bands along the $\Gamma$-$S'$ direction with identical energies but opposite spins are dipole forbidden. These momentum-selective optical transitions for \emph{linearly polarized} pump fields can be traced back to the  orbital and spin character of the states involved in the transitions, which are intertwined by the altermagnetic symmetries. In combination with the altermagnetic d-wave spin texture of RuO$_2$ the momentum selectivity leads to an anisotropic k-space distribution of excited carriers with predominant spin-up character, cf.~section S9 (Supplement). 

For RuO$_2$, drive pulses with photon energies of 1 to 3\,eV  excite carriers into bands that exhibit the altermagnetic spin splitting. In order to characterize the impact of this process on the electronic spins, we plot and discuss in the following the spin polarization of the \emph{excited} electrons 
\begin{equation}
    S^{z}_{\text{exc}} = \sum_{\boldsymbol{k},\mu,\epsilon_{\boldsymbol{k},\mu}\geq E_{\mathrm{F}}}n_{\boldsymbol{k},\mu} s^z_{\boldsymbol{k}\mu}.
    \label{eq:s-z}
\end{equation}
Here, $s^z_{\boldsymbol{k}\mu}$ is the $z$ component of the spin expectation value of the single-particle Bloch state with momentum $\boldsymbol{k}$ and band index $\mu$. The $z$ direction is the direction of the N\'eel vector. A band resolved analysis (not shown) reveals that the \emph{excited} spin polarization effectively contains contributions only from the two altermagnetic bands close to the Fermi energy. We note that the optical excitation process also leads to a \emph{total} electronic spin polarization, which is obtained from summing over all bands below and above $E_{\mathrm{F}}$. It occurs due to spin-orbit induced spin mixing in the initial and final states of the relevant optical transitions.

Figure~\ref{fig:2}F plots the excited spin polarization~(\ref{eq:s-z}) in dependence of the excitation angle for three different pump-photon energies, including $\hbar\omega_{\mathrm{L}} = 1.00$\,eV discussed above. The spin polarization exhibits a sinusoidal dependence on the excitation angle with identical periodicity for all three excitation energies. The $180^{\circ}$-periodicity, which corresponds to a 2-fold symmetry, results from the combination of a linearly polarized optical pump field with the d-wave character of the altermagnetic band structure, as shown in Figure~\ref{fig:2}C~and~D. The changes in amplitude, including the sign of the spin polarization, reflect the details of the RuO$_2$ band structure, i.e., the energy position and the degree of spin-mixing of the bands involved in the optical transitions. In order to demonstrate the decisive importance of the altermagnetic symmetries for this result, we also calculated the optical excitation for a non-magnetic phase of RuO$_2$, and show the resulting spin-polarization for one of the pump-photon energies in Figure~\ref{fig:2}F. In this numerical test case, only a negligibly small spin polarization and, more importantly, no sign change is obtained, cf.~section~S10 (Supplement). The absence of any excitation induced $180^{\circ}$ periodicity for the paramagnetic as compared to the altermagnetic phase shows that an excited spin polarization dependent on the linearly polarized optical pulse is a characteristic property of the altermagnetic phase.

A $180^\circ$-periodicity is also visible in the change in excited carrier distributions in $k$-space shown in the insets (I)--(IV) above Figure~\ref{fig:2}F for different orientations of the light polarization vector and a photon energy of 1\,eV. The brightness of the insets indicates the magnitude of the occupation change and blue/red colors are used for the spin-character of the Bloch states at the corresponding $k$-points. In more detail, dark blue/red indicates a strong excitation of spin-down/spin-up electrons, respectively, while white indicates no excitation of electrons. 
For all excitation angles, we find a pronounced anisotropy of the excited carrier distributions in $k$-space. The anisotropy axes are determined by the excitation angle of the pump pulses. The twofold symmetry of the excited carrier distribution is created by a combination of (i) the altermagnetic band structure and (ii) the momentum-selective optical transition probabilities. Moreover, the characteristic spin-momentum locking of d-wave altermagnets is responsible for the distinct spin polarizations of the $k$-space carrier distributions for different excitation angles, as discussed above. 

Focusing on the insets in Figure~\ref{fig:2}F, we find the largest expected absolute values of spin polarization for excitation angles of $45^\circ$ and $135^\circ$, i.e., insets (I) and (III), respectively.  In either case a different spin character is predominantly excited: spin-up for $45^\circ$ and spin-down for $135^\circ$. The excited carrier distributions in $k$-space consist of carriers with both spin characters for essentially all other excitation angles. In particular, for inset (II) at an excitation angle of $90^\circ$, the momentum-selective excitation leads to an anisotropic change in distribution functions for which spin-up and spin-down contributions are balanced, resulting in a vanishing spin polarization. Thus, our theoretical calculations reveal a clear strategy to control the $k$-space distribution and, by means of altermagnetic spin-momentum locking, the spin polarization of excited electrons in a d-wave altermagnet by tuning the excitation angle and the crystal structure. This prediction is supported by time-resolved all-optical pump-probe experiments that we present next. 

\subsection{Experimental observation of excitation angle dependent transient optically induced spin-polarization}

\begin{figure}[b]
    \centering
    \includegraphics[width=\linewidth]{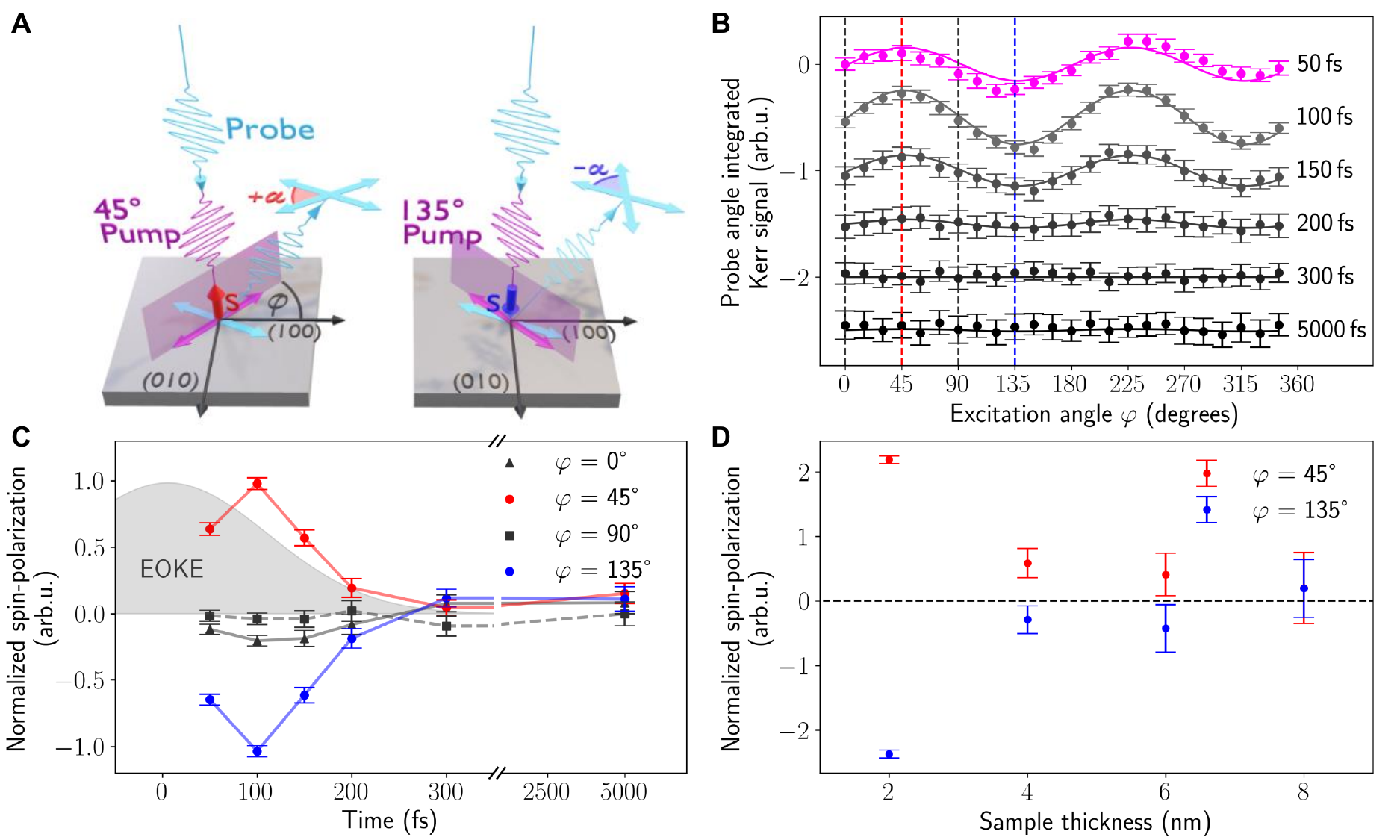}
    \caption{\textbf{Optically induced spin polarization in an altermagnet as detected by MOKE} (A)  Sketch of the time-resolved polar magneto-optical Kerr effect (MOKE) setup. An out-of-plane spin polarization can be induced by the ultrashort pump pulse depending on the excitation angle $\varphi$, which is defined as the angle between the E-field vector of the 1.55\,eV pump beam and the crystallographic [100] direction. The spin polarization is then detected by the 3.10\,eV probe pulse via polar MOKE, with both pulses arriving at the sample under normal incidence.
    The measurement yields an effective rotation of the probe polarization by an angle $\alpha$ which is proportional to the spin polarization. The reflection angle of the probe is exaggerated for clarity. (B)  Magneto-optical signals integrated over probe polarization angle versus the excitation angle $\varphi$ for RuO$_{2}$ (3.5\,nm) at different times after pump. Solid lines represent sinusoidal fits obtained with a fixed 180$^{\circ}$ period. (C) Time dependence of extracted values for 0$^\circ$, 45$0^{\circ}$, 90$^\circ$, 135$^\circ$  from (B) corresponding to the cases in Fig.~\ref{fig:2}F I)-IV). Equivalent angles (e.g., 0$^\circ$ and 180$^\circ$) were averaged to improve the statistics. The grey shaded area indicates pump-probe overlap as measured by the electro-optical Kerr effect (EOKE), cf.~fig.~S1 (Supplement). (D) Spin polarization signal obtained for different thicknesses of RuO$_2$ films at $\approx 100$\,fs for both excitation angles along the directions of maximum spin splitting, i.e., the d-wave loops, in the altermagnetic band structure.}
    \label{fig:3}  
\end{figure}

Figure~\ref{fig:3}A shows a sketch of our experimental procedure and geometries. The polar MOKE geometry provides access to the out-of-plane spin polarization, which is the dominant spin component of the optically excited carriers in RuO$_{2}$. The sample consists of a 3.5\,nm RuO$_{2}$ film grown epitaxially with a (001)-surface orientation by pulsed laser deposition on a TiO${_2}$ substrate~\cite{methods}. The orientation of the RuO$_2$ films is identical to that considered in our theoretical calculations. Similar RuO$_2$ films showed evidence for (altermagnetic) order in the surface region of thin films in recent low-energy muon spin rotation/relaxation \cite{akashdeep2025surface} and core level spectroscopy \cite{Lytvynenko-2026}, which are consistent with the existence of altermagnetic order in the ultrathin-film limit.
We used a time-resolved magneto-optical Kerr technique to investigate signatures of the optically excited spin polarization in RuO$_{2}$. This technique has recently been applied to study spin dynamics in MnTe on ultrafast timescales~\cite{gray2024timeresolved}. All experiments were conducted at room temperature and thus well below the Néel temperature of RuO$_2$ thin films. The sample was optically excited by 1.55\,eV photons, and the transient magneto-optical response was monitored by recording the rotation $\alpha$ of the linearly polarized 3.10\,eV probe beam after reflection from the sample. The key ingredient of our experiment is the near-normal incidence geometry for both the pump and probe beams (less than $2^\circ$ deviation of the pump and probe beam from perfect normal incidence), which allows us to freely rotate the orientation of the light polarization of the pump beam with respect to the crystal axes of the sample. Thus we can precisely control the excitation conditions (represented by the angle $\varphi$) with respect to the altermagnetic spin structure of RuO$_{2}$. Rotating the light polarization relative to the sample is essential, as our theoretical model predicts a clear dependence of the optically generated spin polarization in altermagnetic RuO$_{2}$ on the excitation angle. In particular, as indicated schematically in Figure~\ref{fig:3}A, a complete reversal of the excited spin polarization should occur when the E-field vector of the pump beam is rotated by $\varphi=90^{\circ}$.
Importantly, the free rotation of the probe beam polarization allows us to obtain the true magnetic Kerr response of RuO$_2$ by averaging over several individual Kerr experiments, which are conducted with varying probe-polarization orientations for a fixed direction of the pump polarization. This is necessary due to the strongly anisotropic excited carrier distributions in RuO$_{2}$ after optical excitation. Such anistropic distributions lead to a transient electronic birefringence signal and thus to a non-magnetic Kerr rotation depending on the orientation between the E-field vector of the probe beam with respect to the [100] direction of RuO$_{2}$, cf.~figure~S4 (Supplement). By averaging over several individual Kerr experiments with probe polarization orientations between 0$^\circ$ and 360$^\circ$ for any given direction of the pump polarization, the true magnetic Kerr response is obtained, see also figures~S10 and~S11 (Supplement). Figure~\ref{fig:3}B shows the time dependence of the true magnetic Kerr rotation as obtained from averaging 24 individual Kerr measurements performed at equidistant angles in 15$^\circ$ steps between the E-field of the probe beam and the [100] direction of RuO$_{2}$.
The temporal evolution of the optically induced spin polarization is illustrated in Fig.~\ref{fig:3}B showing the excitation-angle dependent magnetic Kerr signal for different time delays. For short times after the optical excitation, we find a sinusoidal angle dependence of the magnetic Kerr response with fixed angles for the maxima and minima. Crucially, as figure~S5 shows (Supplement), these angles for the minima and maxima are exchanged when changing the position of the sample, which provides clear evidence for the existence of (alter)magnetic domains and thus of the presence of magnetic order in our ultrathin RuO$_2$ films. Irrespective of the individual domains, the amplitudes of these curves decrease for times longer than 100\,fs, indicating a loss of spin polarization via secondary scattering processes~\cite{weber2025newton}. Interestingly, however, these scattering processes do not alter the phase of the sinusoidal spin polarization curves, which suggests a distinct influence of the crystal symmetry and altermagnetic spin structure on the ultrafast scattering processes. 

To visualize the ultrafast dynamics of the spin polarization, we plot in Fig.~\ref{fig:3}C the time dependence of the probe signal for four characteristic excitation angles with respect to the [100] crystallographic direction, as extracted from Figure~\ref{fig:3}B. As a reference, we also display the signature of the time-resolved electro-optical Kerr effect (EOKE), which is a good measure of the temporal broadening of the pump-probe scheme and the overall charge carrier dynamics within RuO$_2$. We observe an increase in spin polarization of opposite sign for excitation angles along the d-wave loops of the altermagnetic band structure (45$^\circ$ and 135$^\circ$) within the first 100\,fs. This is a clear indication of the creation of carrier distributions in RuO$_{2}$ with spin polarizations as given by the states of the altermagnet along this high-symmetry direction. In both cases, the spin polarization decays within the first 300\,fs, which we attribute to the intrinsic timescale of the spin-flip scattering processes in RuO$_{2}$. 
In contrast, no spin polarization is observed for optical excitation along and perpendicular to the [100] direction (0$^\circ$ and 90$^\circ$). This is caused by the equal number of optically excited spin-up and spin-down carriers in these excitation geometries. 

Finally, we investigate the magnitude of the optically generated spin polarization for RuO$_2$ films with thicknesses ranging from $2\,$nm to $8\,$nm. Figure~\ref{fig:3}D summarizes the Kerr signal of the optically generated spin polarization of these films right after optical excitation for excitation angles along the d-wave loops of the altermagnetic band structure (Supplement). We find an identical behavior, i.e., a sign reversal of the spin polarization when changing the excitation geometry for all film thicknesses. However, the magnitude of the spin polarization decreases with increasing film thickness from $2\,$nm to $6\,$nm, before vanishing completely for an $\approx8\,$nm film. This observation is fully consistent with previous experimental reports that altermagnetic order in RuO$_2$ can only be found in the ultrathin film limit below $10\,$nm \cite{he2025evidence,jeong2025metallicity,akashdeep2025interface,akashdeep2025surface} and thus provides additional evidence for the existence of altermagnetic order in our RuO$_2$ samples.

\section{Discussion}

In conclusion, our work has theoretically and experimentally revealed an optical control protocol for generating a transient spin polarization in the otherwise spin-compensated altermagnet RuO$_2$ on ultrafast timescales. The key ingredient is the d-wave nature of the spin-split band structure of the altermagnet RuO$_2$, which allows for a spin response to the band and momentum selective excitation of hot carriers by linearly polarized light pulses, leading to an effective non-equilibrium direct optical spectroscopy of the altermagnetic band structure. This allows us to \emph{control} the sign and magnitude of the optically induced spin polarization on the femtosecond timescale by the orientation of the light polarization vector with respect to the two opposite spin subsystems of the altermagnet. 

The generality of our findings for all d-wave altermagnets will not only set the stage for the realization of new spin functionalities on ultrafast timescales. They are also crucial for the exploration of ultrafast non-equilibrium spin phenomena in the new class of altermagnetic materials. In particular, the ability to directly and efficiently excite spin polarized carriers by optical pulses in compensated magnets is an essential prerequisite for the study of energy and angular momentum dissipation processes in these materials, and ultimately for the transformation of these highly excited non-equilibrium states into the predicted chiral magnon modes that should dominate the magnetization dynamics of altermagnets on longer time scales.

\acknowledgments

This work was funded by the Deutsche Forschungsgemeinschaft (DFG, German Research Foundation) via - TRR 173 - 268565370 Spin + X: spin in its collective environment (projects A01, A02, A03, A08, B02 and B03) and TRR 288 Elasto-Q-Mat  - 422213477 (project A12).
T. Kikkawa and E. Saitoh are supported by JST CREST (JPMJCR20C1 and JPMJCR20T2), Grant-in-Aid for Scientific Research (Grants No. JP19H05600 and JP24K01326), and Grant-in-Aid for Transformative Research Areas (Grant No. JP22H05114) from JSPS KAKENHI, MEXT Initiative to Establish Next-generation Novel Integrated Circuits Centers (X-NICS) (Grant No. JPJ011438), Japan, and the Institute for AI and Beyond of the University of Tokyo.
R. Ramos is supported by grants RYC 2019-026915-I and CNS2024-154679 funded by the MCIN/AEI/10.13039/501100011033 and by the ESF investing in your future and the European Union NextGenerationEU/PRTR,  the Xunta de Galicia (ED431F 2022/04, ED431C 2024/05, Centro Singular de Investigación do Sistema universitario de Galicia Accreditation 2023-2027, ED431G 2023/03) and the European Union (European Regional Development Fund - ERDF).

\appendix

\section{Methods}


\subsection{Density Functional Theory}

The structural optimization is done using the Vienna Ab initio Simulation Package (VASP) \cite{vasp,vasp2,vasp3} within the Perdew-Burke-Ernzerhof (PBE) generalized gradient approximation for the exchange correlation (XC) functional \cite{PBE}. The cutoff energy is set to 500 eV and we use a Monkhorst pack with $22\times 22\times 32$ $k$-points in the irreducible Brillouin zone. Furthermore, we apply a Hubbard correction to account for the strongly correlated $d$-electrons of ruthenium. For this we use the rotationally invariant version by Dudarev et al. \cite{dudarev} with an effective Hubbard parameter of $U_{\mathrm{eff}}=1.6$ eV. We optimize the atomic positions for the ionic configuration with lattice constants $a=b=4.5331\,$\AA and $c=3.1241$\,\AA, as introduced in Ref.~\cite{Smejkal2022GMR}. 

To determine the electronic ground state properties we apply the full potential linearized augmented plane wave method as implemented in the Elk code~\cite{elk-code}. We use the same $k$-grid and XC-functional as introduced before. Cutoff parameters for the angular momentum expansion, additional empty states etc.\ are set to match the preset ``highq'' settings of the Elk code. The DFT+U calculations are carried out in the fully localized limit (FLL). The Hubbard parameters are adjusted to $U=1.85$ eV and $J=0$ eV, which correspond well to the band structure of \v{S}mejkal et al.~\cite{Smejkal2022GMR}. We use the converged result for the 3D-Brillouin zone to interpolate the band dispersion and dipole matrix elements on a two-dimensional slice of the $k$-space in the $(k_x,k_y)$-plane at $k_z=0$.

\subsection{Optical Response from Fermi's Golden Rule}

The optical response is calculated by a Fermi's golden rule approach \cite{essert_electron-phonon_2011,Stiehl2022}, which describes the change of the occupations $n_{\boldsymbol{k},\mu}$ at each $\boldsymbol{k}$-point in band $\mu$ as
\begin{equation}
       \frac{\partial n_{\boldsymbol{k},\mu}}{\partial t} \big\vert_{\mathrm{opt}}=\frac{2 \pi}{\hbar} \sum_{\nu}\big\vert \mathbf{d}^{\mu\nu}_{\boldsymbol{k}} \cdot \mathbf{E}(t) \big \vert^2 g(\epsilon^{\mu}_{\boldsymbol{k}}-\epsilon^{\nu}_{\boldsymbol{k}}-\hbar\omega_{\mathrm{L}})
	\big[n_{\boldsymbol{k},\nu}-n_{\boldsymbol{k},\mu}  \big] \,,  
\label{optical}
\end{equation}
where $\epsilon_{\boldsymbol{k}}^{\mu}$ denotes the energy of a single particle state $\psi_{\boldsymbol{k}}^{\mu}$ and $    \mathbf{d}^{\mu\nu}_{\boldsymbol{k}}= \big \langle \psi_{\boldsymbol{k}}^{\mu}|e \mathbf{r}|\psi_{\boldsymbol{k}}^{\nu}  \big \rangle
$ the dipole-matrix element which connects two states.
The laser frequency $\omega_{\mathrm{L}}$ and the corresponding electric field amplitude $\mathbf{E}(t)$ together with the spectral profile $g(\epsilon)$ will then define the optical properties. To account for the angle dependence of the electric field we choose the expression
\begin{equation}
    \mathbf{E}(t)=E_0\exp\Big[-4\ln(2)\frac{t^2}{\tau_{\mathrm{FWHM}}^2}\Big]\hat{\mathbf{e}}_{\varphi}
\end{equation}
with the polarization vector $\hat{\mathbf{e}}_{\varphi}=(\cos\varphi,\sin\varphi,0)^T$ and the duration $\tau_{\mathrm{FWHM}}=40$ fs for the laser pulse. The pulse's spectral profile is given by the gaussian
\begin{equation}
    g(\epsilon_{\boldsymbol{k}}^{\mu}-\epsilon_{\boldsymbol{k}}^{\nu}-\hbar\omega_{\mathrm{L}})=\frac{1}{\sqrt{2\pi}}\frac{\sqrt{4\ln(2)}}{\Gamma}\exp\left\lbrace-4\ln(2)\frac{(\vert\epsilon_{\boldsymbol{k}}^{\mu}-\epsilon_{\boldsymbol{k}}^{\nu}\vert-\hbar\omega_{\mathrm{L}})^2}{\Gamma^2}\right\rbrace
\end{equation}
where $\Gamma=100$ meV describes the energetic smearing of the fs laser pulse. 

\subsection{Sample growth and characterization}

Epitaxial RuO$_2$(001) films with different thicknesses ranging from $2\,$nm to $8\,$nm were grown by pulsed laser deposition on TiO$_2$(001) substrates that were heated during deposition to 400$^\circ$\,C and 350$^\circ$\,C respectively. The oxygen pressure was 0.02\,mbar, and typical growth rates were 0.2\,nm per minut and 1\,nm per minute with the KrF excimer laser running at 10\,Hz and 150\,mJ pulse energy. All samples have been characterized by in-situ reflective high-energy electron diffraction (RHEED), and the existence of the RHEED scattering pattern shows that crystallinity persists up to the surface. High quality epitaxial NiO(111) thin films were grown on 0.5\,mm thick MgO(11) substrates by reactive sputtering of a Ni target in mixed Ar (flow 15 sccm) and O$_2$ (flow 1.5 sccm) atmosphere, at $430^\circ$. The Pt capping layer was deposited at room temperature after cooling down the samples in a vacuum. See section~S1 for further details (Supplement).

\section*{Author Contributions}

M.W., S.W., K.L., B.S. and H.C.S. conceived the idea and proposed the model and experimental design. M.A., B.S. supervised and S.W., P.H. carried out the magneto-optical measurements. M.W., L.F.H., K.L. performed the band structure and optical absorption calculations with help from L.S. and J.S.; A.A. and G.J. prepared and characterized the RuO$_2$ samples; C.S., T.K., R.R., E.S., and M.K. were responsible for the growth and characterization of the NiO sample. M.W., S.W., L.F.H., P.H., J.S., B.S., H.C.S. co-wrote the manuscript. All authors read and commented on the manuscript.


\begin{thebibliography}{56}%
\makeatletter
\providecommand \@ifxundefined [1]{%
 \@ifx{#1\undefined}
}%
\providecommand \@ifnum [1]{%
 \ifnum #1\expandafter \@firstoftwo
 \else \expandafter \@secondoftwo
 \fi
}%
\providecommand \@ifx [1]{%
 \ifx #1\expandafter \@firstoftwo
 \else \expandafter \@secondoftwo
 \fi
}%
\providecommand \natexlab [1]{#1}%
\providecommand \enquote  [1]{``#1''}%
\providecommand \bibnamefont  [1]{#1}%
\providecommand \bibfnamefont [1]{#1}%
\providecommand \citenamefont [1]{#1}%
\providecommand \href@noop [0]{\@secondoftwo}%
\providecommand \href [0]{\begingroup \@sanitize@url \@href}%
\providecommand \@href[1]{\@@startlink{#1}\@@href}%
\providecommand \@@href[1]{\endgroup#1\@@endlink}%
\providecommand \@sanitize@url [0]{\catcode `\\12\catcode `\$12\catcode
  `\&12\catcode `\#12\catcode `\^12\catcode `\_12\catcode `\%12\relax}%
\providecommand \@@startlink[1]{}%
\providecommand \@@endlink[0]{}%
\providecommand \url  [0]{\begingroup\@sanitize@url \@url }%
\providecommand \@url [1]{\endgroup\@href {#1}{\urlprefix }}%
\providecommand \urlprefix  [0]{URL }%
\providecommand \Eprint [0]{\href }%
\providecommand \doibase [0]{https://doi.org/}%
\providecommand \selectlanguage [0]{\@gobble}%
\providecommand \bibinfo  [0]{\@secondoftwo}%
\providecommand \bibfield  [0]{\@secondoftwo}%
\providecommand \translation [1]{[#1]}%
\providecommand \BibitemOpen [0]{}%
\providecommand \bibitemStop [0]{}%
\providecommand \bibitemNoStop [0]{.\EOS\space}%
\providecommand \EOS [0]{\spacefactor3000\relax}%
\providecommand \BibitemShut  [1]{\csname bibitem#1\endcsname}%
\let\auto@bib@innerbib\@empty
\bibitem [{\citenamefont {Šmejkal}\ \emph {et~al.}(2020)\citenamefont
  {Šmejkal}, \citenamefont {González-Hernández}, \citenamefont {Jungwirth},\
  and\ \citenamefont {Sinova}}]{Smejkal_2020_Crystal}%
  \BibitemOpen
  \bibfield  {author} {\bibinfo {author} {\bibfnamefont {L.}~\bibnamefont
  {Šmejkal}}, \bibinfo {author} {\bibfnamefont {R.}~\bibnamefont
  {González-Hernández}}, \bibinfo {author} {\bibfnamefont {T.}~\bibnamefont
  {Jungwirth}},\ and\ \bibinfo {author} {\bibfnamefont {J.}~\bibnamefont
  {Sinova}},\ }\bibfield  {title} {\bibinfo {title} {Crystal time-reversal
  symmetry breaking and spontaneous hall effect in collinear
  antiferromagnets},\ }\href@noop {} {\bibfield  {journal} {\bibinfo  {journal}
  {Science Advances}\ }\textbf {\bibinfo {volume} {6}},\ \bibinfo {pages}
  {eaaz8809} (\bibinfo {year} {2020})}\BibitemShut {NoStop}%
\bibitem [{\citenamefont {Šmejkal}\ \emph {et~al.}(2022)\citenamefont
  {Šmejkal}, \citenamefont {Sinova},\ and\ \citenamefont
  {Jungwirth}}]{Smejkal2022b}%
  \BibitemOpen
  \bibfield  {author} {\bibinfo {author} {\bibfnamefont {L.}~\bibnamefont
  {Šmejkal}}, \bibinfo {author} {\bibfnamefont {J.}~\bibnamefont {Sinova}},\
  and\ \bibinfo {author} {\bibfnamefont {T.}~\bibnamefont {Jungwirth}},\
  }\bibfield  {title} {\bibinfo {title} {Beyond conventional ferromagnetism and
  antiferromagnetism: A phase with nonrelativistic spin and crystal rotation
  symmetry},\ }\href@noop {} {\bibfield  {journal} {\bibinfo  {journal}
  {Physical Review X}\ }\textbf {\bibinfo {volume} {12}},\ \bibinfo {pages}
  {031042} (\bibinfo {year} {2022})}\BibitemShut {NoStop}%
\bibitem [{\citenamefont {Hayami}\ \emph {et~al.}(2019)\citenamefont {Hayami},
  \citenamefont {Yanagi},\ and\ \citenamefont
  {Kusunose}}]{Hayami_2019_Momentum-Dependent}%
  \BibitemOpen
  \bibfield  {author} {\bibinfo {author} {\bibfnamefont {S.}~\bibnamefont
  {Hayami}}, \bibinfo {author} {\bibfnamefont {Y.}~\bibnamefont {Yanagi}},\
  and\ \bibinfo {author} {\bibfnamefont {H.}~\bibnamefont {Kusunose}},\
  }\bibfield  {title} {\bibinfo {title} {Momentum-dependent spin splitting by
  collinear antiferromagnetic ordering},\ }\href@noop {} {\bibfield  {journal}
  {\bibinfo  {journal} {Journal of the Physical Society of Japan}\ }\textbf
  {\bibinfo {volume} {88}},\ \bibinfo {pages} {123702} (\bibinfo {year}
  {2019})}\BibitemShut {NoStop}%
\bibitem [{\citenamefont {Naka}\ \emph {et~al.}(2019)\citenamefont {Naka},
  \citenamefont {Hayami}, \citenamefont {Kusunose}, \citenamefont {Yanagi},
  \citenamefont {Motome},\ and\ \citenamefont {Seo}}]{Naka2019}%
  \BibitemOpen
  \bibfield  {author} {\bibinfo {author} {\bibfnamefont {M.}~\bibnamefont
  {Naka}}, \bibinfo {author} {\bibfnamefont {S.}~\bibnamefont {Hayami}},
  \bibinfo {author} {\bibfnamefont {H.}~\bibnamefont {Kusunose}}, \bibinfo
  {author} {\bibfnamefont {Y.}~\bibnamefont {Yanagi}}, \bibinfo {author}
  {\bibfnamefont {Y.}~\bibnamefont {Motome}},\ and\ \bibinfo {author}
  {\bibfnamefont {H.}~\bibnamefont {Seo}},\ }\bibfield  {title} {\bibinfo
  {title} {Spin current generation in organic antiferromagnets},\ }\href@noop
  {} {\bibfield  {journal} {\bibinfo  {journal} {Nature Communications}\
  }\textbf {\bibinfo {volume} {10}},\ \bibinfo {pages} {4305} (\bibinfo {year}
  {2019})}\BibitemShut {NoStop}%
\bibitem [{\citenamefont {Ahn}\ \emph {et~al.}(2019)\citenamefont {Ahn},
  \citenamefont {Hariki}, \citenamefont {Lee},\ and\ \citenamefont
  {Kune\ifmmode~\check{s}\else \v{s}\fi{}}}]{Ahn_2019_Antiferromagnetism}%
  \BibitemOpen
  \bibfield  {author} {\bibinfo {author} {\bibfnamefont {K.-H.}\ \bibnamefont
  {Ahn}}, \bibinfo {author} {\bibfnamefont {A.}~\bibnamefont {Hariki}},
  \bibinfo {author} {\bibfnamefont {K.-W.}\ \bibnamefont {Lee}},\ and\ \bibinfo
  {author} {\bibfnamefont {J.}~\bibnamefont {Kune\ifmmode~\check{s}\else
  \v{s}\fi{}}},\ }\bibfield  {title} {\bibinfo {title} {Antiferromagnetism in
  {RuO}$_2$ as $d$-wave {Pomeranchuk} instability},\ }\href@noop {} {\bibfield
  {journal} {\bibinfo  {journal} {Phys. Rev. B}\ }\textbf {\bibinfo {volume}
  {99}},\ \bibinfo {pages} {184432} (\bibinfo {year} {2019})}\BibitemShut
  {NoStop}%
\bibitem [{\citenamefont {Reimers}\ \emph {et~al.}(2024)\citenamefont
  {Reimers}, \citenamefont {Odenbreit}, \citenamefont {{\v{S}}mejkal},
  \citenamefont {Strocov}, \citenamefont {Constantinou}, \citenamefont
  {Hellenes}, \citenamefont {{Jaeschke Ubiergo}}, \citenamefont {Campos},
  \citenamefont {Bharadwaj}, \citenamefont {Chakraborty}, \citenamefont
  {Denneulin}, \citenamefont {Shi}, \citenamefont {Dunin-Borkowski},
  \citenamefont {Das}, \citenamefont {Kl{\"{a}}ui}, \citenamefont {Sinova},\
  and\ \citenamefont {Jourdan}}]{Reimers2024}%
  \BibitemOpen
  \bibfield  {author} {\bibinfo {author} {\bibfnamefont {S.}~\bibnamefont
  {Reimers}}, \bibinfo {author} {\bibfnamefont {L.}~\bibnamefont {Odenbreit}},
  \bibinfo {author} {\bibfnamefont {L.}~\bibnamefont {{\v{S}}mejkal}}, \bibinfo
  {author} {\bibfnamefont {V.~N.}\ \bibnamefont {Strocov}}, \bibinfo {author}
  {\bibfnamefont {P.}~\bibnamefont {Constantinou}}, \bibinfo {author}
  {\bibfnamefont {A.~B.}\ \bibnamefont {Hellenes}}, \bibinfo {author}
  {\bibfnamefont {R.}~\bibnamefont {{Jaeschke Ubiergo}}}, \bibinfo {author}
  {\bibfnamefont {W.~H.}\ \bibnamefont {Campos}}, \bibinfo {author}
  {\bibfnamefont {V.~K.}\ \bibnamefont {Bharadwaj}}, \bibinfo {author}
  {\bibfnamefont {A.}~\bibnamefont {Chakraborty}}, \bibinfo {author}
  {\bibfnamefont {T.}~\bibnamefont {Denneulin}}, \bibinfo {author}
  {\bibfnamefont {W.}~\bibnamefont {Shi}}, \bibinfo {author} {\bibfnamefont
  {R.~E.}\ \bibnamefont {Dunin-Borkowski}}, \bibinfo {author} {\bibfnamefont
  {S.}~\bibnamefont {Das}}, \bibinfo {author} {\bibfnamefont {M.}~\bibnamefont
  {Kl{\"{a}}ui}}, \bibinfo {author} {\bibfnamefont {J.}~\bibnamefont
  {Sinova}},\ and\ \bibinfo {author} {\bibfnamefont {M.}~\bibnamefont
  {Jourdan}},\ }\bibfield  {title} {\bibinfo {title} {Direct observation of
  altermagnetic band splitting in {CrSb} thin films},\ }\href@noop {}
  {\bibfield  {journal} {\bibinfo  {journal} {Nature Communications}\ }\textbf
  {\bibinfo {volume} {15}},\ \bibinfo {pages} {2116} (\bibinfo {year}
  {2024})}\BibitemShut {NoStop}%
\bibitem [{\citenamefont {Bose}\ \emph {et~al.}(2022)\citenamefont {Bose},
  \citenamefont {Schreiber}, \citenamefont {Jain}, \citenamefont {Shao},
  \citenamefont {Nair}, \citenamefont {Sun}, \citenamefont {Zhang},
  \citenamefont {Muller}, \citenamefont {Tsymbal}, \citenamefont {Schlom},\
  and\ \citenamefont {Ralph}}]{Bose2022}%
  \BibitemOpen
  \bibfield  {author} {\bibinfo {author} {\bibfnamefont {A.}~\bibnamefont
  {Bose}}, \bibinfo {author} {\bibfnamefont {N.~J.}\ \bibnamefont {Schreiber}},
  \bibinfo {author} {\bibfnamefont {R.}~\bibnamefont {Jain}}, \bibinfo {author}
  {\bibfnamefont {D.-F.}\ \bibnamefont {Shao}}, \bibinfo {author}
  {\bibfnamefont {H.~P.}\ \bibnamefont {Nair}}, \bibinfo {author}
  {\bibfnamefont {J.}~\bibnamefont {Sun}}, \bibinfo {author} {\bibfnamefont
  {X.~S.}\ \bibnamefont {Zhang}}, \bibinfo {author} {\bibfnamefont {D.~A.}\
  \bibnamefont {Muller}}, \bibinfo {author} {\bibfnamefont {E.~Y.}\
  \bibnamefont {Tsymbal}}, \bibinfo {author} {\bibfnamefont {D.~G.}\
  \bibnamefont {Schlom}},\ and\ \bibinfo {author} {\bibfnamefont {D.~C.}\
  \bibnamefont {Ralph}},\ }\bibfield  {title} {\bibinfo {title} {Tilted spin
  current generated by the collinear antiferromagnet ruthenium dioxide},\
  }\href@noop {} {\bibfield  {journal} {\bibinfo  {journal} {Nature
  Electronics}\ }\textbf {\bibinfo {volume} {5}},\ \bibinfo {pages} {267}
  (\bibinfo {year} {2022})}\BibitemShut {NoStop}%
\bibitem [{\citenamefont {Bai}\ \emph {et~al.}(2022)\citenamefont {Bai},
  \citenamefont {Han}, \citenamefont {Feng}, \citenamefont {Zhou},
  \citenamefont {Su}, \citenamefont {Wang}, \citenamefont {Liao}, \citenamefont
  {Zhu}, \citenamefont {Chen}, \citenamefont {Pan}, \citenamefont {Fan},\ and\
  \citenamefont {Song}}]{Bai2022-new}%
  \BibitemOpen
  \bibfield  {author} {\bibinfo {author} {\bibfnamefont {H.}~\bibnamefont
  {Bai}}, \bibinfo {author} {\bibfnamefont {L.}~\bibnamefont {Han}}, \bibinfo
  {author} {\bibfnamefont {X.}~\bibnamefont {Feng}}, \bibinfo {author}
  {\bibfnamefont {Y.}~\bibnamefont {Zhou}}, \bibinfo {author} {\bibfnamefont
  {R.}~\bibnamefont {Su}}, \bibinfo {author} {\bibfnamefont {Q.}~\bibnamefont
  {Wang}}, \bibinfo {author} {\bibfnamefont {L.}~\bibnamefont {Liao}}, \bibinfo
  {author} {\bibfnamefont {W.}~\bibnamefont {Zhu}}, \bibinfo {author}
  {\bibfnamefont {X.}~\bibnamefont {Chen}}, \bibinfo {author} {\bibfnamefont
  {F.}~\bibnamefont {Pan}}, \bibinfo {author} {\bibfnamefont {X.}~\bibnamefont
  {Fan}},\ and\ \bibinfo {author} {\bibfnamefont {C.}~\bibnamefont {Song}},\
  }\bibfield  {title} {\bibinfo {title} {Observation of spin splitting torque
  in a collinear antiferromagnet {RuO}$_2$},\ }\href@noop {} {\bibfield
  {journal} {\bibinfo  {journal} {Physical Review Letters}\ }\textbf {\bibinfo
  {volume} {128}},\ \bibinfo {pages} {197202} (\bibinfo {year}
  {2022})}\BibitemShut {NoStop}%
\bibitem [{\citenamefont {Karube}\ \emph {et~al.}(2022)\citenamefont {Karube},
  \citenamefont {Tanaka}, \citenamefont {Sugawara}, \citenamefont {Kadoguchi},
  \citenamefont {Kohda},\ and\ \citenamefont {Nitta}}]{Karube2022-new}%
  \BibitemOpen
  \bibfield  {author} {\bibinfo {author} {\bibfnamefont {S.}~\bibnamefont
  {Karube}}, \bibinfo {author} {\bibfnamefont {T.}~\bibnamefont {Tanaka}},
  \bibinfo {author} {\bibfnamefont {D.}~\bibnamefont {Sugawara}}, \bibinfo
  {author} {\bibfnamefont {N.}~\bibnamefont {Kadoguchi}}, \bibinfo {author}
  {\bibfnamefont {M.}~\bibnamefont {Kohda}},\ and\ \bibinfo {author}
  {\bibfnamefont {J.}~\bibnamefont {Nitta}},\ }\bibfield  {title} {\bibinfo
  {title} {{Observation of Spin-Splitter Torque in Collinear Antiferromagnetic
  {RuO}$_2$}},\ }\href@noop {} {\bibfield  {journal} {\bibinfo  {journal}
  {Physical Review Letters}\ }\textbf {\bibinfo {volume} {129}},\ \bibinfo
  {pages} {137201} (\bibinfo {year} {2022})}\BibitemShut {NoStop}%
\bibitem [{\citenamefont {González-Hernández}\ \emph
  {et~al.}(2021)\citenamefont {González-Hernández}, \citenamefont {Šmejkal},
  \citenamefont {Výborný}, \citenamefont {Yahagi}, \citenamefont {Sinova},
  \citenamefont {Jungwirth},\ and\ \citenamefont
  {Železný}}]{GonzalezHernandez2021}%
  \BibitemOpen
  \bibfield  {author} {\bibinfo {author} {\bibfnamefont {R.}~\bibnamefont
  {González-Hernández}}, \bibinfo {author} {\bibfnamefont {L.}~\bibnamefont
  {Šmejkal}}, \bibinfo {author} {\bibfnamefont {K.}~\bibnamefont {Výborný}},
  \bibinfo {author} {\bibfnamefont {Y.}~\bibnamefont {Yahagi}}, \bibinfo
  {author} {\bibfnamefont {J.}~\bibnamefont {Sinova}}, \bibinfo {author}
  {\bibfnamefont {T.}~\bibnamefont {Jungwirth}},\ and\ \bibinfo {author}
  {\bibfnamefont {J.}~\bibnamefont {Železný}},\ }\bibfield  {title} {\bibinfo
  {title} {Efficient electrical spin splitter based on nonrelativistic
  collinear antiferromagnetism},\ }\href@noop {} {\bibfield  {journal}
  {\bibinfo  {journal} {Physical Review Letters}\ }\textbf {\bibinfo {volume}
  {126}},\ \bibinfo {pages} {127701} (\bibinfo {year} {2021})}\BibitemShut
  {NoStop}%
\bibitem [{\citenamefont {Han}\ \emph {et~al.}(2024)\citenamefont {Han},
  \citenamefont {Fu}, \citenamefont {Peng}, \citenamefont {Cheng},
  \citenamefont {Dai}, \citenamefont {Liu}, \citenamefont {Li}, \citenamefont
  {Zhang}, \citenamefont {Zhu}, \citenamefont {Bai}, \citenamefont {Zhou},
  \citenamefont {Liang}, \citenamefont {Chen}, \citenamefont {Wang},
  \citenamefont {Chen}, \citenamefont {Yang}, \citenamefont {Zhang},
  \citenamefont {Song}, \citenamefont {Liu},\ and\ \citenamefont
  {Pan}}]{Han2024}%
  \BibitemOpen
  \bibfield  {author} {\bibinfo {author} {\bibfnamefont {L.}~\bibnamefont
  {Han}}, \bibinfo {author} {\bibfnamefont {X.}~\bibnamefont {Fu}}, \bibinfo
  {author} {\bibfnamefont {R.}~\bibnamefont {Peng}}, \bibinfo {author}
  {\bibfnamefont {X.}~\bibnamefont {Cheng}}, \bibinfo {author} {\bibfnamefont
  {J.}~\bibnamefont {Dai}}, \bibinfo {author} {\bibfnamefont {L.}~\bibnamefont
  {Liu}}, \bibinfo {author} {\bibfnamefont {Y.}~\bibnamefont {Li}}, \bibinfo
  {author} {\bibfnamefont {Y.}~\bibnamefont {Zhang}}, \bibinfo {author}
  {\bibfnamefont {W.}~\bibnamefont {Zhu}}, \bibinfo {author} {\bibfnamefont
  {H.}~\bibnamefont {Bai}}, \bibinfo {author} {\bibfnamefont {Y.}~\bibnamefont
  {Zhou}}, \bibinfo {author} {\bibfnamefont {S.}~\bibnamefont {Liang}},
  \bibinfo {author} {\bibfnamefont {C.}~\bibnamefont {Chen}}, \bibinfo {author}
  {\bibfnamefont {Q.}~\bibnamefont {Wang}}, \bibinfo {author} {\bibfnamefont
  {X.}~\bibnamefont {Chen}}, \bibinfo {author} {\bibfnamefont {L.}~\bibnamefont
  {Yang}}, \bibinfo {author} {\bibfnamefont {Y.}~\bibnamefont {Zhang}},
  \bibinfo {author} {\bibfnamefont {C.}~\bibnamefont {Song}}, \bibinfo {author}
  {\bibfnamefont {J.}~\bibnamefont {Liu}},\ and\ \bibinfo {author}
  {\bibfnamefont {F.}~\bibnamefont {Pan}},\ }\bibfield  {title} {\bibinfo
  {title} {Electrical 180° switching of {Néel} vector in spin-splitting
  antiferromagnet},\ }\href@noop {} {\bibfield  {journal} {\bibinfo  {journal}
  {Science Advances}\ }\textbf {\bibinfo {volume} {10}},\ \bibinfo {pages}
  {eadn0479} (\bibinfo {year} {2024})}\BibitemShut {NoStop}%
\bibitem [{\citenamefont {Naka}\ \emph {et~al.}(2020)\citenamefont {Naka},
  \citenamefont {Hayami}, \citenamefont {Kusunose}, \citenamefont {Yanagi},
  \citenamefont {Motome},\ and\ \citenamefont {Seo}}]{naka_2020_anomalous}%
  \BibitemOpen
  \bibfield  {author} {\bibinfo {author} {\bibfnamefont {M.}~\bibnamefont
  {Naka}}, \bibinfo {author} {\bibfnamefont {S.}~\bibnamefont {Hayami}},
  \bibinfo {author} {\bibfnamefont {H.}~\bibnamefont {Kusunose}}, \bibinfo
  {author} {\bibfnamefont {Y.}~\bibnamefont {Yanagi}}, \bibinfo {author}
  {\bibfnamefont {Y.}~\bibnamefont {Motome}},\ and\ \bibinfo {author}
  {\bibfnamefont {H.}~\bibnamefont {Seo}},\ }\bibfield  {title} {\bibinfo
  {title} {Anomalous {Hall} effect in $\kappa$-type organic antiferromagnets},\
  }\href@noop {} {\bibfield  {journal} {\bibinfo  {journal} {Phys. Rev. B}\
  }\textbf {\bibinfo {volume} {102}},\ \bibinfo {pages} {075112} (\bibinfo
  {year} {2020})}\BibitemShut {NoStop}%
\bibitem [{\citenamefont {Yuan}\ \emph {et~al.}(2020)\citenamefont {Yuan},
  \citenamefont {Wang}, \citenamefont {Luo}, \citenamefont {Rashba},\ and\
  \citenamefont {Zunger}}]{Yuan_2020_Giant}%
  \BibitemOpen
  \bibfield  {author} {\bibinfo {author} {\bibfnamefont {L.-D.}\ \bibnamefont
  {Yuan}}, \bibinfo {author} {\bibfnamefont {Z.}~\bibnamefont {Wang}}, \bibinfo
  {author} {\bibfnamefont {J.-W.}\ \bibnamefont {Luo}}, \bibinfo {author}
  {\bibfnamefont {E.~I.}\ \bibnamefont {Rashba}},\ and\ \bibinfo {author}
  {\bibfnamefont {A.}~\bibnamefont {Zunger}},\ }\bibfield  {title} {\bibinfo
  {title} {Giant momentum-dependent spin splitting in centrosymmetric low-$z$
  antiferromagnets},\ }\href@noop {} {\bibfield  {journal} {\bibinfo  {journal}
  {Phys. Rev. B}\ }\textbf {\bibinfo {volume} {102}},\ \bibinfo {pages}
  {014422} (\bibinfo {year} {2020})}\BibitemShut {NoStop}%
\bibitem [{\citenamefont {Naka}\ \emph {et~al.}(2021)\citenamefont {Naka},
  \citenamefont {Motome},\ and\ \citenamefont {Seo}}]{naka2021}%
  \BibitemOpen
  \bibfield  {author} {\bibinfo {author} {\bibfnamefont {M.}~\bibnamefont
  {Naka}}, \bibinfo {author} {\bibfnamefont {Y.}~\bibnamefont {Motome}},\ and\
  \bibinfo {author} {\bibfnamefont {H.}~\bibnamefont {Seo}},\ }\bibfield
  {title} {\bibinfo {title} {{Perovskite as a spin current generator}},\
  }\href@noop {} {\bibfield  {journal} {\bibinfo  {journal} {Physical Review
  B}\ }\textbf {\bibinfo {volume} {103}},\ \bibinfo {pages} {125114} (\bibinfo
  {year} {2021})}\BibitemShut {NoStop}%
\bibitem [{\citenamefont {Feng}\ \emph {et~al.}(2022)\citenamefont {Feng},
  \citenamefont {Zhou}, \citenamefont {Šmejkal}, \citenamefont {Wu},
  \citenamefont {Zhu}, \citenamefont {Guo}, \citenamefont
  {González-Hernández}, \citenamefont {Wang}, \citenamefont {Yan},
  \citenamefont {Qin}, \citenamefont {Zhang}, \citenamefont {Wu}, \citenamefont
  {Chen}, \citenamefont {Meng}, \citenamefont {Liu}, \citenamefont {Xia},
  \citenamefont {Sinova}, \citenamefont {Jungwirth},\ and\ \citenamefont
  {Liu}}]{Feng2022}%
  \BibitemOpen
  \bibfield  {author} {\bibinfo {author} {\bibfnamefont {Z.}~\bibnamefont
  {Feng}}, \bibinfo {author} {\bibfnamefont {X.}~\bibnamefont {Zhou}}, \bibinfo
  {author} {\bibfnamefont {L.}~\bibnamefont {Šmejkal}}, \bibinfo {author}
  {\bibfnamefont {L.}~\bibnamefont {Wu}}, \bibinfo {author} {\bibfnamefont
  {Z.}~\bibnamefont {Zhu}}, \bibinfo {author} {\bibfnamefont {H.}~\bibnamefont
  {Guo}}, \bibinfo {author} {\bibfnamefont {R.}~\bibnamefont
  {González-Hernández}}, \bibinfo {author} {\bibfnamefont {X.}~\bibnamefont
  {Wang}}, \bibinfo {author} {\bibfnamefont {H.}~\bibnamefont {Yan}}, \bibinfo
  {author} {\bibfnamefont {P.}~\bibnamefont {Qin}}, \bibinfo {author}
  {\bibfnamefont {X.}~\bibnamefont {Zhang}}, \bibinfo {author} {\bibfnamefont
  {H.}~\bibnamefont {Wu}}, \bibinfo {author} {\bibfnamefont {H.}~\bibnamefont
  {Chen}}, \bibinfo {author} {\bibfnamefont {Z.}~\bibnamefont {Meng}}, \bibinfo
  {author} {\bibfnamefont {L.}~\bibnamefont {Liu}}, \bibinfo {author}
  {\bibfnamefont {Z.}~\bibnamefont {Xia}}, \bibinfo {author} {\bibfnamefont
  {J.}~\bibnamefont {Sinova}}, \bibinfo {author} {\bibfnamefont
  {T.}~\bibnamefont {Jungwirth}},\ and\ \bibinfo {author} {\bibfnamefont
  {Z.}~\bibnamefont {Liu}},\ }\bibfield  {title} {\bibinfo {title} {An
  anomalous {Hall} effect in altermagnetic ruthenium dioxide},\ }\href@noop {}
  {\bibfield  {journal} {\bibinfo  {journal} {Nature Electronics}\ }\textbf
  {\bibinfo {volume} {5}},\ \bibinfo {pages} {735} (\bibinfo {year}
  {2022})}\BibitemShut {NoStop}%
\bibitem [{\citenamefont {Reichlová}\ \emph {et~al.}(2024)\citenamefont
  {Reichlová}, \citenamefont {{Lopes Seeger}}, \citenamefont
  {Gonz{\'{a}}lez-Hern{\'{a}}ndez}, \citenamefont {Kounta}, \citenamefont
  {Schlitz}, \citenamefont {Kriegner}, \citenamefont {Ritzinger}, \citenamefont
  {Lammel}, \citenamefont {Leivisk{\"{a}}}, \citenamefont {{Birk Hellenes}},
  \citenamefont {Olejn{\'{i}}k}, \citenamefont {Petři{\v{c}}ek}, \citenamefont
  {Dole{\v{z}}al}, \citenamefont {Horak}, \citenamefont {Schmoranzerova},
  \citenamefont {Badura}, \citenamefont {Bertaina}, \citenamefont {Thomas},
  \citenamefont {Baltz}, \citenamefont {Michez}, \citenamefont {Sinova},
  \citenamefont {Goennenwein}, \citenamefont {Jungwirth},\ and\ \citenamefont
  {{\v{S}}mejkal}}]{Reichlova2024-new}%
  \BibitemOpen
  \bibfield  {author} {\bibinfo {author} {\bibfnamefont {H.}~\bibnamefont
  {Reichlová}}, \bibinfo {author} {\bibfnamefont {R.}~\bibnamefont {{Lopes
  Seeger}}}, \bibinfo {author} {\bibfnamefont {R.}~\bibnamefont
  {Gonz{\'{a}}lez-Hern{\'{a}}ndez}}, \bibinfo {author} {\bibfnamefont
  {I.}~\bibnamefont {Kounta}}, \bibinfo {author} {\bibfnamefont
  {R.}~\bibnamefont {Schlitz}}, \bibinfo {author} {\bibfnamefont
  {D.}~\bibnamefont {Kriegner}}, \bibinfo {author} {\bibfnamefont
  {P.}~\bibnamefont {Ritzinger}}, \bibinfo {author} {\bibfnamefont
  {M.}~\bibnamefont {Lammel}}, \bibinfo {author} {\bibfnamefont
  {M.}~\bibnamefont {Leivisk{\"{a}}}}, \bibinfo {author} {\bibfnamefont
  {A.}~\bibnamefont {{Birk Hellenes}}}, \bibinfo {author} {\bibfnamefont
  {K.}~\bibnamefont {Olejn{\'{i}}k}}, \bibinfo {author} {\bibfnamefont
  {V.}~\bibnamefont {Petři{\v{c}}ek}}, \bibinfo {author} {\bibfnamefont
  {P.}~\bibnamefont {Dole{\v{z}}al}}, \bibinfo {author} {\bibfnamefont
  {L.}~\bibnamefont {Horak}}, \bibinfo {author} {\bibfnamefont
  {E.}~\bibnamefont {Schmoranzerova}}, \bibinfo {author} {\bibfnamefont
  {A.}~\bibnamefont {Badura}}, \bibinfo {author} {\bibfnamefont
  {S.}~\bibnamefont {Bertaina}}, \bibinfo {author} {\bibfnamefont
  {A.}~\bibnamefont {Thomas}}, \bibinfo {author} {\bibfnamefont
  {V.}~\bibnamefont {Baltz}}, \bibinfo {author} {\bibfnamefont
  {L.}~\bibnamefont {Michez}}, \bibinfo {author} {\bibfnamefont
  {J.}~\bibnamefont {Sinova}}, \bibinfo {author} {\bibfnamefont {S.~T.~B.}\
  \bibnamefont {Goennenwein}}, \bibinfo {author} {\bibfnamefont
  {T.}~\bibnamefont {Jungwirth}},\ and\ \bibinfo {author} {\bibfnamefont
  {L.}~\bibnamefont {{\v{S}}mejkal}},\ }\bibfield  {title} {\bibinfo {title}
  {Observation of a spontaneous anomalous {Hall} response in the
  {Mn}$_5${Si}$_3$ d-wave altermagnet candidate},\ }\href@noop {} {\bibfield
  {journal} {\bibinfo  {journal} {Nature Communications}\ }\textbf {\bibinfo
  {volume} {15}},\ \bibinfo {pages} {4961} (\bibinfo {year}
  {2024})}\BibitemShut {NoStop}%
\bibitem [{\citenamefont {Wang}\ \emph {et~al.}(2023)\citenamefont {Wang},
  \citenamefont {Tanaka}, \citenamefont {Sakai}, \citenamefont {Wang},
  \citenamefont {Deng}, \citenamefont {Lyu}, \citenamefont {Li}, \citenamefont
  {Tian}, \citenamefont {Shen}, \citenamefont {Ogawa}, \citenamefont
  {Kanazawa}, \citenamefont {Yu}, \citenamefont {Arita},\ and\ \citenamefont
  {Kagawa}}]{Wang2023a}%
  \BibitemOpen
  \bibfield  {author} {\bibinfo {author} {\bibfnamefont {M.}~\bibnamefont
  {Wang}}, \bibinfo {author} {\bibfnamefont {K.}~\bibnamefont {Tanaka}},
  \bibinfo {author} {\bibfnamefont {S.}~\bibnamefont {Sakai}}, \bibinfo
  {author} {\bibfnamefont {Z.}~\bibnamefont {Wang}}, \bibinfo {author}
  {\bibfnamefont {K.}~\bibnamefont {Deng}}, \bibinfo {author} {\bibfnamefont
  {Y.}~\bibnamefont {Lyu}}, \bibinfo {author} {\bibfnamefont {C.}~\bibnamefont
  {Li}}, \bibinfo {author} {\bibfnamefont {D.}~\bibnamefont {Tian}}, \bibinfo
  {author} {\bibfnamefont {S.}~\bibnamefont {Shen}}, \bibinfo {author}
  {\bibfnamefont {N.}~\bibnamefont {Ogawa}}, \bibinfo {author} {\bibfnamefont
  {N.}~\bibnamefont {Kanazawa}}, \bibinfo {author} {\bibfnamefont
  {P.}~\bibnamefont {Yu}}, \bibinfo {author} {\bibfnamefont {R.}~\bibnamefont
  {Arita}},\ and\ \bibinfo {author} {\bibfnamefont {F.}~\bibnamefont
  {Kagawa}},\ }\bibfield  {title} {\bibinfo {title} {{Emergent zero-field
  anomalous Hall effect in a reconstructed rutile antiferromagnetic metal}},\
  }\href@noop {} {\bibfield  {journal} {\bibinfo  {journal} {Nature
  Communications}\ }\textbf {\bibinfo {volume} {14}},\ \bibinfo {pages} {8240}
  (\bibinfo {year} {2023})}\BibitemShut {NoStop}%
\bibitem [{\citenamefont {Kluczyk}\ \emph {et~al.}(2024)\citenamefont
  {Kluczyk}, \citenamefont {Gas}, \citenamefont {Grzybowski}, \citenamefont
  {Skupi{\'{n}}ski}, \citenamefont {Borysiewicz}, \citenamefont {F{\c{a}}s},
  \citenamefont {Suffczy{\'{n}}ski}, \citenamefont {Domagala}, \citenamefont
  {Grasza}, \citenamefont {Mycielski}, \citenamefont {Baj}, \citenamefont
  {Ahn}, \citenamefont {V{\'{y}}born{\'{y}}}, \citenamefont {Sawicki},\ and\
  \citenamefont {Gryglas-Borysiewicz}}]{Kluczyk2023}%
  \BibitemOpen
  \bibfield  {author} {\bibinfo {author} {\bibfnamefont {K.~P.}\ \bibnamefont
  {Kluczyk}}, \bibinfo {author} {\bibfnamefont {K.}~\bibnamefont {Gas}},
  \bibinfo {author} {\bibfnamefont {M.~J.}\ \bibnamefont {Grzybowski}},
  \bibinfo {author} {\bibfnamefont {P.}~\bibnamefont {Skupi{\'{n}}ski}},
  \bibinfo {author} {\bibfnamefont {M.~A.}\ \bibnamefont {Borysiewicz}},
  \bibinfo {author} {\bibfnamefont {T.}~\bibnamefont {F{\c{a}}s}}, \bibinfo
  {author} {\bibfnamefont {J.}~\bibnamefont {Suffczy{\'{n}}ski}}, \bibinfo
  {author} {\bibfnamefont {J.~Z.}\ \bibnamefont {Domagala}}, \bibinfo {author}
  {\bibfnamefont {K.}~\bibnamefont {Grasza}}, \bibinfo {author} {\bibfnamefont
  {A.}~\bibnamefont {Mycielski}}, \bibinfo {author} {\bibfnamefont
  {M.}~\bibnamefont {Baj}}, \bibinfo {author} {\bibfnamefont {K.~H.}\
  \bibnamefont {Ahn}}, \bibinfo {author} {\bibfnamefont {K.}~\bibnamefont
  {V{\'{y}}born{\'{y}}}}, \bibinfo {author} {\bibfnamefont {M.}~\bibnamefont
  {Sawicki}},\ and\ \bibinfo {author} {\bibfnamefont {M.}~\bibnamefont
  {Gryglas-Borysiewicz}},\ }\href@noop {} {\bibinfo {title} {Coexistence of
  anomalous {Hall} effect and weak net magnetization in collinear
  antiferromagnet {MnTe}}} (\bibinfo {year} {2024})\BibitemShut {NoStop}%
\bibitem [{\citenamefont {Gonzalez~Betancourt}\ \emph
  {et~al.}(2023)\citenamefont {Gonzalez~Betancourt}, \citenamefont
  {Zub{\'{a}}{\v{c}}}, \citenamefont {Gonzalez-Hernandez}, \citenamefont
  {Geishendorf}, \citenamefont {{\v{S}}ob{\'{a}}ň}, \citenamefont
  {Springholz}, \citenamefont {Olejn{\'{i}}k}, \citenamefont {{\v{S}}mejkal},
  \citenamefont {Sinova}, \citenamefont {Jungwirth}, \citenamefont
  {Goennenwein}, \citenamefont {Thomas}, \citenamefont {Reichlov{\'{a}}},
  \citenamefont {{\v{Z}}elezn{\'{y}}},\ and\ \citenamefont
  {Kriegner}}]{Betancourt2023}%
  \BibitemOpen
  \bibfield  {author} {\bibinfo {author} {\bibfnamefont {R.~D.}\ \bibnamefont
  {Gonzalez~Betancourt}}, \bibinfo {author} {\bibfnamefont {J.}~\bibnamefont
  {Zub{\'{a}}{\v{c}}}}, \bibinfo {author} {\bibfnamefont {R.}~\bibnamefont
  {Gonzalez-Hernandez}}, \bibinfo {author} {\bibfnamefont {K.}~\bibnamefont
  {Geishendorf}}, \bibinfo {author} {\bibfnamefont {Z.}~\bibnamefont
  {{\v{S}}ob{\'{a}}ň}}, \bibinfo {author} {\bibfnamefont {G.}~\bibnamefont
  {Springholz}}, \bibinfo {author} {\bibfnamefont {K.}~\bibnamefont
  {Olejn{\'{i}}k}}, \bibinfo {author} {\bibfnamefont {L.}~\bibnamefont
  {{\v{S}}mejkal}}, \bibinfo {author} {\bibfnamefont {J.}~\bibnamefont
  {Sinova}}, \bibinfo {author} {\bibfnamefont {T.}~\bibnamefont {Jungwirth}},
  \bibinfo {author} {\bibfnamefont {S.~T.~B.}\ \bibnamefont {Goennenwein}},
  \bibinfo {author} {\bibfnamefont {A.}~\bibnamefont {Thomas}}, \bibinfo
  {author} {\bibfnamefont {H.}~\bibnamefont {Reichlov{\'{a}}}}, \bibinfo
  {author} {\bibfnamefont {J.}~\bibnamefont {{\v{Z}}elezn{\'{y}}}},\ and\
  \bibinfo {author} {\bibfnamefont {D.}~\bibnamefont {Kriegner}},\ }\bibfield
  {title} {\bibinfo {title} {Spontaneous anomalous {Hall} effect arising from
  an unconventional compensated magnetic phase in a semiconductor},\
  }\href@noop {} {\bibfield  {journal} {\bibinfo  {journal} {Physical Review
  Letters}\ }\textbf {\bibinfo {volume} {130}},\ \bibinfo {pages} {036702}
  (\bibinfo {year} {2023})}\BibitemShut {NoStop}%
\bibitem [{\citenamefont {Samanta}\ \emph {et~al.}(2020)\citenamefont
  {Samanta}, \citenamefont {Le{\v{z}}ai{\'{c}}}, \citenamefont {Merte},
  \citenamefont {Freimuth}, \citenamefont {Bl{\"{u}}gel},\ and\ \citenamefont
  {Mokrousov}}]{Samanta2020-new}%
  \BibitemOpen
  \bibfield  {author} {\bibinfo {author} {\bibfnamefont {K.}~\bibnamefont
  {Samanta}}, \bibinfo {author} {\bibfnamefont {M.}~\bibnamefont
  {Le{\v{z}}ai{\'{c}}}}, \bibinfo {author} {\bibfnamefont {M.}~\bibnamefont
  {Merte}}, \bibinfo {author} {\bibfnamefont {F.}~\bibnamefont {Freimuth}},
  \bibinfo {author} {\bibfnamefont {S.}~\bibnamefont {Bl{\"{u}}gel}},\ and\
  \bibinfo {author} {\bibfnamefont {Y.}~\bibnamefont {Mokrousov}},\ }\bibfield
  {title} {\bibinfo {title} {Crystal {Hall} and crystal magneto-optical effect
  in thin films of {SrRuO}$_3$},\ }\href@noop {} {\bibfield  {journal}
  {\bibinfo  {journal} {Journal of Applied Physics}\ }\textbf {\bibinfo
  {volume} {127}},\ \bibinfo {pages} {213904} (\bibinfo {year}
  {2020})}\BibitemShut {NoStop}%
\bibitem [{\citenamefont {Shao}\ \emph {et~al.}(2021)\citenamefont {Shao},
  \citenamefont {Zhang}, \citenamefont {Li}, \citenamefont {Eom},\ and\
  \citenamefont {Tsymbal}}]{shao2021}%
  \BibitemOpen
  \bibfield  {author} {\bibinfo {author} {\bibfnamefont {D.-F.}\ \bibnamefont
  {Shao}}, \bibinfo {author} {\bibfnamefont {S.-H.}\ \bibnamefont {Zhang}},
  \bibinfo {author} {\bibfnamefont {M.}~\bibnamefont {Li}}, \bibinfo {author}
  {\bibfnamefont {C.-B.}\ \bibnamefont {Eom}},\ and\ \bibinfo {author}
  {\bibfnamefont {E.~Y.}\ \bibnamefont {Tsymbal}},\ }\bibfield  {title}
  {\bibinfo {title} {Spin-neutral currents for spintronics},\ }\href@noop {}
  {\bibfield  {journal} {\bibinfo  {journal} {Nature Communications}\ }\textbf
  {\bibinfo {volume} {12}},\ \bibinfo {pages} {7061} (\bibinfo {year}
  {2021})}\BibitemShut {NoStop}%
\bibitem [{\citenamefont {Hariki}\ \emph {et~al.}(2024)\citenamefont {Hariki},
  \citenamefont {Takahashi},\ and\ \citenamefont
  {Kune{\v{s}}}}]{hariki2024xray}%
  \BibitemOpen
  \bibfield  {author} {\bibinfo {author} {\bibfnamefont {A.}~\bibnamefont
  {Hariki}}, \bibinfo {author} {\bibfnamefont {Y.}~\bibnamefont {Takahashi}},\
  and\ \bibinfo {author} {\bibfnamefont {J.}~\bibnamefont {Kune{\v{s}}}},\
  }\bibfield  {title} {\bibinfo {title} {X-ray magnetic circular dichroism in
  {RuO}$_2$},\ }\href@noop {} {\bibfield  {journal} {\bibinfo  {journal}
  {Physical Review B}\ }\textbf {\bibinfo {volume} {109}},\ \bibinfo {pages}
  {094413} (\bibinfo {year} {2024})}\BibitemShut {NoStop}%
\bibitem [{\citenamefont {Krempask{\'{y}}}\ \emph {et~al.}(2024)\citenamefont
  {Krempask{\'{y}}}, \citenamefont {{\v{S}}mejkal}, \citenamefont {D'Souza},
  \citenamefont {Hajlaoui}, \citenamefont {Springholz}, \citenamefont
  {Uhl{\'{i}}řov{\'{a}}}, \citenamefont {Alarab}, \citenamefont
  {Constantinou}, \citenamefont {Strocov}, \citenamefont {Usanov},
  \citenamefont {Pudelko}, \citenamefont {Gonz{\'{a}}lez-Hern{\'{a}}ndez},
  \citenamefont {{Birk Hellenes}}, \citenamefont {Jansa}, \citenamefont
  {Reichlov{\'{a}}}, \citenamefont {{\v{S}}ob{\'{a}}ň}, \citenamefont
  {{Gonzalez Betancourt}}, \citenamefont {Wadley}, \citenamefont {Sinova},
  \citenamefont {Kriegner}, \citenamefont {Min{\'{a}}r}, \citenamefont {Dil},\
  and\ \citenamefont {Jungwirth}}]{Krempasky2024}%
  \BibitemOpen
  \bibfield  {author} {\bibinfo {author} {\bibfnamefont {J.}~\bibnamefont
  {Krempask{\'{y}}}}, \bibinfo {author} {\bibfnamefont {L.}~\bibnamefont
  {{\v{S}}mejkal}}, \bibinfo {author} {\bibfnamefont {S.~W.}\ \bibnamefont
  {D'Souza}}, \bibinfo {author} {\bibfnamefont {M.}~\bibnamefont {Hajlaoui}},
  \bibinfo {author} {\bibfnamefont {G.}~\bibnamefont {Springholz}}, \bibinfo
  {author} {\bibfnamefont {K.}~\bibnamefont {Uhl{\'{i}}řov{\'{a}}}}, \bibinfo
  {author} {\bibfnamefont {F.}~\bibnamefont {Alarab}}, \bibinfo {author}
  {\bibfnamefont {P.~C.}\ \bibnamefont {Constantinou}}, \bibinfo {author}
  {\bibfnamefont {V.}~\bibnamefont {Strocov}}, \bibinfo {author} {\bibfnamefont
  {D.}~\bibnamefont {Usanov}}, \bibinfo {author} {\bibfnamefont {W.~R.}\
  \bibnamefont {Pudelko}}, \bibinfo {author} {\bibfnamefont {R.}~\bibnamefont
  {Gonz{\'{a}}lez-Hern{\'{a}}ndez}}, \bibinfo {author} {\bibfnamefont
  {A.}~\bibnamefont {{Birk Hellenes}}}, \bibinfo {author} {\bibfnamefont
  {Z.}~\bibnamefont {Jansa}}, \bibinfo {author} {\bibfnamefont
  {H.}~\bibnamefont {Reichlov{\'{a}}}}, \bibinfo {author} {\bibfnamefont
  {Z.}~\bibnamefont {{\v{S}}ob{\'{a}}ň}}, \bibinfo {author} {\bibfnamefont
  {R.~D.}\ \bibnamefont {{Gonzalez Betancourt}}}, \bibinfo {author}
  {\bibfnamefont {P.}~\bibnamefont {Wadley}}, \bibinfo {author} {\bibfnamefont
  {J.}~\bibnamefont {Sinova}}, \bibinfo {author} {\bibfnamefont
  {D.}~\bibnamefont {Kriegner}}, \bibinfo {author} {\bibfnamefont
  {J.}~\bibnamefont {Min{\'{a}}r}}, \bibinfo {author} {\bibfnamefont {J.~H.}\
  \bibnamefont {Dil}},\ and\ \bibinfo {author} {\bibfnamefont {T.}~\bibnamefont
  {Jungwirth}},\ }\bibfield  {title} {\bibinfo {title} {Altermagnetic lifting
  of {Kramers} spin degeneracy},\ }\href@noop {} {\bibfield  {journal}
  {\bibinfo  {journal} {Nature}\ }\textbf {\bibinfo {volume} {626}},\ \bibinfo
  {pages} {517} (\bibinfo {year} {2024})}\BibitemShut {NoStop}%
\bibitem [{\citenamefont {Lee}\ \emph {et~al.}(2024)\citenamefont {Lee},
  \citenamefont {Lee}, \citenamefont {Jung}, \citenamefont {Jung},
  \citenamefont {Kim}, \citenamefont {Lee}, \citenamefont {Seok}, \citenamefont
  {Kim}, \citenamefont {Park}, \citenamefont {{\v{S}}mejkal}, \citenamefont
  {Kang},\ and\ \citenamefont {Kim}}]{Lee2024a}%
  \BibitemOpen
  \bibfield  {author} {\bibinfo {author} {\bibfnamefont {S.}~\bibnamefont
  {Lee}}, \bibinfo {author} {\bibfnamefont {S.}~\bibnamefont {Lee}}, \bibinfo
  {author} {\bibfnamefont {S.}~\bibnamefont {Jung}}, \bibinfo {author}
  {\bibfnamefont {J.}~\bibnamefont {Jung}}, \bibinfo {author} {\bibfnamefont
  {D.}~\bibnamefont {Kim}}, \bibinfo {author} {\bibfnamefont {Y.}~\bibnamefont
  {Lee}}, \bibinfo {author} {\bibfnamefont {B.}~\bibnamefont {Seok}}, \bibinfo
  {author} {\bibfnamefont {J.}~\bibnamefont {Kim}}, \bibinfo {author}
  {\bibfnamefont {B.~G.}\ \bibnamefont {Park}}, \bibinfo {author}
  {\bibfnamefont {L.}~\bibnamefont {{\v{S}}mejkal}}, \bibinfo {author}
  {\bibfnamefont {C.-J.}\ \bibnamefont {Kang}},\ and\ \bibinfo {author}
  {\bibfnamefont {C.}~\bibnamefont {Kim}},\ }\bibfield  {title} {\bibinfo
  {title} {{Broken Kramers Degeneracy in Altermagnetic MnTe}},\ }\href@noop {}
  {\bibfield  {journal} {\bibinfo  {journal} {Physical Review Letters}\
  }\textbf {\bibinfo {volume} {132}},\ \bibinfo {pages} {036702} (\bibinfo
  {year} {2024})}\BibitemShut {NoStop}%
\bibitem [{\citenamefont {Osumi}\ \emph {et~al.}(2024)\citenamefont {Osumi},
  \citenamefont {Souma}, \citenamefont {Aoyama}, \citenamefont {Yamauchi},
  \citenamefont {Honma}, \citenamefont {Nakayama}, \citenamefont {Takahashi},
  \citenamefont {Ohgushi},\ and\ \citenamefont {Sato}}]{Osumi2024}%
  \BibitemOpen
  \bibfield  {author} {\bibinfo {author} {\bibfnamefont {T.}~\bibnamefont
  {Osumi}}, \bibinfo {author} {\bibfnamefont {S.}~\bibnamefont {Souma}},
  \bibinfo {author} {\bibfnamefont {T.}~\bibnamefont {Aoyama}}, \bibinfo
  {author} {\bibfnamefont {K.}~\bibnamefont {Yamauchi}}, \bibinfo {author}
  {\bibfnamefont {A.}~\bibnamefont {Honma}}, \bibinfo {author} {\bibfnamefont
  {K.}~\bibnamefont {Nakayama}}, \bibinfo {author} {\bibfnamefont
  {T.}~\bibnamefont {Takahashi}}, \bibinfo {author} {\bibfnamefont
  {K.}~\bibnamefont {Ohgushi}},\ and\ \bibinfo {author} {\bibfnamefont
  {T.}~\bibnamefont {Sato}},\ }\bibfield  {title} {\bibinfo {title}
  {Observation of a giant band splitting in altermagnetic {MnTe}},\ }\href@noop
  {} {\bibfield  {journal} {\bibinfo  {journal} {Physical Review B}\ }\textbf
  {\bibinfo {volume} {109}},\ \bibinfo {pages} {115102} (\bibinfo {year}
  {2024})}\BibitemShut {NoStop}%
\bibitem [{\citenamefont {Hajlaoui}\ \emph {et~al.}(2024)\citenamefont
  {Hajlaoui}, \citenamefont {{Wilfred D'Souza}}, \citenamefont {{\v{S}}mejkal},
  \citenamefont {Kriegner}, \citenamefont {Krizman}, \citenamefont {Zakusylo},
  \citenamefont {Olszowska}, \citenamefont {Caha}, \citenamefont
  {Michali{\v{c}}ka}, \citenamefont {S{\'{a}}nchez‐Barriga}, \citenamefont
  {Marmodoro}, \citenamefont {V{\'{y}}born{\'{y}}}, \citenamefont {Ernst},
  \citenamefont {Cinchetti}, \citenamefont {Minar}, \citenamefont {Jungwirth},\
  and\ \citenamefont {Springholz}}]{Hajlaoui2024}%
  \BibitemOpen
  \bibfield  {author} {\bibinfo {author} {\bibfnamefont {M.}~\bibnamefont
  {Hajlaoui}}, \bibinfo {author} {\bibfnamefont {S.}~\bibnamefont {{Wilfred
  D'Souza}}}, \bibinfo {author} {\bibfnamefont {L.}~\bibnamefont
  {{\v{S}}mejkal}}, \bibinfo {author} {\bibfnamefont {D.}~\bibnamefont
  {Kriegner}}, \bibinfo {author} {\bibfnamefont {G.}~\bibnamefont {Krizman}},
  \bibinfo {author} {\bibfnamefont {T.}~\bibnamefont {Zakusylo}}, \bibinfo
  {author} {\bibfnamefont {N.}~\bibnamefont {Olszowska}}, \bibinfo {author}
  {\bibfnamefont {O.}~\bibnamefont {Caha}}, \bibinfo {author} {\bibfnamefont
  {J.}~\bibnamefont {Michali{\v{c}}ka}}, \bibinfo {author} {\bibfnamefont
  {J.}~\bibnamefont {S{\'{a}}nchez‐Barriga}}, \bibinfo {author}
  {\bibfnamefont {A.}~\bibnamefont {Marmodoro}}, \bibinfo {author}
  {\bibfnamefont {K.}~\bibnamefont {V{\'{y}}born{\'{y}}}}, \bibinfo {author}
  {\bibfnamefont {A.}~\bibnamefont {Ernst}}, \bibinfo {author} {\bibfnamefont
  {M.}~\bibnamefont {Cinchetti}}, \bibinfo {author} {\bibfnamefont
  {J.}~\bibnamefont {Minar}}, \bibinfo {author} {\bibfnamefont
  {T.}~\bibnamefont {Jungwirth}},\ and\ \bibinfo {author} {\bibfnamefont
  {G.}~\bibnamefont {Springholz}},\ }\bibfield  {title} {\bibinfo {title}
  {{Temperature Dependence of Relativistic Valence Band Splitting Induced by an
  Altermagnetic Phase Transition}},\ }\href@noop {} {\bibfield  {journal}
  {\bibinfo  {journal} {Advanced Materials}\ } (\bibinfo {year}
  {2024})}\BibitemShut {NoStop}%
\bibitem [{\citenamefont {Liu}\ \emph {et~al.}(2023)\citenamefont {Liu},
  \citenamefont {Bai}, \citenamefont {Song}, \citenamefont {Ji}, \citenamefont
  {Lou}, \citenamefont {Zhang}, \citenamefont {Song},\ and\ \citenamefont
  {Jin}}]{Liu2023-new}%
  \BibitemOpen
  \bibfield  {author} {\bibinfo {author} {\bibfnamefont {Y.}~\bibnamefont
  {Liu}}, \bibinfo {author} {\bibfnamefont {H.}~\bibnamefont {Bai}}, \bibinfo
  {author} {\bibfnamefont {Y.}~\bibnamefont {Song}}, \bibinfo {author}
  {\bibfnamefont {Z.}~\bibnamefont {Ji}}, \bibinfo {author} {\bibfnamefont
  {S.}~\bibnamefont {Lou}}, \bibinfo {author} {\bibfnamefont {Z.}~\bibnamefont
  {Zhang}}, \bibinfo {author} {\bibfnamefont {C.}~\bibnamefont {Song}},\ and\
  \bibinfo {author} {\bibfnamefont {Q.}~\bibnamefont {Jin}},\ }\bibfield
  {title} {\bibinfo {title} {Inverse altermagnetic spin splitting
  effect‐induced terahertz emission in {RuO$_2$}},\ }\bibfield  {journal}
  {\bibinfo  {journal} {Advanced Optical Materials}\ }\textbf {\bibinfo
  {volume} {11}},\ \href {https://doi.org/10.1002/adom.202300177}
  {10.1002/adom.202300177} (\bibinfo {year} {2023})\BibitemShut {NoStop}%
\bibitem [{\citenamefont {Fedchenko}\ \emph {et~al.}(2024)\citenamefont
  {Fedchenko}, \citenamefont {Minar}, \citenamefont {Akashdeep}, \citenamefont
  {D'Souza}, \citenamefont {Vasilyev}, \citenamefont {Tkach}, \citenamefont
  {Odenbreit}, \citenamefont {Nguyen}, \citenamefont {Kutnyakhov},
  \citenamefont {Wind}, \citenamefont {Wenthaus}, \citenamefont {Scholz},
  \citenamefont {Rossnagel}, \citenamefont {Hoesch}, \citenamefont
  {Aeschlimann}, \citenamefont {Stadtmueller}, \citenamefont {Klaeui},
  \citenamefont {Schoenhense}, \citenamefont {Jakob}, \citenamefont
  {Jungwirth}, \citenamefont {Smejkal}, \citenamefont {Sinova},\ and\
  \citenamefont {Elmers}}]{Fedchenko2024}%
  \BibitemOpen
  \bibfield  {author} {\bibinfo {author} {\bibfnamefont {O.}~\bibnamefont
  {Fedchenko}}, \bibinfo {author} {\bibfnamefont {J.}~\bibnamefont {Minar}},
  \bibinfo {author} {\bibfnamefont {A.}~\bibnamefont {Akashdeep}}, \bibinfo
  {author} {\bibfnamefont {S.~W.}\ \bibnamefont {D'Souza}}, \bibinfo {author}
  {\bibfnamefont {D.}~\bibnamefont {Vasilyev}}, \bibinfo {author}
  {\bibfnamefont {O.}~\bibnamefont {Tkach}}, \bibinfo {author} {\bibfnamefont
  {L.}~\bibnamefont {Odenbreit}}, \bibinfo {author} {\bibfnamefont {Q.~L.}\
  \bibnamefont {Nguyen}}, \bibinfo {author} {\bibfnamefont {D.}~\bibnamefont
  {Kutnyakhov}}, \bibinfo {author} {\bibfnamefont {N.}~\bibnamefont {Wind}},
  \bibinfo {author} {\bibfnamefont {L.}~\bibnamefont {Wenthaus}}, \bibinfo
  {author} {\bibfnamefont {M.}~\bibnamefont {Scholz}}, \bibinfo {author}
  {\bibfnamefont {K.}~\bibnamefont {Rossnagel}}, \bibinfo {author}
  {\bibfnamefont {M.}~\bibnamefont {Hoesch}}, \bibinfo {author} {\bibfnamefont
  {M.}~\bibnamefont {Aeschlimann}}, \bibinfo {author} {\bibfnamefont
  {B.}~\bibnamefont {Stadtmueller}}, \bibinfo {author} {\bibfnamefont
  {M.}~\bibnamefont {Klaeui}}, \bibinfo {author} {\bibfnamefont
  {G.}~\bibnamefont {Schoenhense}}, \bibinfo {author} {\bibfnamefont
  {G.}~\bibnamefont {Jakob}}, \bibinfo {author} {\bibfnamefont
  {T.}~\bibnamefont {Jungwirth}}, \bibinfo {author} {\bibfnamefont
  {L.}~\bibnamefont {Smejkal}}, \bibinfo {author} {\bibfnamefont
  {J.}~\bibnamefont {Sinova}},\ and\ \bibinfo {author} {\bibfnamefont {H.~J.}\
  \bibnamefont {Elmers}},\ }\bibfield  {title} {\bibinfo {title} {{Observation
  of time-reversal symmetry breaking in the band structure of altermagnetic
  RuO$_2$}},\ }\href@noop {} {\bibfield  {journal} {\bibinfo  {journal}
  {Science Advances}\ }\textbf {\bibinfo {volume} {10}},\ \bibinfo {pages} {31}
  (\bibinfo {year} {2024})}\BibitemShut {NoStop}%
\bibitem [{\citenamefont {Zhu}\ \emph {et~al.}(2024)\citenamefont {Zhu},
  \citenamefont {Chen}, \citenamefont {Liu}, \citenamefont {Liu}, \citenamefont
  {Liu}, \citenamefont {Zha}, \citenamefont {Qu}, \citenamefont {Hong},
  \citenamefont {Li}, \citenamefont {Jiang}, \citenamefont {Ma}, \citenamefont
  {Hao}, \citenamefont {Zhu}, \citenamefont {Liu}, \citenamefont {Zeng},
  \citenamefont {Jayaram}, \citenamefont {Lenger}, \citenamefont {Ding},
  \citenamefont {Mo}, \citenamefont {Tanaka}, \citenamefont {Arita},
  \citenamefont {Liu}, \citenamefont {Ye}, \citenamefont {Shen}, \citenamefont
  {Wrachtrup}, \citenamefont {Huang}, \citenamefont {He}, \citenamefont {Qiao},
  \citenamefont {Liu},\ and\ \citenamefont {Liu}}]{Zhu2024}%
  \BibitemOpen
  \bibfield  {author} {\bibinfo {author} {\bibfnamefont {Y.-P.}\ \bibnamefont
  {Zhu}}, \bibinfo {author} {\bibfnamefont {X.}~\bibnamefont {Chen}}, \bibinfo
  {author} {\bibfnamefont {X.-R.}\ \bibnamefont {Liu}}, \bibinfo {author}
  {\bibfnamefont {Y.}~\bibnamefont {Liu}}, \bibinfo {author} {\bibfnamefont
  {P.}~\bibnamefont {Liu}}, \bibinfo {author} {\bibfnamefont {H.}~\bibnamefont
  {Zha}}, \bibinfo {author} {\bibfnamefont {G.}~\bibnamefont {Qu}}, \bibinfo
  {author} {\bibfnamefont {C.}~\bibnamefont {Hong}}, \bibinfo {author}
  {\bibfnamefont {J.}~\bibnamefont {Li}}, \bibinfo {author} {\bibfnamefont
  {Z.}~\bibnamefont {Jiang}}, \bibinfo {author} {\bibfnamefont {X.-M.}\
  \bibnamefont {Ma}}, \bibinfo {author} {\bibfnamefont {Y.-J.}\ \bibnamefont
  {Hao}}, \bibinfo {author} {\bibfnamefont {M.-Y.}\ \bibnamefont {Zhu}},
  \bibinfo {author} {\bibfnamefont {W.}~\bibnamefont {Liu}}, \bibinfo {author}
  {\bibfnamefont {M.}~\bibnamefont {Zeng}}, \bibinfo {author} {\bibfnamefont
  {S.}~\bibnamefont {Jayaram}}, \bibinfo {author} {\bibfnamefont
  {M.}~\bibnamefont {Lenger}}, \bibinfo {author} {\bibfnamefont
  {J.}~\bibnamefont {Ding}}, \bibinfo {author} {\bibfnamefont {S.}~\bibnamefont
  {Mo}}, \bibinfo {author} {\bibfnamefont {K.}~\bibnamefont {Tanaka}}, \bibinfo
  {author} {\bibfnamefont {M.}~\bibnamefont {Arita}}, \bibinfo {author}
  {\bibfnamefont {Z.}~\bibnamefont {Liu}}, \bibinfo {author} {\bibfnamefont
  {M.}~\bibnamefont {Ye}}, \bibinfo {author} {\bibfnamefont {D.}~\bibnamefont
  {Shen}}, \bibinfo {author} {\bibfnamefont {J.}~\bibnamefont {Wrachtrup}},
  \bibinfo {author} {\bibfnamefont {Y.}~\bibnamefont {Huang}}, \bibinfo
  {author} {\bibfnamefont {R.-H.}\ \bibnamefont {He}}, \bibinfo {author}
  {\bibfnamefont {S.}~\bibnamefont {Qiao}}, \bibinfo {author} {\bibfnamefont
  {Q.}~\bibnamefont {Liu}},\ and\ \bibinfo {author} {\bibfnamefont
  {C.}~\bibnamefont {Liu}},\ }\bibfield  {title} {\bibinfo {title}
  {{Observation of plaid-like spin splitting in a noncoplanar
  antiferromagnet}},\ }\href {https://doi.org/10.1038/s41586-024-07023-w}
  {\bibfield  {journal} {\bibinfo  {journal} {Nature}\ }\textbf {\bibinfo
  {volume} {626}},\ \bibinfo {pages} {523} (\bibinfo {year}
  {2024})}\BibitemShut {NoStop}%
\bibitem [{\citenamefont {Lin}\ \emph {et~al.}(2024)\citenamefont {Lin},
  \citenamefont {Chen}, \citenamefont {Lu}, \citenamefont {Liang},
  \citenamefont {Feng}, \citenamefont {Yamagami}, \citenamefont {Osiecki},
  \citenamefont {Leandersson}, \citenamefont {Thiagarajan}, \citenamefont
  {Liu}, \citenamefont {Felser},\ and\ \citenamefont {Ma}}]{Lin2024-arxiv}%
  \BibitemOpen
  \bibfield  {author} {\bibinfo {author} {\bibfnamefont {Z.}~\bibnamefont
  {Lin}}, \bibinfo {author} {\bibfnamefont {D.}~\bibnamefont {Chen}}, \bibinfo
  {author} {\bibfnamefont {W.}~\bibnamefont {Lu}}, \bibinfo {author}
  {\bibfnamefont {X.}~\bibnamefont {Liang}}, \bibinfo {author} {\bibfnamefont
  {S.}~\bibnamefont {Feng}}, \bibinfo {author} {\bibfnamefont {K.}~\bibnamefont
  {Yamagami}}, \bibinfo {author} {\bibfnamefont {J.}~\bibnamefont {Osiecki}},
  \bibinfo {author} {\bibfnamefont {M.}~\bibnamefont {Leandersson}}, \bibinfo
  {author} {\bibfnamefont {B.}~\bibnamefont {Thiagarajan}}, \bibinfo {author}
  {\bibfnamefont {J.}~\bibnamefont {Liu}}, \bibinfo {author} {\bibfnamefont
  {C.}~\bibnamefont {Felser}},\ and\ \bibinfo {author} {\bibfnamefont
  {J.}~\bibnamefont {Ma}},\ }\href {https://doi.org/10.48550} {\bibinfo {title}
  {{Observation of Giant Spin Splitting and d-wave Spin Texture in Room
  Temperature Altermagnet {RuO}$_2$}}} (\bibinfo {year} {2024})\BibitemShut
  {NoStop}%
\bibitem [{\citenamefont {{\v{S}}mejkal}\ \emph
  {et~al.}(2022{\natexlab{a}})\citenamefont {{\v{S}}mejkal}, \citenamefont
  {Hellenes}, \citenamefont {Gonz{\'{a}}lez-Hern{\'{a}}ndez}, \citenamefont
  {Sinova},\ and\ \citenamefont {Jungwirth}}]{Smejkal2022GMR}%
  \BibitemOpen
  \bibfield  {author} {\bibinfo {author} {\bibfnamefont {L.}~\bibnamefont
  {{\v{S}}mejkal}}, \bibinfo {author} {\bibfnamefont {A.~B.}\ \bibnamefont
  {Hellenes}}, \bibinfo {author} {\bibfnamefont {R.}~\bibnamefont
  {Gonz{\'{a}}lez-Hern{\'{a}}ndez}}, \bibinfo {author} {\bibfnamefont
  {J.}~\bibnamefont {Sinova}},\ and\ \bibinfo {author} {\bibfnamefont
  {T.}~\bibnamefont {Jungwirth}},\ }\bibfield  {title} {\bibinfo {title}
  {{Giant and Tunneling Magnetoresistance in Unconventional Collinear
  Antiferromagnets with Nonrelativistic Spin-Momentum Coupling}},\ }\href@noop
  {} {\bibfield  {journal} {\bibinfo  {journal} {Physical Review X}\ }\textbf
  {\bibinfo {volume} {12}},\ \bibinfo {pages} {011028} (\bibinfo {year}
  {2022}{\natexlab{a}})}\BibitemShut {NoStop}%
\bibitem [{\citenamefont {{\v{S}}mejkal}\ \emph
  {et~al.}(2022{\natexlab{b}})\citenamefont {{\v{S}}mejkal}, \citenamefont
  {Sinova},\ and\ \citenamefont {Jungwirth}}]{Smejkal2022a}%
  \BibitemOpen
  \bibfield  {author} {\bibinfo {author} {\bibfnamefont {L.}~\bibnamefont
  {{\v{S}}mejkal}}, \bibinfo {author} {\bibfnamefont {J.}~\bibnamefont
  {Sinova}},\ and\ \bibinfo {author} {\bibfnamefont {T.}~\bibnamefont
  {Jungwirth}},\ }\bibfield  {title} {\bibinfo {title} {{Emerging Research
  Landscape of Altermagnetism}},\ }\href@noop {} {\bibfield  {journal}
  {\bibinfo  {journal} {Physical Review X}\ }\textbf {\bibinfo {volume} {12}},\
  \bibinfo {pages} {040501} (\bibinfo {year} {2022}{\natexlab{b}})}\BibitemShut
  {NoStop}%
\bibitem [{\citenamefont {Smolyanyuk}\ \emph {et~al.}(2024)\citenamefont
  {Smolyanyuk}, \citenamefont {Mazin}, \citenamefont {Garcia-Gassull},\ and\
  \citenamefont {Valentí}}]{Smolyanyuk2024}%
  \BibitemOpen
  \bibfield  {author} {\bibinfo {author} {\bibfnamefont {A.}~\bibnamefont
  {Smolyanyuk}}, \bibinfo {author} {\bibfnamefont {I.~I.}\ \bibnamefont
  {Mazin}}, \bibinfo {author} {\bibfnamefont {L.}~\bibnamefont
  {Garcia-Gassull}},\ and\ \bibinfo {author} {\bibfnamefont {R.}~\bibnamefont
  {Valentí}},\ }\bibfield  {title} {\bibinfo {title} {Fragility of the
  magnetic order in the prototypical altermagnet {RuO$_2$}},\ }\href@noop {}
  {\bibfield  {journal} {\bibinfo  {journal} {Physical Review B}\ }\textbf
  {\bibinfo {volume} {109}},\ \bibinfo {pages} {134424} (\bibinfo {year}
  {2024})}\BibitemShut {NoStop}%
\bibitem [{\citenamefont {Liu}\ \emph {et~al.}(2024)\citenamefont {Liu},
  \citenamefont {Zhan}, \citenamefont {Li}, \citenamefont {Liu}, \citenamefont
  {Cheng}, \citenamefont {Shi}, \citenamefont {Deng}, \citenamefont {Zhang},
  \citenamefont {Li}, \citenamefont {Ding}, \citenamefont {Jiang},
  \citenamefont {Ye}, \citenamefont {Liu}, \citenamefont {Jiang}, \citenamefont
  {Wang}, \citenamefont {Li}, \citenamefont {Xie}, \citenamefont {Wang},
  \citenamefont {Qiao}, \citenamefont {Wen}, \citenamefont {Sun},\ and\
  \citenamefont {Shen}}]{Liu2024a}%
  \BibitemOpen
  \bibfield  {author} {\bibinfo {author} {\bibfnamefont {J.}~\bibnamefont
  {Liu}}, \bibinfo {author} {\bibfnamefont {J.}~\bibnamefont {Zhan}}, \bibinfo
  {author} {\bibfnamefont {T.}~\bibnamefont {Li}}, \bibinfo {author}
  {\bibfnamefont {J.}~\bibnamefont {Liu}}, \bibinfo {author} {\bibfnamefont
  {S.}~\bibnamefont {Cheng}}, \bibinfo {author} {\bibfnamefont
  {Y.}~\bibnamefont {Shi}}, \bibinfo {author} {\bibfnamefont {L.}~\bibnamefont
  {Deng}}, \bibinfo {author} {\bibfnamefont {M.}~\bibnamefont {Zhang}},
  \bibinfo {author} {\bibfnamefont {C.}~\bibnamefont {Li}}, \bibinfo {author}
  {\bibfnamefont {J.}~\bibnamefont {Ding}}, \bibinfo {author} {\bibfnamefont
  {Q.}~\bibnamefont {Jiang}}, \bibinfo {author} {\bibfnamefont
  {M.}~\bibnamefont {Ye}}, \bibinfo {author} {\bibfnamefont {Z.}~\bibnamefont
  {Liu}}, \bibinfo {author} {\bibfnamefont {Z.}~\bibnamefont {Jiang}}, \bibinfo
  {author} {\bibfnamefont {S.}~\bibnamefont {Wang}}, \bibinfo {author}
  {\bibfnamefont {Q.}~\bibnamefont {Li}}, \bibinfo {author} {\bibfnamefont
  {Y.}~\bibnamefont {Xie}}, \bibinfo {author} {\bibfnamefont {Y.}~\bibnamefont
  {Wang}}, \bibinfo {author} {\bibfnamefont {S.}~\bibnamefont {Qiao}}, \bibinfo
  {author} {\bibfnamefont {J.}~\bibnamefont {Wen}}, \bibinfo {author}
  {\bibfnamefont {Y.}~\bibnamefont {Sun}},\ and\ \bibinfo {author}
  {\bibfnamefont {D.}~\bibnamefont {Shen}},\ }\bibfield  {title} {\bibinfo
  {title} {Absence of altermagnetic spin splitting character in rutile oxide},\
  }\href@noop {} {\bibfield  {journal} {\bibinfo  {journal} {Phys. Rev. Lett.}\
  }\textbf {\bibinfo {volume} {133}},\ \bibinfo {pages} {176401} (\bibinfo
  {year} {2024})}\BibitemShut {NoStop}%
\bibitem [{\citenamefont {Keßler}\ \emph {et~al.}(2024)\citenamefont
  {Keßler}, \citenamefont {Garcia-Gassull}, \citenamefont {Suter},
  \citenamefont {Prokscha}, \citenamefont {Salman}, \citenamefont {Khalyavin},
  \citenamefont {Manuel}, \citenamefont {Orlandi}, \citenamefont {Mazin},
  \citenamefont {Valentı},\ and\ \citenamefont {Moser}}]{Kessler2024}%
  \BibitemOpen
  \bibfield  {author} {\bibinfo {author} {\bibfnamefont {P.}~\bibnamefont
  {Keßler}}, \bibinfo {author} {\bibfnamefont {L.}~\bibnamefont
  {Garcia-Gassull}}, \bibinfo {author} {\bibfnamefont {A.}~\bibnamefont
  {Suter}}, \bibinfo {author} {\bibfnamefont {T.}~\bibnamefont {Prokscha}},
  \bibinfo {author} {\bibfnamefont {Z.}~\bibnamefont {Salman}}, \bibinfo
  {author} {\bibfnamefont {D.}~\bibnamefont {Khalyavin}}, \bibinfo {author}
  {\bibfnamefont {P.}~\bibnamefont {Manuel}}, \bibinfo {author} {\bibfnamefont
  {F.}~\bibnamefont {Orlandi}}, \bibinfo {author} {\bibfnamefont {I.~I.}\
  \bibnamefont {Mazin}}, \bibinfo {author} {\bibfnamefont {R.}~\bibnamefont
  {Valentı}},\ and\ \bibinfo {author} {\bibfnamefont {S.}~\bibnamefont
  {Moser}},\ }\bibfield  {title} {\bibinfo {title} {Absence of magnetic order
  in {RuO$_2$}: insights from $\mu$sr spectroscopy and neutron diffraction},\
  }\href@noop {} {\bibfield  {journal} {\bibinfo  {journal} {npj Spintronics}\
  }\textbf {\bibinfo {volume} {2}},\ \bibinfo {pages} {50} (\bibinfo {year}
  {2024})}\BibitemShut {NoStop}%
\bibitem [{\citenamefont {Hiraishi}\ \emph {et~al.}(2024)\citenamefont
  {Hiraishi}, \citenamefont {Okabe}, \citenamefont {Koda}, \citenamefont
  {Kadono}, \citenamefont {Muroi}, \citenamefont {Hirai},\ and\ \citenamefont
  {Hiroi}}]{Hiraishi2024a}%
  \BibitemOpen
  \bibfield  {author} {\bibinfo {author} {\bibfnamefont {M.}~\bibnamefont
  {Hiraishi}}, \bibinfo {author} {\bibfnamefont {H.}~\bibnamefont {Okabe}},
  \bibinfo {author} {\bibfnamefont {A.}~\bibnamefont {Koda}}, \bibinfo {author}
  {\bibfnamefont {R.}~\bibnamefont {Kadono}}, \bibinfo {author} {\bibfnamefont
  {T.}~\bibnamefont {Muroi}}, \bibinfo {author} {\bibfnamefont
  {D.}~\bibnamefont {Hirai}},\ and\ \bibinfo {author} {\bibfnamefont
  {Z.}~\bibnamefont {Hiroi}},\ }\bibfield  {title} {\bibinfo {title}
  {Nonmagnetic ground state in {RuO$_2$} revealed by muon spin rotation},\
  }\href {https://doi.org/10.1103/PhysRevLett.132.166702} {\bibfield  {journal}
  {\bibinfo  {journal} {Physical Review Letters}\ }\textbf {\bibinfo {volume}
  {132}},\ \bibinfo {pages} {166702} (\bibinfo {year} {2024})}\BibitemShut
  {NoStop}%
\bibitem [{\citenamefont {He}\ \emph {et~al.}(2025)\citenamefont {He},
  \citenamefont {Wen}, \citenamefont {Okabayashi}, \citenamefont {Miura},
  \citenamefont {Ma}, \citenamefont {Ohkubo}, \citenamefont {Seki},
  \citenamefont {Sukegawa},\ and\ \citenamefont {Mitani}}]{he2025evidence}%
  \BibitemOpen
  \bibfield  {author} {\bibinfo {author} {\bibfnamefont {C.}~\bibnamefont
  {He}}, \bibinfo {author} {\bibfnamefont {Z.}~\bibnamefont {Wen}}, \bibinfo
  {author} {\bibfnamefont {J.}~\bibnamefont {Okabayashi}}, \bibinfo {author}
  {\bibfnamefont {Y.}~\bibnamefont {Miura}}, \bibinfo {author} {\bibfnamefont
  {T.}~\bibnamefont {Ma}}, \bibinfo {author} {\bibfnamefont {T.}~\bibnamefont
  {Ohkubo}}, \bibinfo {author} {\bibfnamefont {T.}~\bibnamefont {Seki}},
  \bibinfo {author} {\bibfnamefont {H.}~\bibnamefont {Sukegawa}},\ and\
  \bibinfo {author} {\bibfnamefont {S.}~\bibnamefont {Mitani}},\ }\bibfield
  {title} {\bibinfo {title} {Evidence for single variant in altermagnetic
  {RuO$_2$} (101) thin films},\ }\href@noop {} {\bibfield  {journal} {\bibinfo
  {journal} {Nature Communications}\ }\textbf {\bibinfo {volume} {16}},\
  \bibinfo {pages} {8235} (\bibinfo {year} {2025})}\BibitemShut {NoStop}%
\bibitem [{\citenamefont {Jeong}\ \emph {et~al.}(2025)\citenamefont {Jeong},
  \citenamefont {Lee}, \citenamefont {Lin}, \citenamefont {Yang}, \citenamefont
  {Choi}, \citenamefont {Oh}, \citenamefont {Song}, \citenamefont {Lee},
  \citenamefont {Nair}, \citenamefont {Choudhary} \emph
  {et~al.}}]{jeong2025metallicity}%
  \BibitemOpen
  \bibfield  {author} {\bibinfo {author} {\bibfnamefont {S.~G.}\ \bibnamefont
  {Jeong}}, \bibinfo {author} {\bibfnamefont {S.}~\bibnamefont {Lee}}, \bibinfo
  {author} {\bibfnamefont {B.}~\bibnamefont {Lin}}, \bibinfo {author}
  {\bibfnamefont {Z.}~\bibnamefont {Yang}}, \bibinfo {author} {\bibfnamefont
  {I.~H.}\ \bibnamefont {Choi}}, \bibinfo {author} {\bibfnamefont {J.~Y.}\
  \bibnamefont {Oh}}, \bibinfo {author} {\bibfnamefont {S.}~\bibnamefont
  {Song}}, \bibinfo {author} {\bibfnamefont {S.~w.}\ \bibnamefont {Lee}},
  \bibinfo {author} {\bibfnamefont {S.}~\bibnamefont {Nair}}, \bibinfo {author}
  {\bibfnamefont {R.}~\bibnamefont {Choudhary}}, \emph {et~al.},\ }\bibfield
  {title} {\bibinfo {title} {Metallicity and anomalous hall effect in
  epitaxially strained, atomically thin ruo2 films},\ }\href@noop {} {\bibfield
   {journal} {\bibinfo  {journal} {Proceedings of the National Academy of
  Sciences}\ }\textbf {\bibinfo {volume} {122}},\ \bibinfo {pages}
  {e2500831122} (\bibinfo {year} {2025})}\BibitemShut {NoStop}%
\bibitem [{\citenamefont {Akashdeep}\ \emph {et~al.}(2025)\citenamefont
  {Akashdeep}, \citenamefont {Ababneh}, \citenamefont {Schmitt}, \citenamefont
  {Gal{\'\i}ndez-Ruales}, \citenamefont {Fuhrmann}, \citenamefont {Kuschel},
  \citenamefont {Kl{\"a}ui}, \citenamefont {Amin},\ and\ \citenamefont
  {Jakob}}]{akashdeep2025interface}%
  \BibitemOpen
  \bibfield  {author} {\bibinfo {author} {\bibfnamefont {A.}~\bibnamefont
  {Akashdeep}}, \bibinfo {author} {\bibfnamefont {E.~M.}\ \bibnamefont
  {Ababneh}}, \bibinfo {author} {\bibfnamefont {C.}~\bibnamefont {Schmitt}},
  \bibinfo {author} {\bibfnamefont {E.}~\bibnamefont {Gal{\'\i}ndez-Ruales}},
  \bibinfo {author} {\bibfnamefont {F.}~\bibnamefont {Fuhrmann}}, \bibinfo
  {author} {\bibfnamefont {T.}~\bibnamefont {Kuschel}}, \bibinfo {author}
  {\bibfnamefont {M.}~\bibnamefont {Kl{\"a}ui}}, \bibinfo {author}
  {\bibfnamefont {V.}~\bibnamefont {Amin}},\ and\ \bibinfo {author}
  {\bibfnamefont {G.}~\bibnamefont {Jakob}},\ }\bibfield  {title} {\bibinfo
  {title} {Interface-generated spin current induced magnetoresistance in
  {RuO}$_2$/{Py} heterostructures},\ }\href@noop {} {\bibfield  {journal}
  {\bibinfo  {journal} {Phys. Rev. Appl.}\ }\textbf {\bibinfo {volume} {24}},\
  \bibinfo {pages} {054018} (\bibinfo {year} {2025})}\BibitemShut {NoStop}%
\bibitem [{\citenamefont {Akashdeep}\ \emph {et~al.}(2026)\citenamefont
  {Akashdeep}, \citenamefont {Krishnia}, \citenamefont {Ha}, \citenamefont
  {An}, \citenamefont {Gaerner}, \citenamefont {Prokscha}, \citenamefont
  {Suter}, \citenamefont {Janka}, \citenamefont {Reiss}, \citenamefont
  {Kuschel}, \citenamefont {Han}, \citenamefont {Di~Bernardo}, \citenamefont
  {Salman}, \citenamefont {Jakob},\ and\ \citenamefont
  {Kl{\"a}ui}}]{akashdeep2025surface}%
  \BibitemOpen
  \bibfield  {author} {\bibinfo {author} {\bibfnamefont {A.}~\bibnamefont
  {Akashdeep}}, \bibinfo {author} {\bibfnamefont {S.}~\bibnamefont {Krishnia}},
  \bibinfo {author} {\bibfnamefont {J.-H.}\ \bibnamefont {Ha}}, \bibinfo
  {author} {\bibfnamefont {S.}~\bibnamefont {An}}, \bibinfo {author}
  {\bibfnamefont {M.}~\bibnamefont {Gaerner}}, \bibinfo {author} {\bibfnamefont
  {T.}~\bibnamefont {Prokscha}}, \bibinfo {author} {\bibfnamefont
  {A.}~\bibnamefont {Suter}}, \bibinfo {author} {\bibfnamefont
  {G.}~\bibnamefont {Janka}}, \bibinfo {author} {\bibfnamefont
  {G.}~\bibnamefont {Reiss}}, \bibinfo {author} {\bibfnamefont
  {T.}~\bibnamefont {Kuschel}}, \bibinfo {author} {\bibfnamefont {D.-S.}\
  \bibnamefont {Han}}, \bibinfo {author} {\bibfnamefont {A.}~\bibnamefont
  {Di~Bernardo}}, \bibinfo {author} {\bibfnamefont {Z.}~\bibnamefont {Salman}},
  \bibinfo {author} {\bibfnamefont {G.}~\bibnamefont {Jakob}},\ and\ \bibinfo
  {author} {\bibfnamefont {M.}~\bibnamefont {Kl{\"a}ui}},\ }\bibfield  {title}
  {\bibinfo {title} {Surface-localized magnetic order in {RuO$_2$} thin films
  revealed by low-energy muon probes},\ }\href@noop {} {\bibfield  {journal}
  {\bibinfo  {journal} {Appl. Phys. Lett.}\ }\textbf {\bibinfo {volume}
  {128}},\ \bibinfo {pages} {022406} (\bibinfo {year} {2026})}\BibitemShut
  {NoStop}%
\bibitem [{\citenamefont {Beaurepaire}\ \emph {et~al.}(1996)\citenamefont
  {Beaurepaire}, \citenamefont {Merle}, \citenamefont {Daunois},\ and\
  \citenamefont {Bigot}}]{Beaurepaire.1996}%
  \BibitemOpen
  \bibfield  {author} {\bibinfo {author} {\bibfnamefont {E.}~\bibnamefont
  {Beaurepaire}}, \bibinfo {author} {\bibfnamefont {J.-C.}\ \bibnamefont
  {Merle}}, \bibinfo {author} {\bibfnamefont {A.}~\bibnamefont {Daunois}},\
  and\ \bibinfo {author} {\bibfnamefont {J.-Y.}\ \bibnamefont {Bigot}},\
  }\bibfield  {title} {\bibinfo {title} {Ultrafast spin dynamics in
  ferromagnetic nickel},\ }\href@noop {} {\bibfield  {journal} {\bibinfo
  {journal} {Phys. Rev. Lett.}\ }\textbf {\bibinfo {volume} {76}},\ \bibinfo
  {pages} {4250} (\bibinfo {year} {1996})}\BibitemShut {NoStop}%
\bibitem [{\citenamefont {Boeglin}\ \emph {et~al.}(2010)\citenamefont
  {Boeglin}, \citenamefont {Beaurepaire}, \citenamefont {Halté}, \citenamefont
  {López-Flores}, \citenamefont {Stamm}, \citenamefont {Pontius},
  \citenamefont {Dürr},\ and\ \citenamefont
  {Bigot}}]{boeglin_distinguishing_2010}%
  \BibitemOpen
  \bibfield  {author} {\bibinfo {author} {\bibfnamefont {C.}~\bibnamefont
  {Boeglin}}, \bibinfo {author} {\bibfnamefont {E.}~\bibnamefont
  {Beaurepaire}}, \bibinfo {author} {\bibfnamefont {V.}~\bibnamefont {Halté}},
  \bibinfo {author} {\bibfnamefont {V.}~\bibnamefont {López-Flores}}, \bibinfo
  {author} {\bibfnamefont {C.}~\bibnamefont {Stamm}}, \bibinfo {author}
  {\bibfnamefont {N.}~\bibnamefont {Pontius}}, \bibinfo {author} {\bibfnamefont
  {H.~A.}\ \bibnamefont {Dürr}},\ and\ \bibinfo {author} {\bibfnamefont
  {J.-Y.}\ \bibnamefont {Bigot}},\ }\bibfield  {title} {\bibinfo {title}
  {Distinguishing the ultrafast dynamics of spin and orbital moments in
  solids},\ }\href@noop {} {\bibfield  {journal} {\bibinfo  {journal} {Nature}\
  }\textbf {\bibinfo {volume} {465}},\ \bibinfo {pages} {458} (\bibinfo {year}
  {2010})}\BibitemShut {NoStop}%
\bibitem [{\citenamefont {Tengdin}\ \emph {et~al.}(2018)\citenamefont
  {Tengdin}, \citenamefont {You}, \citenamefont {Chen}, \citenamefont {Shi},
  \citenamefont {Zusin}, \citenamefont {Zhang}, \citenamefont {Gentry},
  \citenamefont {Blonsky}, \citenamefont {Keller}, \citenamefont {Oppeneer},
  \citenamefont {Kapteyn}, \citenamefont {Tao},\ and\ \citenamefont
  {Murnane}}]{tengdin_critical_2018}%
  \BibitemOpen
  \bibfield  {author} {\bibinfo {author} {\bibfnamefont {P.}~\bibnamefont
  {Tengdin}}, \bibinfo {author} {\bibfnamefont {W.}~\bibnamefont {You}},
  \bibinfo {author} {\bibfnamefont {C.}~\bibnamefont {Chen}}, \bibinfo {author}
  {\bibfnamefont {X.}~\bibnamefont {Shi}}, \bibinfo {author} {\bibfnamefont
  {D.}~\bibnamefont {Zusin}}, \bibinfo {author} {\bibfnamefont
  {Y.}~\bibnamefont {Zhang}}, \bibinfo {author} {\bibfnamefont
  {C.}~\bibnamefont {Gentry}}, \bibinfo {author} {\bibfnamefont
  {A.}~\bibnamefont {Blonsky}}, \bibinfo {author} {\bibfnamefont
  {M.}~\bibnamefont {Keller}}, \bibinfo {author} {\bibfnamefont {P.~M.}\
  \bibnamefont {Oppeneer}}, \bibinfo {author} {\bibfnamefont {H.~C.}\
  \bibnamefont {Kapteyn}}, \bibinfo {author} {\bibfnamefont {Z.}~\bibnamefont
  {Tao}},\ and\ \bibinfo {author} {\bibfnamefont {M.~M.}\ \bibnamefont
  {Murnane}},\ }\bibfield  {title} {\bibinfo {title} {Critical behavior within
  20 fs drives the out-of-equilibrium laser-induced magnetic phase transition
  in nickel},\ }\href@noop {} {\bibfield  {journal} {\bibinfo  {journal}
  {Science Advances}\ }\textbf {\bibinfo {volume} {4}},\ \bibinfo {pages}
  {eaap9744} (\bibinfo {year} {2018})}\BibitemShut {NoStop}%
\bibitem [{\citenamefont {Koopmans}\ \emph {et~al.}(2010)\citenamefont
  {Koopmans}, \citenamefont {Malinowski}, \citenamefont {Dalla~Longa},
  \citenamefont {Steiauf}, \citenamefont {F{\"a}hnle}, \citenamefont {Roth},
  \citenamefont {Cinchetti},\ and\ \citenamefont
  {Aeschlimann}}]{koopmans_explaining_2010}%
  \BibitemOpen
  \bibfield  {author} {\bibinfo {author} {\bibfnamefont {B.}~\bibnamefont
  {Koopmans}}, \bibinfo {author} {\bibfnamefont {G.}~\bibnamefont
  {Malinowski}}, \bibinfo {author} {\bibfnamefont {F.}~\bibnamefont
  {Dalla~Longa}}, \bibinfo {author} {\bibfnamefont {D.}~\bibnamefont
  {Steiauf}}, \bibinfo {author} {\bibfnamefont {M.}~\bibnamefont {F{\"a}hnle}},
  \bibinfo {author} {\bibfnamefont {T.}~\bibnamefont {Roth}}, \bibinfo {author}
  {\bibfnamefont {M.}~\bibnamefont {Cinchetti}},\ and\ \bibinfo {author}
  {\bibfnamefont {M.}~\bibnamefont {Aeschlimann}},\ }\bibfield  {title}
  {\bibinfo {title} {Explaining the paradoxical diversity of ultrafast
  laser-induced demagnetization},\ }\href@noop {} {\bibfield  {journal}
  {\bibinfo  {journal} {Nature Materials}\ }\textbf {\bibinfo {volume} {9}},\
  \bibinfo {pages} {259} (\bibinfo {year} {2010})}\BibitemShut {NoStop}%
\bibitem [{\citenamefont {Weber}\ \emph {et~al.}(2025)\citenamefont {Weber},
  \citenamefont {Haag}, \citenamefont {Leckron}, \citenamefont
  {Jaeschke-Ubiergo}, \citenamefont {Smejkal}, \citenamefont {Sinova},\ and\
  \citenamefont {Schneider}}]{weber2025newton}%
  \BibitemOpen
  \bibfield  {author} {\bibinfo {author} {\bibfnamefont {M.}~\bibnamefont
  {Weber}}, \bibinfo {author} {\bibfnamefont {L.}~\bibnamefont {Haag}},
  \bibinfo {author} {\bibfnamefont {K.}~\bibnamefont {Leckron}}, \bibinfo
  {author} {\bibfnamefont {R.}~\bibnamefont {Jaeschke-Ubiergo}}, \bibinfo
  {author} {\bibfnamefont {L.}~\bibnamefont {Smejkal}}, \bibinfo {author}
  {\bibfnamefont {J.}~\bibnamefont {Sinova}},\ and\ \bibinfo {author}
  {\bibfnamefont {H.~C.}\ \bibnamefont {Schneider}},\ }\bibfield  {title}
  {\bibinfo {title} {Ultrafast electron dynamics in a planar d-wave
  altermagnet},\ }\href {https://doi.org/10.1016/j.newton.2025.100266}
  {\bibfield  {journal} {\bibinfo  {journal} {Newton}\ }\textbf {\bibinfo
  {volume} {1}},\ \bibinfo {pages} {100266} (\bibinfo {year}
  {2025})}\BibitemShut {NoStop}%
\bibitem [{met()}]{methods}%
  \BibitemOpen
  \href@noop {} {}\bibinfo {note} {Materials and methods are available as
  supplementary material}\BibitemShut {NoStop}%
\bibitem [{\citenamefont {Lytvynenko}\ \emph {et~al.}(2026)\citenamefont
  {Lytvynenko}, \citenamefont {Akashdeep}, \citenamefont {Vo}, \citenamefont
  {Tkach}, \citenamefont {Chernov}, \citenamefont {Gloskovskii}, \citenamefont
  {Schlueter}, \citenamefont {Luo}, \citenamefont {Ukleev}, \citenamefont
  {Radu}, \citenamefont {Kronast}, \citenamefont {Hiroto}, \citenamefont
  {Winkelmann}, \citenamefont {Min\'ar}, \citenamefont {Kl\"aui}, \citenamefont
  {Sch\"onhense}, \citenamefont {Jakob}, \citenamefont {Elmers},\ and\
  \citenamefont {Fedchenko}}]{Lytvynenko-2026}%
  \BibitemOpen
  \bibfield  {author} {\bibinfo {author} {\bibfnamefont {Y.}~\bibnamefont
  {Lytvynenko}}, \bibinfo {author} {\bibfnamefont {A.}~\bibnamefont
  {Akashdeep}}, \bibinfo {author} {\bibfnamefont {T.~P.}\ \bibnamefont {Vo}},
  \bibinfo {author} {\bibfnamefont {O.}~\bibnamefont {Tkach}}, \bibinfo
  {author} {\bibfnamefont {S.~V.}\ \bibnamefont {Chernov}}, \bibinfo {author}
  {\bibfnamefont {A.}~\bibnamefont {Gloskovskii}}, \bibinfo {author}
  {\bibfnamefont {C.}~\bibnamefont {Schlueter}}, \bibinfo {author}
  {\bibfnamefont {C.}~\bibnamefont {Luo}}, \bibinfo {author} {\bibfnamefont
  {V.}~\bibnamefont {Ukleev}}, \bibinfo {author} {\bibfnamefont
  {F.}~\bibnamefont {Radu}}, \bibinfo {author} {\bibfnamefont {F.}~\bibnamefont
  {Kronast}}, \bibinfo {author} {\bibfnamefont {T.}~\bibnamefont {Hiroto}},
  \bibinfo {author} {\bibfnamefont {A.}~\bibnamefont {Winkelmann}}, \bibinfo
  {author} {\bibfnamefont {J.}~\bibnamefont {Min\'ar}}, \bibinfo {author}
  {\bibfnamefont {M.}~\bibnamefont {Kl\"aui}}, \bibinfo {author} {\bibfnamefont
  {G.}~\bibnamefont {Sch\"onhense}}, \bibinfo {author} {\bibfnamefont
  {G.}~\bibnamefont {Jakob}}, \bibinfo {author} {\bibfnamefont {H.~J.}\
  \bibnamefont {Elmers}},\ and\ \bibinfo {author} {\bibfnamefont
  {O.}~\bibnamefont {Fedchenko}},\ }\bibfield  {title} {\bibinfo {title}
  {Magnetic circular dichroism in core-level {X-ray} photoelectron spectroscopy
  of altermagnetic {RuO}$_{2}$ films},\ }\href@noop {} {\bibfield  {journal}
  {\bibinfo  {journal} {Phys. Rev. B}\ }\textbf {\bibinfo {volume} {113}},\
  \bibinfo {pages} {014403} (\bibinfo {year} {2026})}\BibitemShut {NoStop}%
\bibitem [{\citenamefont {Gray}\ \emph {et~al.}(2024)\citenamefont {Gray},
  \citenamefont {Deng}, \citenamefont {Tian}, \citenamefont {Chilcote},
  \citenamefont {Dodge}, \citenamefont {Brahlek},\ and\ \citenamefont
  {Wu}}]{gray2024timeresolved}%
  \BibitemOpen
  \bibfield  {author} {\bibinfo {author} {\bibfnamefont {I.}~\bibnamefont
  {Gray}}, \bibinfo {author} {\bibfnamefont {Q.}~\bibnamefont {Deng}}, \bibinfo
  {author} {\bibfnamefont {Q.}~\bibnamefont {Tian}}, \bibinfo {author}
  {\bibfnamefont {M.}~\bibnamefont {Chilcote}}, \bibinfo {author}
  {\bibfnamefont {J.~S.}\ \bibnamefont {Dodge}}, \bibinfo {author}
  {\bibfnamefont {M.}~\bibnamefont {Brahlek}},\ and\ \bibinfo {author}
  {\bibfnamefont {L.}~\bibnamefont {Wu}},\ }\bibfield  {title} {\bibinfo
  {title} {Time-resolved magneto-optical effects in the altermagnet candidate
  {MnTe}},\ }\href@noop {} {\bibfield  {journal} {\bibinfo  {journal} {Appl.
  Phys. Lett.}\ }\textbf {\bibinfo {volume} {125}},\ \bibinfo {pages} {212404}
  (\bibinfo {year} {2024})}\BibitemShut {NoStop}%
\bibitem [{\citenamefont {Kresse}\ and\ \citenamefont {Hafner}(1993)}]{vasp}%
  \BibitemOpen
  \bibfield  {author} {\bibinfo {author} {\bibfnamefont {G.}~\bibnamefont
  {Kresse}}\ and\ \bibinfo {author} {\bibfnamefont {J.}~\bibnamefont
  {Hafner}},\ }\bibfield  {title} {\bibinfo {title} {Ab initio molecular
  dynamics for liquid metals},\ }\href@noop {} {\bibfield  {journal} {\bibinfo
  {journal} {Phys. Rev. B}\ }\textbf {\bibinfo {volume} {47}},\ \bibinfo
  {pages} {558} (\bibinfo {year} {1993})}\BibitemShut {NoStop}%
\bibitem [{\citenamefont {Kresse}\ and\ \citenamefont {Hafner}(1994)}]{vasp2}%
  \BibitemOpen
  \bibfield  {author} {\bibinfo {author} {\bibfnamefont {G.}~\bibnamefont
  {Kresse}}\ and\ \bibinfo {author} {\bibfnamefont {J.}~\bibnamefont
  {Hafner}},\ }\bibfield  {title} {\bibinfo {title} {Ab initio
  molecular-dynamics simulation of the liquid-metal--amorphous-semiconductor
  transition in germanium},\ }\href@noop {} {\bibfield  {journal} {\bibinfo
  {journal} {Phys. Rev. B}\ }\textbf {\bibinfo {volume} {49}},\ \bibinfo
  {pages} {14251} (\bibinfo {year} {1994})}\BibitemShut {NoStop}%
\bibitem [{\citenamefont {Kresse}\ and\ \citenamefont
  {Furthmüller}(1996)}]{vasp3}%
  \BibitemOpen
  \bibfield  {author} {\bibinfo {author} {\bibfnamefont {G.}~\bibnamefont
  {Kresse}}\ and\ \bibinfo {author} {\bibfnamefont {J.}~\bibnamefont
  {Furthmüller}},\ }\bibfield  {title} {\bibinfo {title} {Efficiency of
  ab-initio total energy calculations for metals and semiconductors using a
  plane-wave basis set},\ }\href@noop {} {\bibfield  {journal} {\bibinfo
  {journal} {Computational Materials Science}\ }\textbf {\bibinfo {volume}
  {6}},\ \bibinfo {pages} {15} (\bibinfo {year} {1996})}\BibitemShut {NoStop}%
\bibitem [{\citenamefont {Perdew}\ \emph {et~al.}(1996)\citenamefont {Perdew},
  \citenamefont {Burke},\ and\ \citenamefont {Ernzerhof}}]{PBE}%
  \BibitemOpen
  \bibfield  {author} {\bibinfo {author} {\bibfnamefont {J.~P.}\ \bibnamefont
  {Perdew}}, \bibinfo {author} {\bibfnamefont {K.}~\bibnamefont {Burke}},\ and\
  \bibinfo {author} {\bibfnamefont {M.}~\bibnamefont {Ernzerhof}},\ }\bibfield
  {title} {\bibinfo {title} {Generalized gradient approximation made simple},\
  }\href@noop {} {\bibfield  {journal} {\bibinfo  {journal} {Phys. Rev. Lett.}\
  }\textbf {\bibinfo {volume} {77}},\ \bibinfo {pages} {3865} (\bibinfo {year}
  {1996})}\BibitemShut {NoStop}%
\bibitem [{\citenamefont {Dudarev}\ \emph {et~al.}(1998)\citenamefont
  {Dudarev}, \citenamefont {Botton}, \citenamefont {Savrasov}, \citenamefont
  {Humphreys},\ and\ \citenamefont {Sutton}}]{dudarev}%
  \BibitemOpen
  \bibfield  {author} {\bibinfo {author} {\bibfnamefont {S.~L.}\ \bibnamefont
  {Dudarev}}, \bibinfo {author} {\bibfnamefont {G.~A.}\ \bibnamefont {Botton}},
  \bibinfo {author} {\bibfnamefont {S.~Y.}\ \bibnamefont {Savrasov}}, \bibinfo
  {author} {\bibfnamefont {C.~J.}\ \bibnamefont {Humphreys}},\ and\ \bibinfo
  {author} {\bibfnamefont {A.~P.}\ \bibnamefont {Sutton}},\ }\bibfield  {title}
  {\bibinfo {title} {Electron-energy-loss spectra and the structural stability
  of nickel oxide: {An LSDA+U} study},\ }\href@noop {} {\bibfield  {journal}
  {\bibinfo  {journal} {Phys. Rev. B}\ }\textbf {\bibinfo {volume} {57}},\
  \bibinfo {pages} {1505} (\bibinfo {year} {1998})}\BibitemShut {NoStop}%
\bibitem [{elk(0 04)}]{elk-code}%
  \BibitemOpen
  \href@noop {} {\bibinfo {title} {{The Elk Code}}},\ \bibinfo {howpublished}
  {\url{http://elk.sourceforge.net/}} (\bibinfo {year} {Accessed:
  2024-10-04})\BibitemShut {NoStop}%
\bibitem [{\citenamefont {Essert}\ and\ \citenamefont
  {Schneider}(2011)}]{essert_electron-phonon_2011}%
  \BibitemOpen
  \bibfield  {author} {\bibinfo {author} {\bibfnamefont {S.}~\bibnamefont
  {Essert}}\ and\ \bibinfo {author} {\bibfnamefont {H.~C.}\ \bibnamefont
  {Schneider}},\ }\bibfield  {title} {\bibinfo {title} {Electron-phonon
  scattering dynamics in ferromagnetic metals and their influence on ultrafast
  demagnetization processes},\ }\href@noop {} {\bibfield  {journal} {\bibinfo
  {journal} {Phys. Rev. B}\ }\textbf {\bibinfo {volume} {84}},\ \bibinfo
  {pages} {224405} (\bibinfo {year} {2011})}\BibitemShut {NoStop}%
\bibitem [{\citenamefont {Stiehl}\ \emph {et~al.}(2022)\citenamefont {Stiehl},
  \citenamefont {Weber}, \citenamefont {Seibel}, \citenamefont {Hoefer},
  \citenamefont {Weber}, \citenamefont {Nenno}, \citenamefont {Schneider},
  \citenamefont {Rethfeld}, \citenamefont {Stadtmüller},\ and\ \citenamefont
  {Aeschlimann}}]{Stiehl2022}%
  \BibitemOpen
  \bibfield  {author} {\bibinfo {author} {\bibfnamefont {M.}~\bibnamefont
  {Stiehl}}, \bibinfo {author} {\bibfnamefont {M.}~\bibnamefont {Weber}},
  \bibinfo {author} {\bibfnamefont {C.}~\bibnamefont {Seibel}}, \bibinfo
  {author} {\bibfnamefont {J.}~\bibnamefont {Hoefer}}, \bibinfo {author}
  {\bibfnamefont {S.~T.}\ \bibnamefont {Weber}}, \bibinfo {author}
  {\bibfnamefont {D.~M.}\ \bibnamefont {Nenno}}, \bibinfo {author}
  {\bibfnamefont {H.~C.}\ \bibnamefont {Schneider}}, \bibinfo {author}
  {\bibfnamefont {B.}~\bibnamefont {Rethfeld}}, \bibinfo {author}
  {\bibfnamefont {B.}~\bibnamefont {Stadtmüller}},\ and\ \bibinfo {author}
  {\bibfnamefont {M.}~\bibnamefont {Aeschlimann}},\ }\bibfield  {title}
  {\bibinfo {title} {Role of primary and secondary processes in the ultrafast
  spin dynamics of nickel},\ }\href@noop {} {\bibfield  {journal} {\bibinfo
  {journal} {Applied Physics Letters}\ }\textbf {\bibinfo {volume} {120}}
  (\bibinfo {year} {2022})}\BibitemShut {NoStop}%
\end{thebibliography}
%

\end{document}


\title{All Optical Excitation of Spin Polarization in d-wave Altermagnets---Supplement}
\author{Marius Weber}
       \thanks{these authors contributed equally}
       \affiliation{Department of Physics and Research Center OPTIMAS, RPTU University Kaiserslautern-Landau, 67663 Kaiserslautern, Germany}
       \affiliation{Institute of Physics, Johannes Gutenberg University Mainz, 55099 Mainz, Germany}
\author{Stephan Wust}
        \thanks{these authors contributed equally}
        \affiliation{Department of Physics and Research Center OPTIMAS, RPTU University Kaiserslautern-Landau, 67663 Kaiserslautern, Germany}
\author{Luca Felipe Haag}
    \affiliation{Department of Physics and Research Center OPTIMAS, RPTU University Kaiserslautern-Landau, 67663 Kaiserslautern, Germany}
\author{Paul Herrgen}
    \affiliation{Department of Physics and Research Center OPTIMAS, RPTU University Kaiserslautern-Landau, 67663 Kaiserslautern, Germany}
\author{Akashdeep Akashdeep}
    \affiliation{Institute of Physics, Johannes Gutenberg University Mainz, 55099 Mainz, Germany}
\author{Kai Leckron}
    \affiliation{Department of Physics and Research Center OPTIMAS, RPTU University Kaiserslautern-Landau, 67663 Kaiserslautern, Germany}
\author{Christin Schmitt}
    \affiliation{Institute of Physics, Johannes Gutenberg University Mainz, 55099 Mainz, Germany}
\author{Rafael~\surname{Ramos}}
\affiliation{Centro Singular de Investigación en Química Biolóxica e Materiais Moleculares (CIQUS), Universidade de Santiago de Compostela, Santiago de Compostela, Spain}
\affiliation{Departamento de Química-Física, Universidade de Santiago de Compostela, Santiago de Compostela, Spain}
\affiliation{WPI Advanced Institute for Materials Research, Tohoku University, Sendai 980-8577, Japan}
    
\author{Takashi~\surname{Kikkawa}}
    \affiliation{Department of Applied Physics, The University of Tokyo, Tokyo 113-8656, Japan}
\affiliation{Advanced Science Research Center, Japan Atomic Energy Agency, Tokai 319-1195, Japan}

\author{Eiji~\surname{Saitoh}}
    \affiliation{WPI Advanced Institute for Materials Research, Tohoku University, Sendai 980-8577, Japan}
    \affiliation{Department of Applied Physics, The University of Tokyo, Tokyo 113-8656, Japan}
    \affiliation{Institute for AI and Beyond, The University of Tokyo, Tokyo 113-8656, Japan}
    \affiliation{RIKEN Center for Emergent Matter Science (CEMS), Wako 351–0198, Japan}
\author{Mathias Kläui}
       \affiliation{Institute of Physics, Johannes Gutenberg University Mainz, 55099 Mainz, Germany}
\author{Libor \v{S}mejkal}
    \affiliation{Max Planck Institute for the Physics of Complex Systems, 01187 Dresden, Germany }
    \affiliation{Max Planck Institute for Chemical Physics of Solids, 01187 Dresden, Germany}
\author{Jairo Sinova}
       \affiliation{Institute of Physics, Johannes Gutenberg University Mainz, 55099 Mainz, Germany}
    \affiliation{Department of Physics, Texas A\&M University, College Station, Texas 77843-4242, USA}
\author{Martin Aeschlimann}
    \affiliation{Department of Physics and Research Center OPTIMAS, RPTU University Kaiserslautern-Landau, 67663 Kaiserslautern, Germany}
\author{Gerhard Jakob}
       \affiliation{Institute of Physics, Johannes Gutenberg University Mainz, 55099 Mainz, Germany}
\author{Benjamin Stadtmüller}
    \affiliation{Department of Physics and Research Center OPTIMAS, RPTU University Kaiserslautern-Landau, 67663 Kaiserslautern, Germany}
    \affiliation{Experimentalphysik II, Institute of Physics, Augsburg University, 86159 Augsburg, Germany}
\author{Hans Christian Schneider}
    \affiliation{Department of Physics and Research Center OPTIMAS, RPTU University Kaiserslautern-Landau, 67663 Kaiserslautern, Germany}

\date{2026-06-04}
\maketitle

\section{Measurement principle\label{ch:measurement-principle}}

The data presented in the main manuscript were recorded using an all-optical pump-probe setup using an amplified Ti:Sapphire laser system 1.55\,eV, 35\,fs, 1\,kHz, 7\,mJ per pulse) to generate the pump and probe beams by using a polar Kerr geometry under normal incidence and without an external magnetic field. The angle of incidence of the pump beam is $<2^\circ$, the one of the probe beam is $<1^\circ$ with respect to the surface normal.
For RuO$_{2}$, TiO$_{2}$ and Cu/TiO$_{2}$ the pump energy was set to 1.55\,eV with a temporal pulse duration of 40$\pm$ 3,fs, while the probe energy was set to 3.10\,eV with a temporal pulse duration of 47$\pm$5\,fs. The values for the applied fluence $\Phi_{\text{app}}$ and absorbed fluence $\Phi_{\text{absorb}}$ for each sample used in this work are listed in Tab.~\ref{tab:fluence}. The total applied peak fluence $\Phi_{\text{app}}$ was calculated using the definition
\begin{equation}
    \Phi_{\text{app}} = \frac{2P_{\text{app}}}{f\pi A}
\end{equation}
where $P_{\text{app}}$ is the applied pump power, $f$ is the laser repetition rate and $A$ is the corresponding $1/e^{2}$ spot size. For the absorbed fluence $\Phi_{\text{absorb}}$, the absorption $\alpha$ in each sample was obtained by the transfer matrix method as implemented in the Python package \textsc{TMM}~\cite{TMM}.

\begin{table}[h]
\caption{\textbf{Optical excitation strength} Characteristic values of the optical excitation along with calculated values for $\Phi_{\text{app}}$ and $\Phi_{\text{absorb}}$ for each sample used in this work. The calculated absorbed fluence for the NiO thin film capped with $2\,$nm of Pt refers to the absorption of the NiO layer.}
\begin{tabular}{@{}cccccccc@{}}
\hline
Sample &
  \begin{tabular}[c]{@{}c@{}}$A$\\ \mbox{} [$\times10^{-3}$\,cm$^{2}]$\end{tabular} &
  \begin{tabular}[c]{@{}c@{}}$E_{\text{Pump}}$\\  \mbox{}[eV]\end{tabular} &
  \begin{tabular}[c]{@{}c@{}}$P_{\text{app}}$\\ \mbox{}[mW]\end{tabular} &
  \begin{tabular}[c]{@{}c@{}}$\Phi_{\text{app}}$\\ \mbox{} [mJ/cm$^{2}$]\end{tabular} &
  \begin{tabular}[c]{@{}c@{}}$\alpha$\\  \mbox{} [\%]\end{tabular} &
  \begin{tabular}[c]{@{}c@{}}$\Phi_{\text{absorb}}$\\ \mbox{} [$\mu $J/cm$^{2}$]\end{tabular} &
  \begin{tabular}[c]{@{}c@{}}Refs.\\for $n$\end{tabular} \\ \hline
RuO$_{2}$(3.5\,nm) & 3.04 & 1.55 & 40 & 26.32 & 1.6   & 421 & \cite{goel1981optical} \\
TiO$_{2}$ (substrate) & 3.04 & 1.55 & 80 & 52.70 & 0.2 & 105 & \cite{zhukovsky2015experimental} \\
TiO$_{2}$ (substrate) & 8.82 & 4.65 & 3 & 0.68 & 67.8 & 461 & \cite{zhukovsky2015experimental} \\
MgO(3$\,$nm)/Cu(3$\,$nm)/TiO$_{2}$ & 3.54 & 1.55 & 80 & 45.11 & 4.2 & 1895 & \cite{stephens1952index, Babar2015, zhukovsky2015experimental} \\
Pt(2\,nm)/NiO(10\,nm) & 8.82 & 4.65 & 3  & 0.68  & 4.34  & 301 & \cite{franta2005optical, Tselin2024} \\ \hline
\label{tab:fluence}
\end{tabular}
\end{table}

The probe pulse was generated by second harmonic generation in a beta barium borate (BBO) crystal. For the NiO(111) measurements and a second measurement on TiO$_{2}$, the pump energy was set to 4.65\,eV with a pulse duration of 87.3$\pm$10\,fs to drive excitations above the band gap. In our setup, we can freely adjust the pump and probe polarization axes with respect to the sample geometry. This is achieved by rotating the pump polarization with a $\lambda$/2 plate and the sample azimuthal angle. The probe polarization was fixed to an s-polarization state. The probe beam was adjusted to the center of rotation to avoid any beam drift onto the sample. The reflected probe beam was measured in a balancing photodetector, which is sensitive to the rotation of the probe polarization.

\section{Electro-optical Kerr effect (EOKE) signal}
\begin{figure}
    \centering
    \includegraphics[width=.8\linewidth]{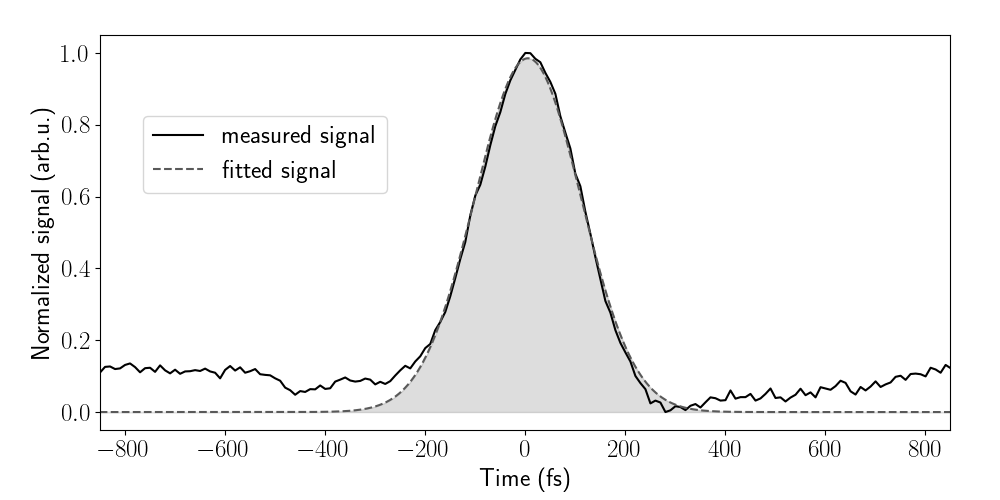}
    \caption{\textbf{Measured EOKE signal and fit} Black solid line is the measurement obtained with a fixed relative polarization angle of $45^\circ$ between the pump and probe beams. The EOKE signal around the temporal overlap of the pump and probe at $0\,$fs was fitted with a Gaussian curve. The fitting envelope is shown in Fig.~3B of the main manuscript.}
    \label{fig:SI_EOKE}  
\end{figure}

In the main text, we used the electro-optical Kerr effect (or EOKE) trace in Fig.~3C to estimate the cross correlation between the pump and probe pulse. The EOKE effect in a metal such as RuO$_2$ is a higher-order nonlinear optical response that originates from the momentum distribution of the (optically excited) conduction electrons and thus contains contributions from both the pump and probe beam as well as from the ultrafast electron thermalization. The EOKE signal of RuO$_2$ was recorded in a time-resolved experiment with a fixed relative polarization angle of $45^\circ$ between pump and probe beams, ensuring a maximal contribution from the EOKE effect, see also figure~\ref{fig:probe_pol}. This experimental configuration ensures that the time-dependent signal is dominated by the EOKE response. The resulting EOKE trace is shown in figure~\ref{fig:SI_EOKE}. The envelope of the EOKE signal plotted in Fig.~3C of the main part of the manuscript was obtained by fitting the EOKE trace near time zero (the temporal overlap of the pump and probe pulses) with a Gaussian function. 

\section{Reference measurements on TiO$_{2}$ and Cu/TiO$_{2}$}

\begin{figure}[h]
    \centering
    \includegraphics[width=\linewidth]{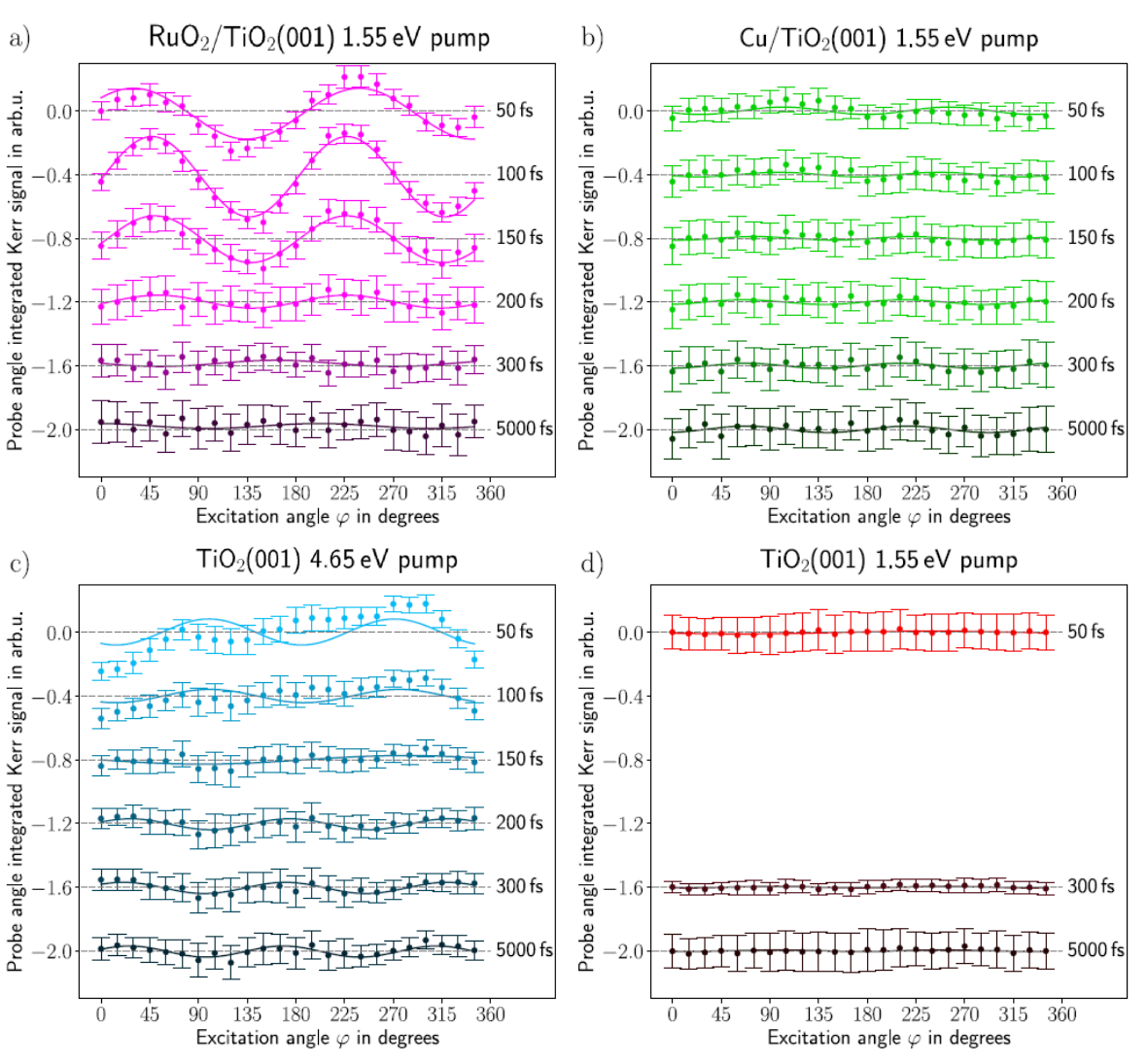}
    \caption{\textbf{Magneto-optical response of RuO$_2$ compared to reference experiments} Magneto-optical signals integrated over probe polarization angle versus the excitation angle $\varphi$ for a) RuO$_{2}$(3.5\,nm)/TiO$_2$ pumped with 1.55\,eV photons, b) Cu(3\,nm)/TiO$_{2}$ pumped with 1.55\,eV photons, c) a TiO$_{2}$ substrate pumped with 4.65\,eV photons and d) pumped with 1.55\,eV photons for different time steps after the optical excitation.}
    \label{fig:ref_exp}  
\end{figure}

The key experimental result of our work is shown in Figs.~3B and C of the main manuscript, where we present the transient polar Kerr response of a 3.5\,nm thin RuO$_2$ film, which is attributed to the optically generated spin polarization in the altermagnet RuO$_2$. These data are again included here as figure~\ref{fig:ref_exp} a). 

To unambiguously demonstrate that this signal originates from the optically excited RuO$_2$ layer and not from the TiO$_2$ substrate, we performed several reference experiments for two sample systems. 
First, we investigated the optical Kerr response of a thin metallic film (3\,nm of Cu) grown on TiO$_2$ in polar geometry depending on the excitation angle. The Cu/TiO$_2$ sample was optically excited with $1.55\,$eV photons in analogy to the RuO$_2$/TiO$_2$ sample discussed in the main manuscript. In both cases, the TiO$_2$ substrate is excited only indirectly by interfacial energy transfer from the optically excited metal layer (i.e., the RuO$_2$ and the copper layer).   
The results for the Cu/TiO$_2$ sample are shown in Fig.~\ref{fig:ref_exp} b). We do not observe a significant polar Kerr response for this sample for any excitation geometry (excitation angle) or time step, despite the approximately 4 times larger light absorption in the Cu/TiO$_2$ sample compared to RuO$_2$/TiO$_2$, see Table~\ref{tab:fluence}.  

In addition, we also investigated the magneto-optical response of the bare TiO$_2$ substrate for two different pump photon energies of $4.65\,$eV and $1.55\,$eV. The photon energy of $4.65\,$eV is larger than the band gap of TiO$_2$ and thus leads to a direct optical excitation of TiO$_2$ resulting in the generation of excited carriers. In contrast, the photon energy of $1.55\,$eV is below the band gap, but corresponds to the optical excitation conditions used for the experiments presented in the main part of our manuscript. The excitation-angle dependence of the Kerr response for optically excited TiO$_2$ at different times are shown in Figs.~\ref{fig:ref_exp} c) and d) for photon energies of $4.65\,$eV and $1.55\,$eV, respectively. The corresponding incident and absorbed fluences are summarized in table~\ref{tab:fluence}.

None of these data show a magneto-optical Kerr response comparable to our results for the 3.5\,nm thin RuO$_2$ film discussed in the main manuscript (and repeated in Fig.~\ref{fig:ref_exp}~a)). TiO$_{2}$ pumped at 1.55 eV shows a clear flat line behavior, while TiO$_2$ excited at $4.65\,$eV shows a slight dip in the signal near time zero. However, the corresponding signal strength decays within the time of the temporal pump pulse width of $87.3\pm 10\,$fs without observing a clear periodicity, so that this small signal can be attributed to an excitation effect in TiO$_2$ different from that observed in the RuO$_{2}$/TiO$_{2}$ sample.

\begin{figure}
    \centering
    \includegraphics[width=0.6\linewidth]{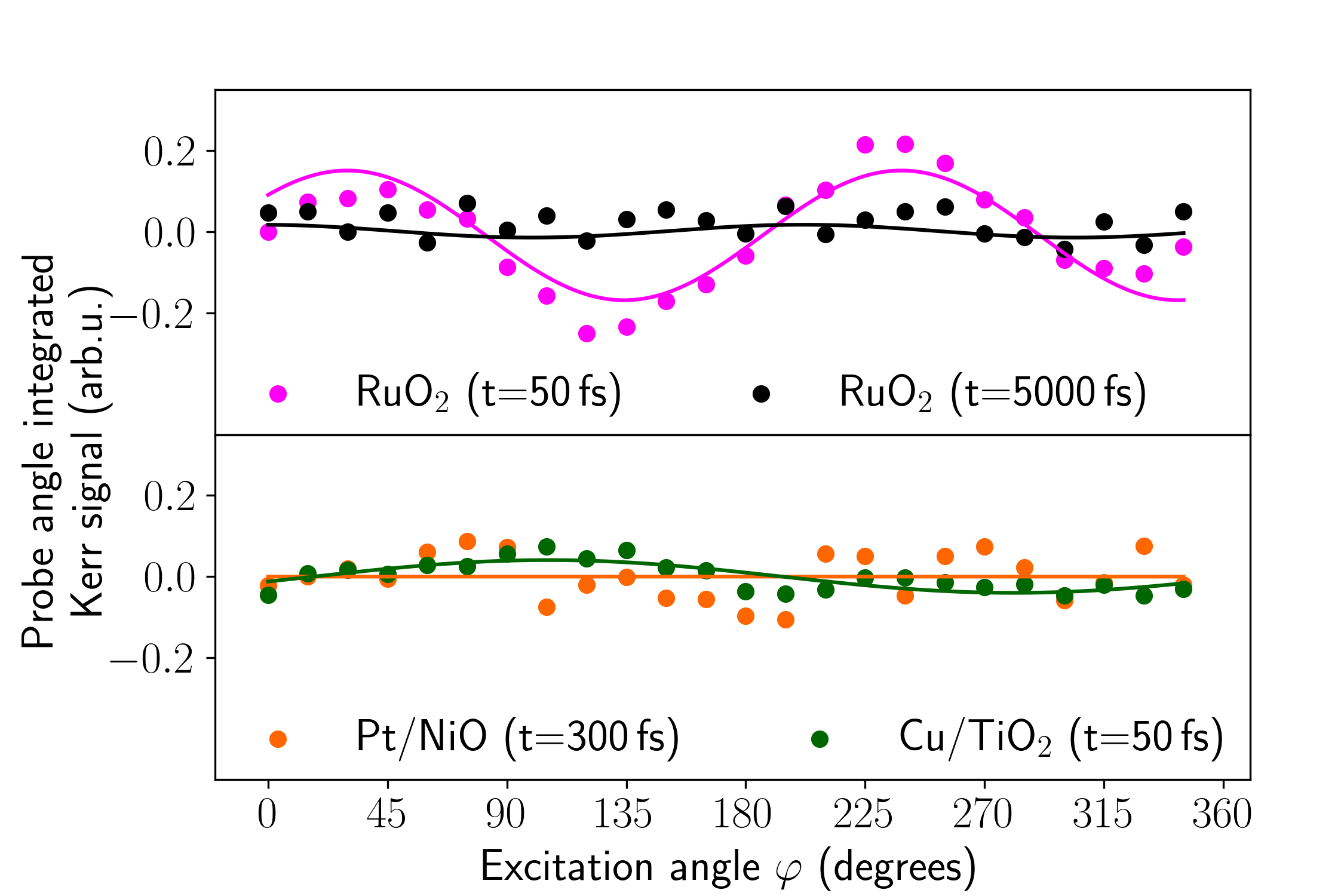}
    \caption{\textbf{Magneto-optical response of RuO$_2$ compared to TiO$_2$ and NiO} Magneto-optical signals integrated over probe polarization versus the excitation angle $\varphi$ for RuO$_2$ at $50\,$fs and $5000\,$fs and (bottom) for Cu/TiO$_2$ at $50\,$fs and NiO capped with Pt at $300\,$fs. Solid lines are the best sinusoidal fits with an arbitrary period.}
    \label{fig:SIold3b}  
\end{figure}

Next we directly compare the time- and excitation-angle resolved magnetic Kerr response of the different material systems in figure~\ref{fig:SIold3b}. Immediately after the optical excitation ($t=50$\,fs), the magnetic Kerr response of RuO$_2$ shows a distinct 180$^\circ$ periodicity with maximum and minimum at $\varphi=45^\circ$ and 135$^\circ$, respectively. The different signs of the Kerr rotation at $\varphi=45^\circ$ and 135$^\circ$ clearly indicate a reversal of the spin polarization for different excitation angles. Crucially, the characteristic signature of the excitation-angle dependent spin polarization vanishes for large time delays ($\Delta t=5000$\,fs) after the optical excitation. This clearly indicates that the twofold magnetic response is not part of the quasi-static structure of RuO$_{2}$, but is indeed the optically induced spin polarization that varies with the excitation angle. 

In contrast, we observe no excitation-angle dependent Kerr rotation as shown in the lower panel of figure~\ref{fig:SIold3b} for both reference systems TiO$_2$ and NiO. Thus there is clear evidence that the magnetic Kerr response of RuO$_{2}$ does not originate from the substrate, but represents an extraordinary effect for an optically excited compensated magnet that arises from the alternating  spin structure in momentum space.

\section{Birefringence signal in RuO\(_2\), TiO\(_2\) and NiO/Pt}

\begin{figure}
    \centering
    \includegraphics[width=\linewidth]{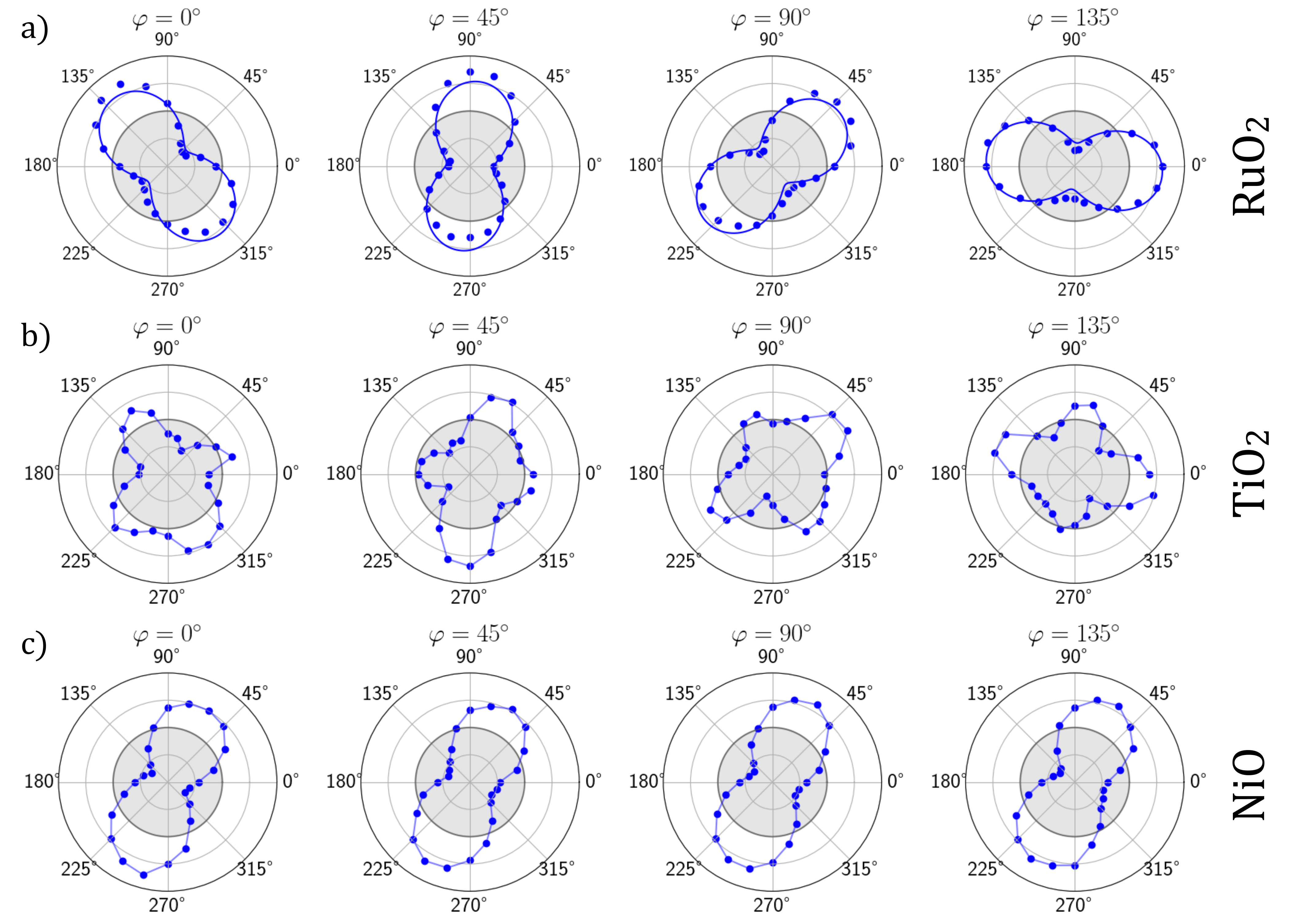}
    \caption{\textbf{Measured probe-polarization anisotropy} Shown are results for  different excitation angles $\varphi$ for a) RuO$_{2}$ pumped with $\Phi_{\text{absorb}}=5.24$\,mJ/cm$^2$, b) TiO$_{2}$ pumped with $\Phi_{\text{absorb}}=0.09$\,mJ/cm$^2$, and c) 10\,nm NiO(111) film capped with $2\,$nm Pt and pumped with $\Phi_{\text{absorb}}=0.27$\,mJ/cm$^2$.}
    \label{fig:probe_pol}  
\end{figure}

In addition to the pump-induced spin polarization described in the main manuscript, RuO$_{2}$ also exhibits an excitation-angle dependent birefringence signal that can be made visible by fitting and subtracting the background signal. The probe Kerr rotation for RuO$_{2}$ excited at different excitation angles is shown in the polar plots in Fig.~\ref{fig:probe_pol}~a). The plots show a clear twofold symmetrical signal where the phase of the sinusoid is determined by the chosen excitation angle $\varphi$. This two-fold symmetry and its excitation-angle dependence is the direct experimental observable of an optically induced transient birefringence of RuO$_2$ that is rooted in the strongly anisotropic excited carrier distribution. This two fold symmetric signal hence contains contributions from the linear optical response due to the anisotropic carrier population as well as the higher-order nonlinear optical response of the electro optical Kerr effect (EOKE) which hence all follow the same pump polarization dependence.     

The two-fold symmetric signal in Fig.~\ref{fig:probe_pol}~a) is in stark contrast to the probe Kerr rotation in TiO$_{2}$ in Fig.~\ref{fig:probe_pol}~b), which shows a 4-fold symmetry after optical excitation with $1.55\,$eV photons corresponding to an in-gap excitation without excitation of charge carriers.  However, the measured two-fold symmetry in RuO$_{2}$ shows that we are creating a pump-induced optical asymmetry that is independent of the crystallographic orientation of the sample and dominates over the 4-fold signal of the substrate. The origin of this 2-fold symmetry can be related to the calculated spin-resolved excitation maps shown in the insets of Fig.~2F in the main manuscript. If one focuses on the charge distribution of electrons above $E_{\text{Fermi}}$, i.e., ignoring their spin character, it becomes apparent that the excited-carrier distribution is characterized by a 2-fold symmetry. As in Fig.~\ref{fig:probe_pol}~a), the actual angle of the carrier distribution is controlled by the chosen excitation angle. This shows that the birefringence observed in Fig.~\ref{fig:probe_pol}~a) can be a sign of the pump-induced anisotropic carrier distribution. For further comparison, Fig.~\ref{fig:probe_pol}~c) shows the results of the same experiment on the prototypical antiferromagnet NiO/Pt bilayer. The $10\,$nm NiO(111) layer was optically excited with a photon energy of 4.65\,eV. This photon energy exceeds the optical band gap of NiO (4.3\,eV), leading to the generation of optically excited carriers, similar to the excitation of RuO$_2$. We selected a thin film of NiO for its distinct domain structure compared to the bulk system. In thin films, additional strain from the substrate reduces the number of S-domains per T-domain, ultimately resulting in a single domain within the probe beam spot on the sample surface. The ultrafast response of the antiferromagnetic order of NiO was monitored experimentally via changes in the magneto-optical birefringence (MOBF) signal. This signal arises from strong magneto-elastic coupling and the corresponding distortion of the crystal lattice, which is strongly correlated with the orientation of the sample Néel vector. For NiO(111), we found a T domain with a substantial out-of-plane orientation of the Néel vector and an in-plane distortion of the NiO crystal structure accessible in our experiment geometry (almost normal incidence of pump and probe beam with respect to the sample surface). As in Fig.~\ref{fig:probe_pol}~a), a twofold symmetry can be observed as a result of pumping NiO. An important difference, however, is that the orientation of this 2-fold symmetric signal is \emph{independent} of the chosen excitation angle because the pump acts only as a heating pulse \cite{Wust2022,Kholid2023,Rongione2023}. Since the orientation of the Néel vector is fixed with respect to the crystallographic axes, a rotating probe will always measure the same anisotropy. Apart from this structural effect, no angle-dependent signals can be observed for this prototypical antiferromagnet as in the case of RuO$_{2}$.

\section{Signatures of altermagnetic domains in RuO$_2$}

\begin{figure}
    \centering
    \includegraphics[width=0.6\linewidth]{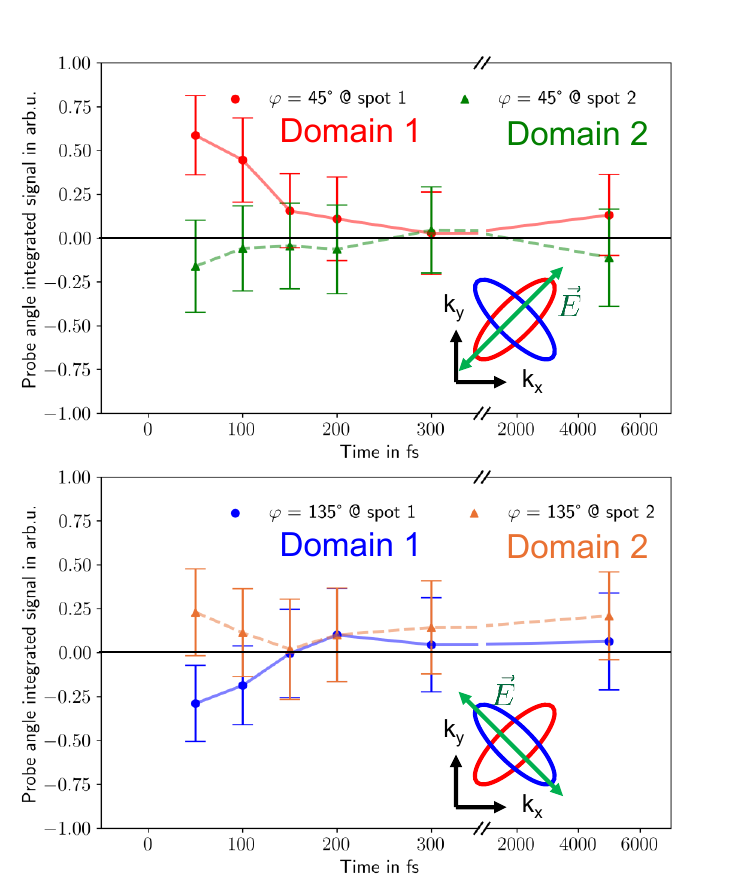}
    \caption{\textbf{Signatures of magnetic domains in the Kerr spectra} Transient MOKE response of a 4\,nm RuO$_2$ film for excitation angles along the d-wave loops of the altermagnetic band structure (45$^\circ$: top and 135$^\circ$: bottom) and two different spots on the sample. The sign reversal indicates two distinct majority domains within the laser beam spot on the sample.}
    \label{fig:domains}
\end{figure}

All experimental data were collected from an as-grown RuO$_2$ film in its pristine magnetic multi-domain state. It is likely that the domain sizes are similar to those of other compensated magnets and thus on the order of a few micrometers, as recently determined for MnTe~\cite{amin2024nanoscale}. Our probe beam with a spot size of approximately $300\pm25\,\mu$m averages signals from multiple individual domains. To evaluate how these multiple domains affect the magnetic Kerr signal, we performed the experiment at different positions on the sample. The transient Kerr signals for two points on a $4\,$nm thick RuO$_2$ film are shown in Fig.~\ref{fig:domains}. The pump beam’s excitation angle was aligned along the d-wave loops of the altermagnetic band structure (45$^\circ$ for the top panel and 135$^\circ$ for the lower panel). We clearly see a change in the sign of the transient Kerr signal between the two locations on the surface. This sign change provides strong evidence for the presence of (alter)magnetic domains, thereby confirming altermagnetic order in our ultrathin RuO$_2$ samples. Furthermore, it indicates the existence of a dominant majority domain within each probe spot on the surface, enabling the detection of a non-zero spin polarization of optically excited charge carriers in RuO$_2$.

\section{Thickness dependence of the spin polarization signal for ultrathin RuO$_2$ films}

\begin{figure}[h]
    \centering
    \includegraphics[width=0.9\linewidth]{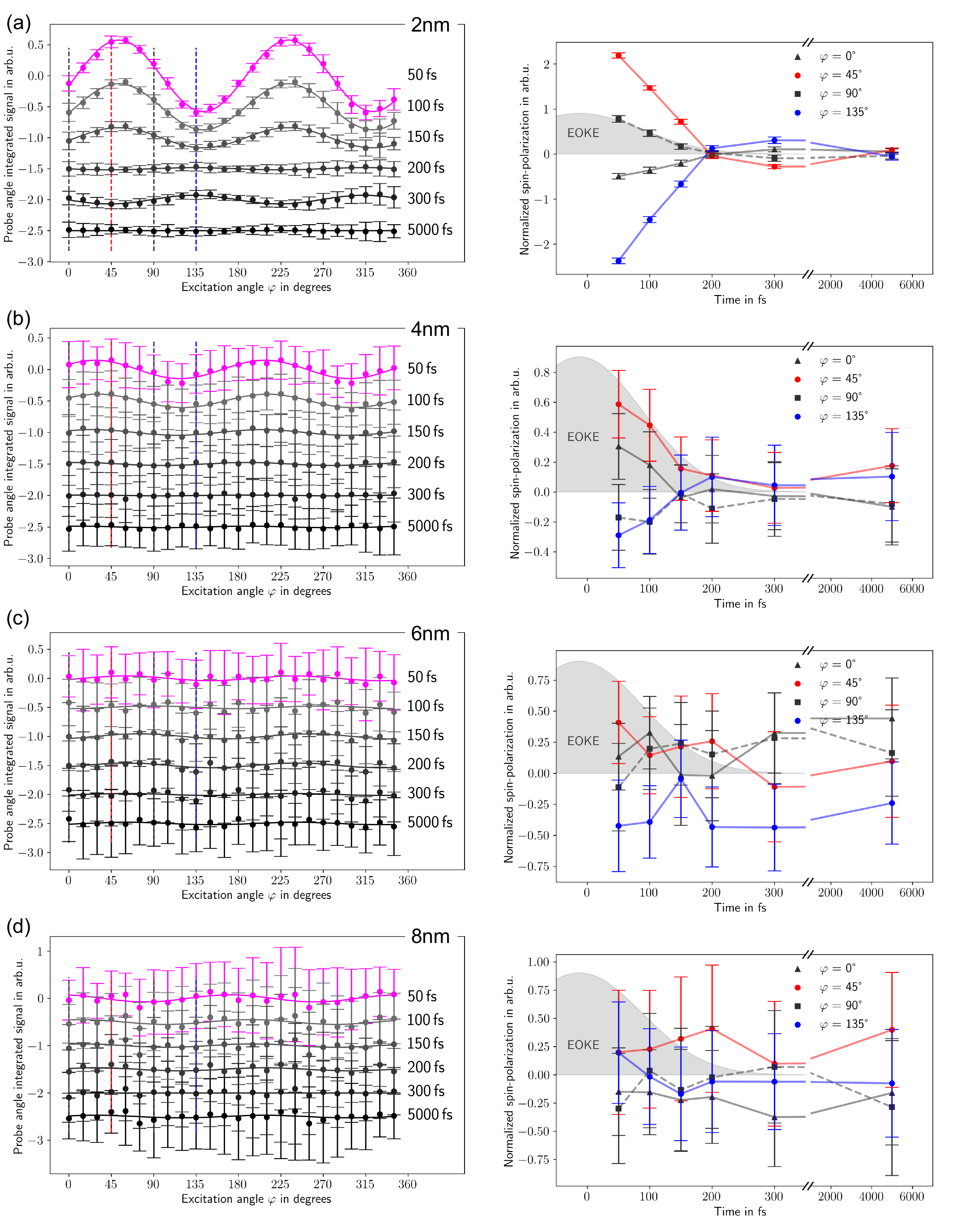}
    \caption{\textbf{Thickness dependence of the spin polarization signal for ultrathin RuO$_2$ films} Left column: Magneto-optical signals integrated over probe polarization angle vs. the excitation angle $\varphi$ at different times after pump for ultrathin RuO$_2$ films with thickness of $2\,$nm (a), $4\,$nm (b), $6\,$nm (c), and $8\,$nm (d). Right column: The corresponding time-dependent magneto-optical signals for excitation angles of $0^\circ$, $45^\circ$, $90^\circ$, $135^\circ$. Note that the excitation along one of the main d-wave loops corresponds to excitation angles of $45^\circ$ and $135^\circ$.}
    \label{fig:LabFS4}
\end{figure}

We conducted time-resolved polar MOKE experiments on a series of RuO$_2$(001) films with thicknesses between 2\,nm and 8\,nm. All samples were prepared on TiO$_2$ using the same procedure, and we recorded the averaged probe-polarization Kerr signal for excitation angles (pump-polarization orientations) between $0^\circ$ and $360^\circ$ in steps of $15^\circ$, at selected time steps. The corresponding data are shown in the left column of Fig.~\ref{fig:LabFS4}. For the $2\,$nm RuO$_2$ film, we find a clear sinusoidal angular dependence of the magnetic Kerr response on short timescales ($50\,$fs - $100\,$fs), with fixed maxima and minima, in analogy to the results of the $3.5\,$nm RuO$_2$ film discussed in Fig.~3 of the main manuscript. Importantly, the amplitude of the sinusoidal angular dependence of the Kerr signal decreases significantly as the film thickness increases to $4\,$nm and is barely visible at 6\,nm. In contrast, no sinusoidal angular dependence of the Kerr signal can be observed for a film thickness of $8\,$nm indicating the absence of any optically generated spin polarization in RuO$_2$ films with thicknesses between $6\,$nm and $8\,$nm. This conclusion is further supported by the temporal evolution of the Kerr signal at selected excitation angles, as shown in the right column of Fig.~\ref{fig:LabFS4}. Here, we find a clear and continuous reduction of the transient Kerr signal for excitation angles along the d-wave loop for film thicknesses of $2\,$nm and $4\,$nm. This characteristic transient spin-polarization signal is again substantially less pronounced for a thickness of $6\,$nm and is no longer visible for $8\,$nm. The continuous decrease of the spin polarization signal with increasing film thickness is a clear indication of the loss of altermagnetic order with increased film thickness. This trend, as well as the quantitative film thicknesses, are fully in line with previous reports that provide experimental evidence for the existence of altermagnetic order in RuO$_2$ only in the ultrathin film limit below $\approx 10\,$nm \cite{he2025evidence,jeong2025metallicity,akashdeep2025interface,akashdeep2025surface}.

\section{Photon-energy dependence of the optical response of RuO$_2$}

\begin{figure}
    \centering
    \includegraphics[width=0.95\linewidth]{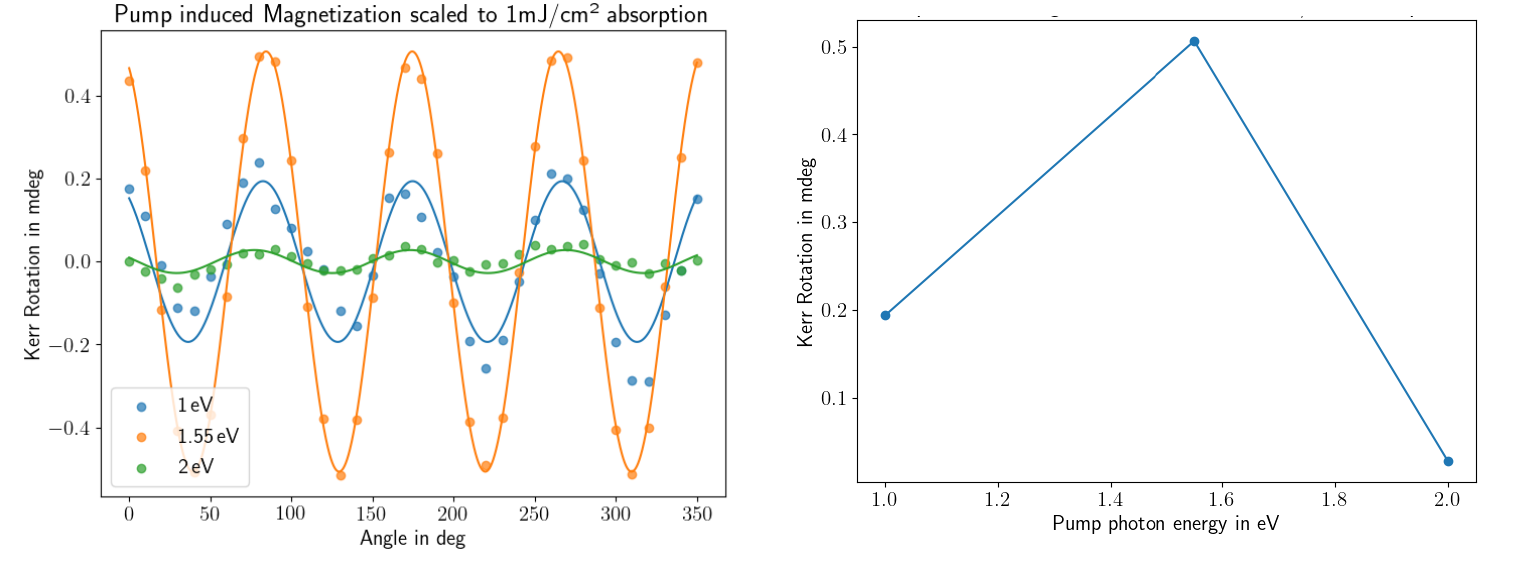}
    \caption{\textbf{Dependence on pump-photon energy} Kerr response of RuO$_2$ for different pump pulse photon energies and different orientations of the pump pulse polarization, but identical probe pulse polarization orientation. We observe a clear photon-energy dependence in the magnitude of the Kerr response, as predicted by theory.}
    \label{fig:LabFS5}
\end{figure}

The the photon-energy dependence of the magneto-optical response of RuO$_2$ is shown in Fig.~\ref{fig:LabFS5} for the same excitation energies used in our theoretical calculations. The Kerr response was recorded for different orientations of the pump-pulse polarization, with identical probe-pulse polarization, yielding a different periodicity in the sinusoidal modulation of the Kerr signal compared with the experiments discussed in our main manuscript. Our data reveal a substantially stronger Kerr response for pump photons at $1.55\,$eV than at $1.00\,$eV or $2.00\,$eV. This demonstrates a pronounced dependence of the MOKE signal on the pump-photon energy, thereby confirming the general trend predicted by theory: the magneto-optical response of RuO$_2$ is photon-energy dependent.

Note that our experiment yields a photon-energy-dependent magnitude of the Kerr signal that differs slightly from our theoretical predictions. This discrepancy likely arises because the theoretical results are based on a bandstructure calculation for bulk RuO$_2$ that uses a suitable effective Hubbard correction $U$. In contrast, our experiments were performed on ultrathin RuO$_2$ films, which have been shown to exhibit altermagnetic order.

\section{Calculated birefringence and magneto-optical signatures in RuO$_2$ \label{ch:birefringence_theory}}

In this section, we present calculations of the pump-induced changes in the dielectric tensor and the resulting birefringence and magneto-optical signatures. These calculation are intended to complement our experimental results. Here we take into account the full three-dimensional Brillouin zone to capture all of the transitions involved in the optical excitation. We apply the same model as described in the Methods section of the main manuscript to calculate the electronic occupation numbers after pump excitation with a photon energy of 1.55 eV and focus on the most important polarization directions, i.e. 0°, 45°, 90° and 135° with respect to the $[100]$-axis of the RuO$_2$ crystal. Using these occupation numbers after laser excitation as input, we calculate the optical conductivity tensor after pump-excitation via the Kubo-Greenwood formalism implemented in the Elk Code \cite{kubo_elk}. The momentum matrix elements and band energies are obtained from the ground state DFT calculation. The Lorentzian smearing used for the calculation of the dielectric tensor and optical conductivity tensor is set to 0.003 Ha for all calculations. 

\begin{figure}
    \centering
    \includegraphics[width=\linewidth]{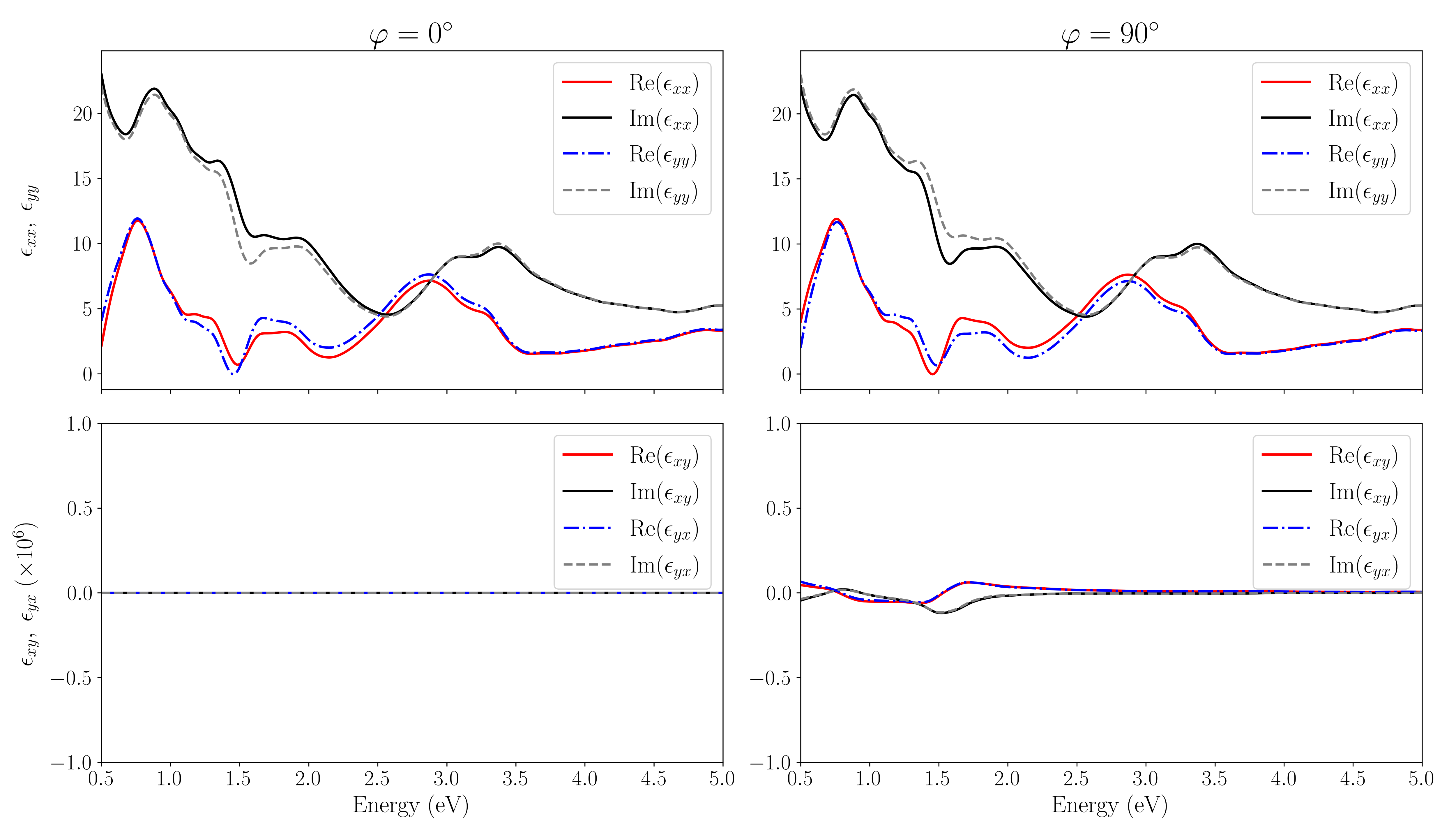}
    \caption{\textbf{Dielectric tensor components computed for optically excited RuO$_2$ exhibiting no spin polarization} Diagonal (upper row) and off-diagonal (lower row) components of the dielectric tensor for a pump excitation with an excitation angle of $\varphi=0$° (left side) and $\varphi=90$° (right side). Note that the off-diagonal elements are scaled by $10^{6}$.}
    \label{fig:dielec_0_90}
\end{figure}
\begin{figure}
    \centering
    \includegraphics[width=\linewidth]{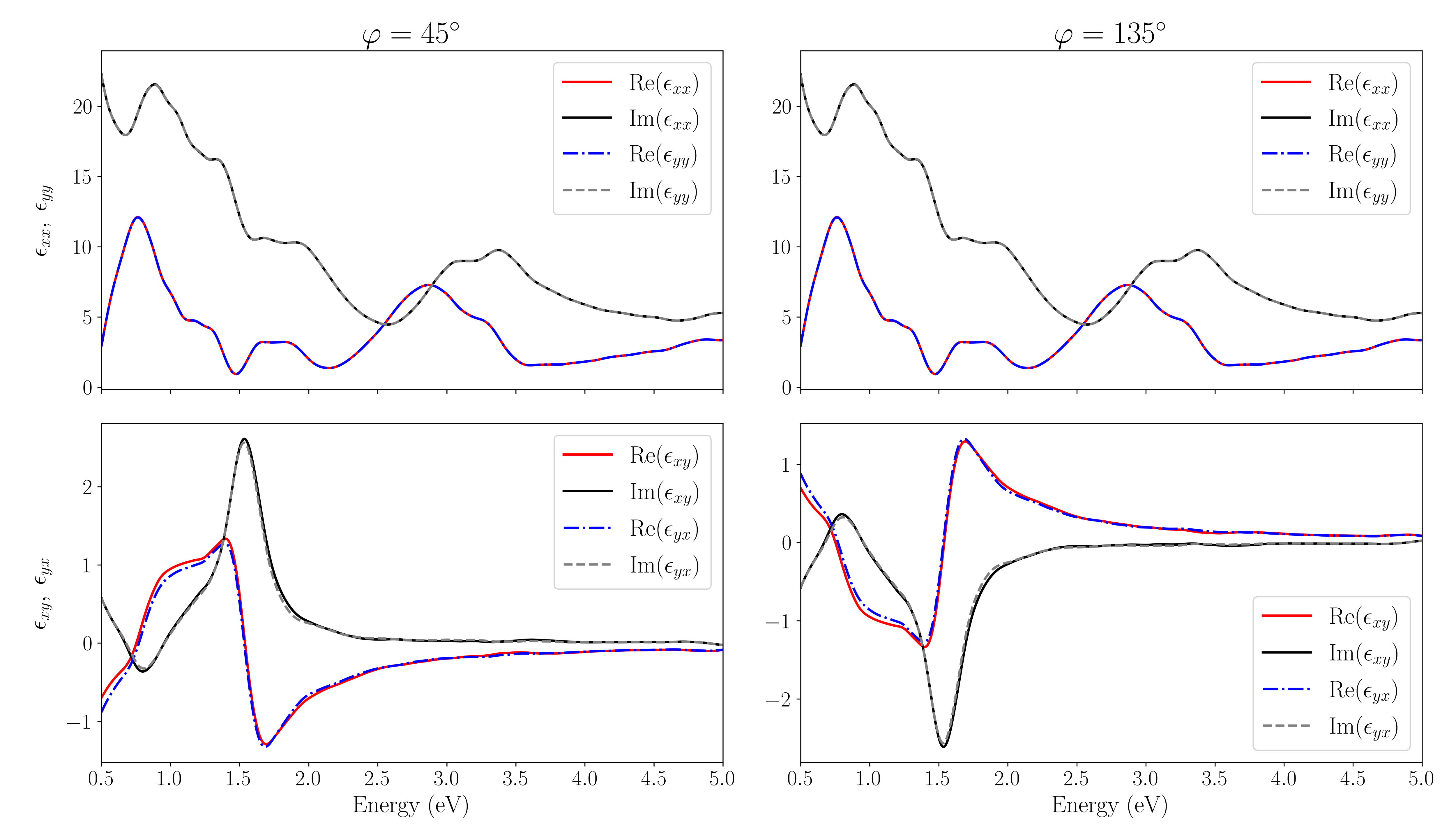}
    \caption{\textbf{Dielectric tensor components computed for optically excited spin-polarized RuO$_2$} Diagonal (upper row) and off-diagonal (lower row) components of the dielectric tensor for a pump excitation with an excitation angle of $\varphi=45$° (left side) and $\varphi=135$° (right side).}
    \label{fig:dielec_45_135}
\end{figure}
Figure~\ref{fig:dielec_0_90} shows the diagonal $\epsilon_{xx}$ and $\epsilon_{yy}$ components as well as the off-diagonal $\epsilon_{xy}$ and $\epsilon_{yx}$ components of the dielectric tensor for irradiation with an excitation angle of 0° and 90°. Both excitations introduce a birefringence due to the unequal diagonal components ($\epsilon_{xx}\neq\epsilon_{yy}$), which is is absent in the ground state. In addition, the excitation does not introduce any appreciable off-diagonal components $\epsilon_{xy}$ or $\epsilon_{yx}$. In these cases one would not expect any signal to be present besides the birefringence induced by the anisotropic carrier distributions. We will further analyze this at the end of this section.

Figure~\ref{fig:dielec_45_135} illustrates the dielectric tensor components for an excitation angle of 45° and 135°. In this case the diagonal components are equal ($\epsilon_{xx}=\epsilon_{yy}$) and sizable off-diagonal components $\epsilon_{xy}$ and $\epsilon_{yx}$ appear. The dominant symmetric part of these off-diagonal components is related to the anisotropic excitation of electrons, which is strongest along or perpendicular to the polarization direction of the pump pulse. This leads again to a birefringence for the probe beam. However, one can clearly observe an asymmetry $\epsilon_{xy}\ne\epsilon_{yx}$ in the off-diagonal components, which causes an additional rotation and ellipticity of the probe beam that originates from magnetic signatures.

To complete our description of the pump-induced changes we compute the probe angle dependent rotation and ellipticity. To this end, we write the dielectric tensor after pump excitation in the form
\begin{equation}
    \boldsymbol{\epsilon}_{\varphi}(\omega_{\mathrm{probe}})=\begin{pmatrix}
        \epsilon_{xx} & \epsilon_{xy} & 0 \\
        \epsilon_{yx} & \epsilon_{yy} & 0 \\
        0 & 0 & \epsilon_{zz}
    \end{pmatrix}\equiv\begin{pmatrix}
        \bar{\epsilon}+\Delta & \tilde{\epsilon}+\delta & 0 \\
        \tilde{\epsilon}-\delta & \bar{\epsilon}-\Delta & 0 \\
        0 & 0 & \epsilon_{zz}
    \end{pmatrix}, \label{eq:general_tensor}
\end{equation}
with  symmetric components, $\bar{\epsilon}=(\epsilon_{xx}+\epsilon_{yy})/2$ and $\tilde{\epsilon}=(\epsilon_{xy}+\epsilon_{yx})/2$ and antisymmetric ones $\Delta=(\epsilon_{xx}-\epsilon_{yy})/2$ and $\delta=(\epsilon_{xy}-\epsilon_{yx})/2$. 
Note that the tensor depends on the pump excitation angle $\varphi$ and the probe energy $\omega_{\mathrm{probe}}$. Solving the Fresnel equation \cite{kerr_openeer} for a probe pulse incident from the $z$-direction yields the refractive indices
\begin{equation}
    n_{\pm}^2=\bar{\epsilon}\pm\sqrt{\Delta^2+\tilde{\epsilon}^2-\delta^2}=:\bar{\epsilon}\pm C
\end{equation}
and eigenmodes
\begin{equation}
    \mathbf{u}_{\pm}e^{i\omega t}=\begin{pmatrix}
        \tilde{\epsilon}+\delta \\ \pm C-\Delta \\ 0
    \end{pmatrix}e^{i\omega( t-n_\pm z/c)}.
\end{equation}
The reflection coefficients for the two different eigenmodes at normal incidence are given by $r_{\pm}=(n_{\pm}-n_0)/(n_\pm+n_0)$, where $n_0\approx1$ is the refractive index of the surrounding medium (here: air) \cite{ciddor96}. Since we are not interested in the propagation through the material, the $z$-dependence of the eigenmodes is neglected. 
\begin{figure}[t]
    \centering
    \includegraphics[width=\linewidth]{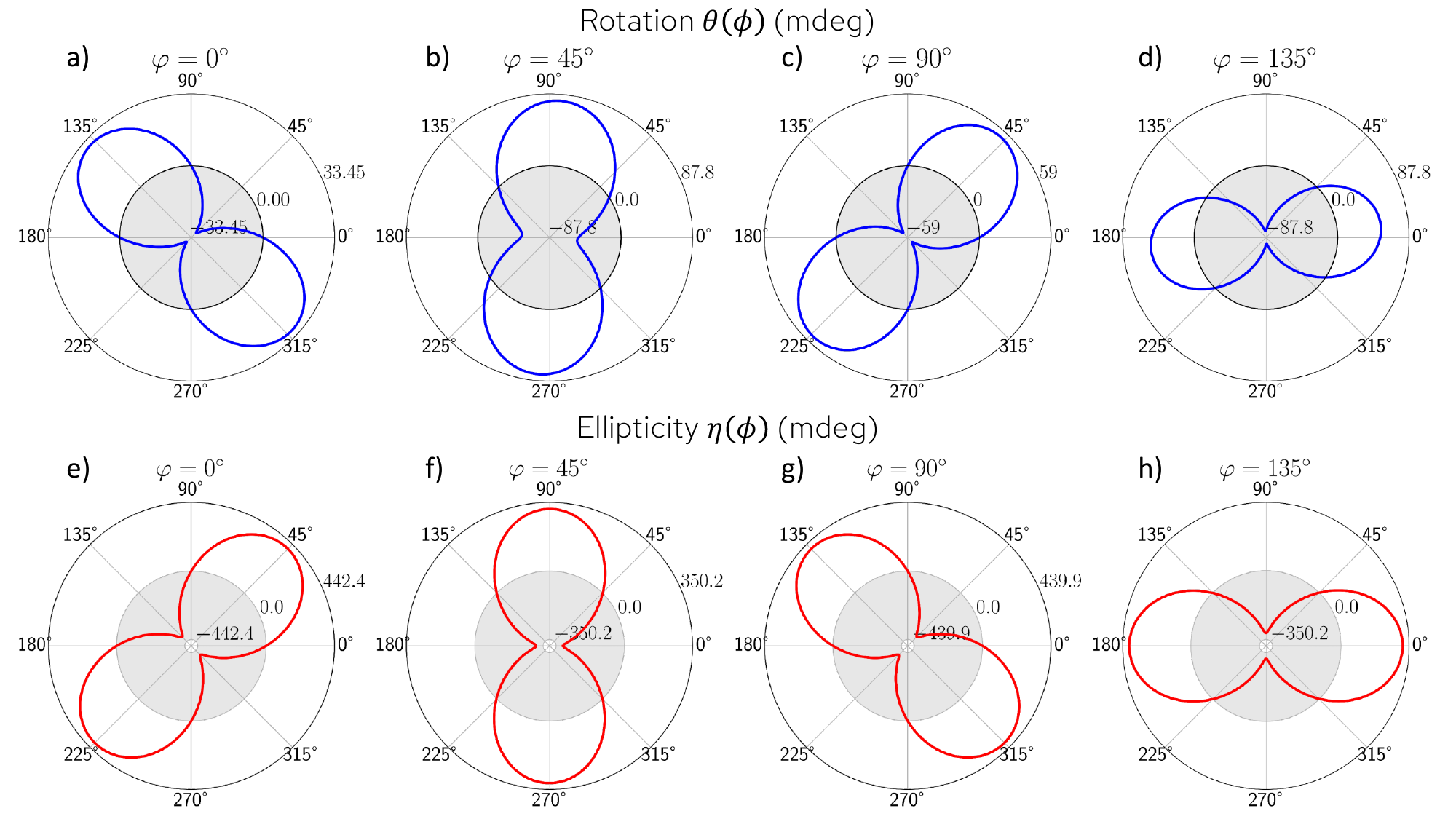}
    \caption{\textbf{Computed probe-angle dependence of Kerr response} Rotation a)-d) and ellipticity e)-h) of the probe beam in $10^{-3}$ degrees vs.\ probe angle for different pump excitation angles after pumping RuO$_2$ with a photon energy of $\hbar\omega_{\mathrm{L}}=1.55$ eV and probing at $\hbar\omega_{\mathrm{probe}}=3.10$ eV.}
    \label{fig:rot_and_elip_theory}
\end{figure}
Decomposing the initial probe beam that is linearly polarized along $\hat{\mathbf{e}}_{\phi}=(\cos\phi,\sin\phi,0)^T$ into the eigenmodes yields
\begin{equation}
    \mathbf{E}_{\mathrm{probe}}^{\phi}=E_0e^{i\omega t}(\alpha_{+}^{\phi}\mathbf{u}_{+}+\alpha_{-}^{\phi}\mathbf{u}_{-}),
\end{equation}
with complex expansion coefficients~$\alpha_{\pm}^{\phi}$. From the reflected probe beam
\begin{equation}
    \mathbf{E}^{\phi}_{r}=E_0e^{i\omega t}(r_{+}\alpha_{+}^{\phi}\mathbf{u}_{+}+r_{-}\alpha_{-}^{\phi}\mathbf{u}_{-})
\end{equation}
which is, in general, elliptically polarized, we compute the two major axis $\mathbf{A}^{\phi}$ and $\mathbf{B}^{\phi}$ ($\vert \mathbf{A}^{\phi}\vert\geq\vert \mathbf{B}^{\phi}\vert$) of the polarization ellipse via the relations \cite{Berry_2004}
\begin{equation}
        \mathbf{A}^\phi = \frac{1}{\vert\sqrt{\mathbf{E}_{r}^{\phi}\cdot\mathbf{E}_r^{\phi}}\vert}\mathrm{Re}\Big[\mathbf{E}_r^{\phi}\sqrt{\big(\mathbf{E}_r^{\phi}\big)^{\ast}\cdot\big(\mathbf{E}_r^{\phi}\big)^{\ast}}\Big]
        \quad \text{and} \quad
        \mathbf{B}^\phi = \frac{1}{\vert\sqrt{\mathbf{E}_{r}^{\phi}\cdot\mathbf{E}_r^{\phi}}\vert}
        \mathrm{Im}\Big[\mathbf{E}_r^{\phi}\sqrt{\big(\mathbf{E}_r^{\phi}\big)^{\ast}\cdot\big(\mathbf{E}_r^{\phi}\big)^{\ast}}\Big].
\end{equation}
The probe beam rotation can now be evaluated as the angle between the main axis $\mathbf{A}^{\phi}$ and the probe beam polarization direction before reflection, which is defined as
\begin{equation}\theta(\phi)=\arccos\big(\hat{\mathbf{e}}_{\phi}\cdot\hat{\mathbf{A}}^{\phi}\big)\mathrm{sgn}\big(\big[\hat{\mathbf{e}}_{\phi}\times\hat{\mathbf{A}}^{\phi}\big]_{z}\big).
\end{equation}
The ellipticity of the probe beam after reflection is given by 
\begin{equation}
    \eta(\phi)=\arctan\Big(\frac{\vert\mathbf{B}^{\phi}\vert}{\vert\mathbf{A}^{\phi}\vert}\Big)\mathrm{sgn}\big(\mathrm{Im}\big[\big(\mathbf{E}_r^{\phi}\big)^{\ast}\times\mathbf{E}_{r}^{\phi}\big]_{z}\big),
\end{equation}
where the ``sign'' functions account for a clockwise (negative sign) or counter-clockwise (positive sign) rotation, i.e., left- or right-handed ellipticity, respectively. 

\begin{figure}[t]
    \centering
    \includegraphics[width=\linewidth]{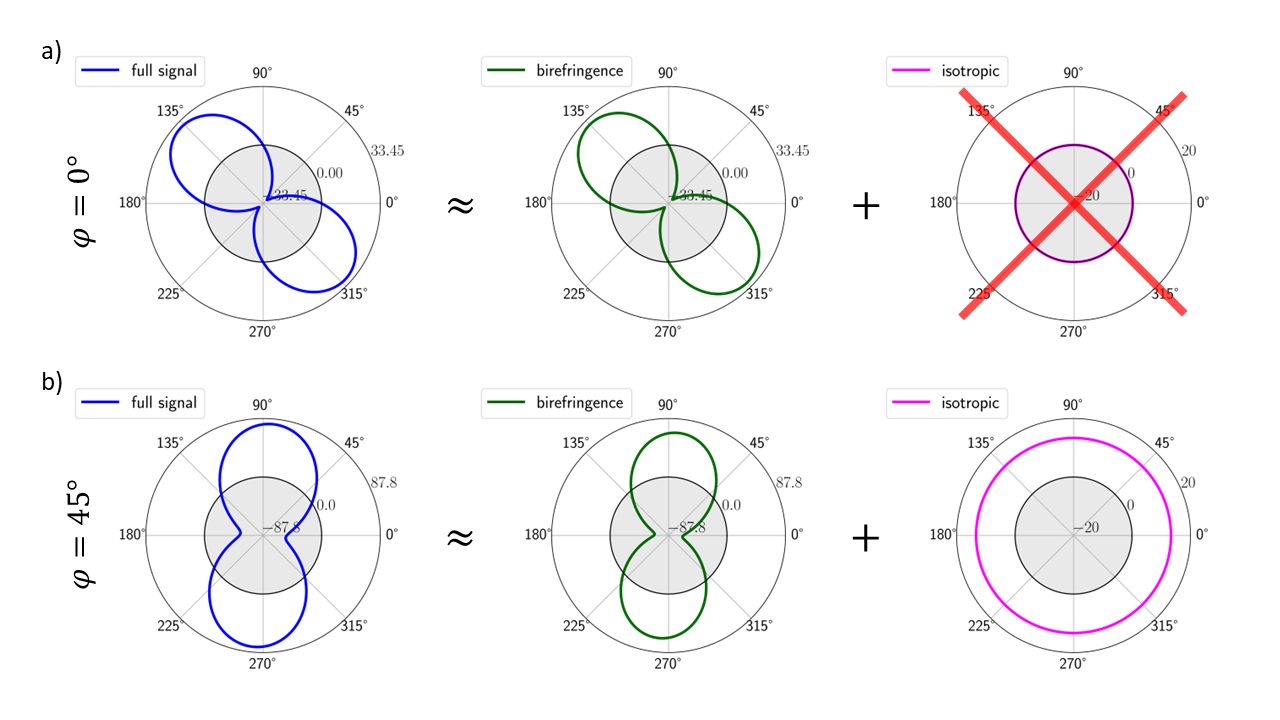}
    \caption{\textbf{Birefringent and magnetic contributions to computed magneto-optical signal} Separation of the computed signal into  for an excitation angle of a) $\varphi=0$° (no excited spin polarization) and b) $\varphi=45$° (maximum excited spin polarization).}
    \label{fig:separation}
\end{figure}
In the following we analyze this angle dependence for the conditions used in the experiment, i.e., a pump photon energy of 1.55 eV and a probe photon energy of 3.10 eV. Figure~\ref{fig:rot_and_elip_theory} shows the computed rotation and ellipticity of the probe beam after reflection upon the sample for all possible probe polarization-angles $\phi$. Figures~\ref{fig:rot_and_elip_theory}~a)-d) depict the angle dependent rotation of the probe beam for the four pump excitation angles plotted in the same fashion as the experimentally observed ones in Fig.~\ref{fig:probe_pol}~a).
\begin{figure}
    \centering
    \includegraphics[width=\linewidth]{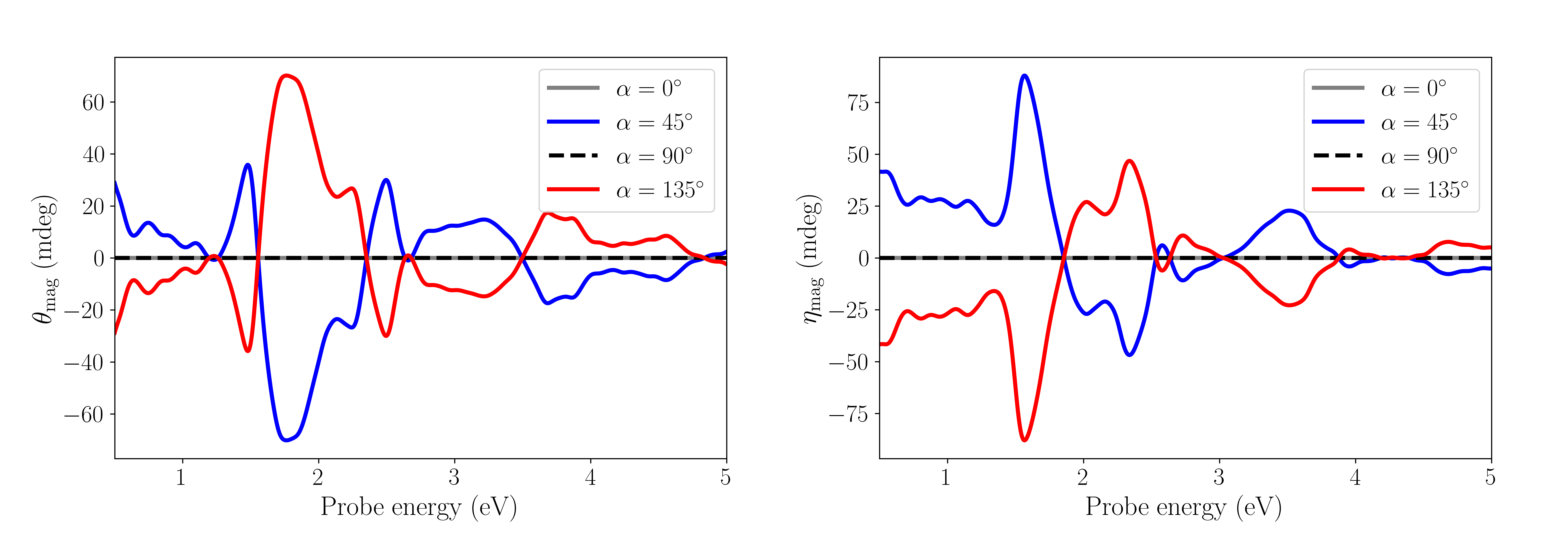}
    \caption{\textbf{Energy-dependence of the true magnetic Kerr response} Magnetic Kerr rotation $\theta_{\mathrm{mag}}$ (left side) and magnetic Kerr ellipticity $\eta_{\mathrm{mag}}$ (right side) obtained by integrating over probe angle as a function of probe photon energies in the visible range. The system was pumped with an energy of $\hbar\omega_{\mathrm{L}}=1.55$ eV with the excitation angles $\varphi=0$° (gray line), 45° (blue line), 90° (black dashed line) and 135° (red line).}
    \label{fig:angle_integrated}
\end{figure}
The calculated shapes match the experiment well and show the same clockwise rotational behavior in $\theta(\phi)$. In all cases the birefringence due to the excited carriers is the dominant effect. For completeness, we also show in Figs.~\ref{fig:rot_and_elip_theory}~e)-h) the computed Kerr ellipticity.  \\
Figure~\ref{fig:separation} separates the full signal into pure birefringence parts and pure magnetic signatures. We fit the signal with the function $\theta(\phi)=A\sin(2\phi-\xi)+C$, which separates into a pure birefringence part proportional to $\sin(2\phi)$ and a constant isotropic offset caused by magnetic signatures. Fig.~\ref{fig:separation}~a) shows the separation for the excitation with $\varphi=0$°. Here one only gets a pure birefringence signal (vanishing isotropic part) since no spin polarization is excited. Fig.~\ref{fig:separation}~b) illustrates the $\varphi=45$° excitation. In this case the full signal has to be decomposed in an angle-dependent birefringence part and an isotropic contribution, thus showing the signatures of the spin polarized excitation. The separation of the isotropic magnetic contribution can therefore also be achieved by an angle integration of the signals, since all the birefringent contributions will cancel out. The integrated signals
\begin{equation}
\theta_{\mathrm{mag}}=\int_{0}^{2\pi}\theta(\phi)\mathrm{d}\phi \quad \text{and} \quad\eta_{\mathrm{mag}}=\int_{0}^{2\pi}\eta(\phi)\mathrm{d}\phi
\end{equation}
are shown in Fig.~\ref{fig:angle_integrated} for the spectrum of probe photon energies in the visible range. If we excite with an angle of $\varphi=0$° or 90° the angle integrated signals for the rotation and ellipticity vanish completely over the entire range of probe energies. In these cases the pump excitation only causes a birefringence and no magnetic signatures, as shown before. For the excitation angles of 45° and 135°, however, the probe angle integration shows sizable signals up to 80~mdeg. Importantly, the angle integrated rotation and ellipticity flip signs when comparing the 45° and 135° pump excitation, which is in accordance with our measurements. The calculation also shows that the pump induced effect should be visible over a broad spectrum of probe energies and that the sign of the observed Kerr rotation/ellipticity depends on the probe photon energy.

\section{Spin dependence of optical transitions\label{ch:spin-orbital-locking}}
\begin{figure}[t]
    \centering
    \includegraphics[width=\linewidth]{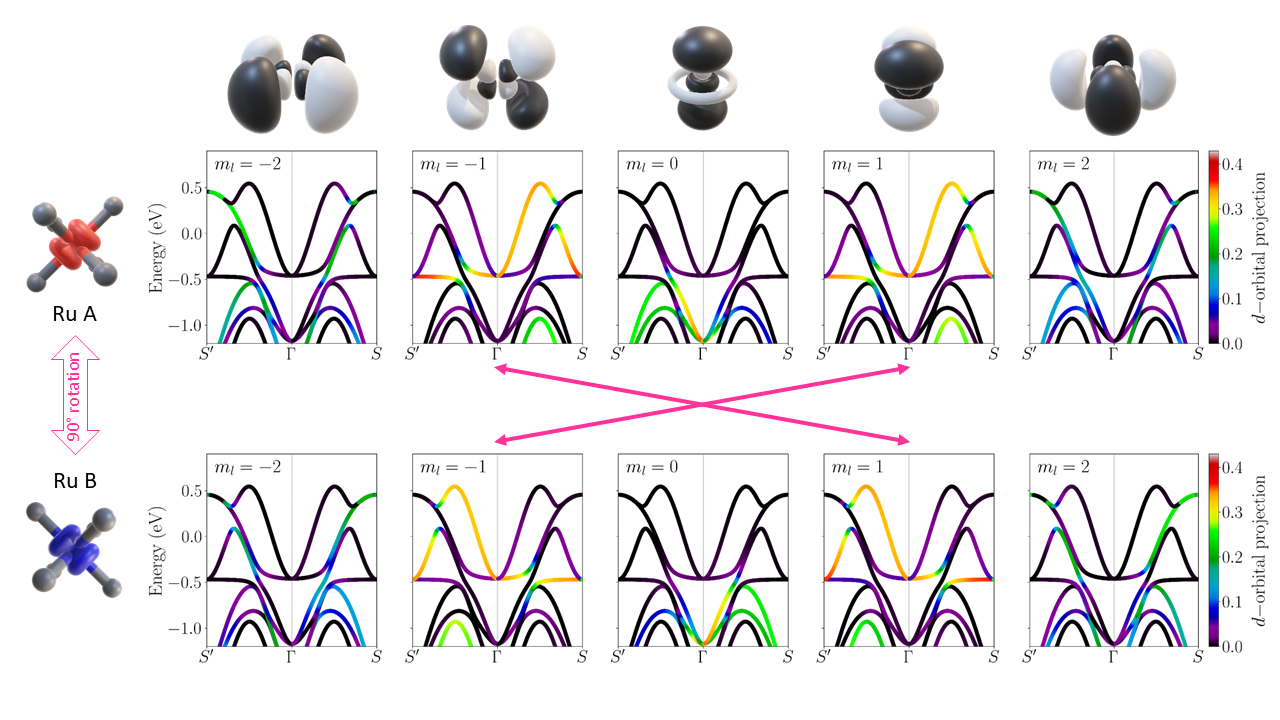}
    \caption{\textbf{Computed orbital character of the bands near the Fermi edge} The upper row shows the projection of $d$-orbitals onto the Ruthenium atom on sublattice A, i.e. spin-up, for $m_l=-2,\dots,2$. The lower row shows the projections considering the Ruthenium atom on sublattice B (spin-down). The octahedra formed by the surrounding oxygen atoms between the two sublattices are connected by a 90° rotation, indicated by the scheme on the left, which also swaps around the orbital projections for the sublattices (pink arrows).}
    \label{fig:orbital-projection}
\end{figure}
The optical transitions considered in the main manuscript are to a large extent determined by the altermagnetic $[C_2\Vert C_{4z}]$ symmetry of the material under consideration. This can be seen from the orbital character of the bands involved in the optical excitations, which are influenced by the octahedral environment around the magnetic atoms. As the $d$ orbitals are the ones affected by the crystal field splitting mechanism, we compute the sublattice resolved projection of the $d$ orbitals from the DFT states, which are shown in figure~\ref{fig:orbital-projection} for quantum numbers $m_l=-2,\ldots,2$. The upper (lower) row shows the projections for the ruthenium atoms on the sublattice A (B) with a color code for the contribution from the respective $d$-orbitals. The graph for $m_l^{(A)}=-1$ shows the same projection as the graph for $m_l^{(B)}=1$, just mirrored around the vertical line at $\Gamma$ (indicated by the pink arrows), showing that for the spin-up sublattice the transition occurs for the $d_{xz}$-orbitals, while for the spin-down sublattice the same transition should occur for the 90° rotated $d_{yz}$-orbitals. It now depends on the orientation of the electric field vector whether electrons are excited to the $d_{xz}$- or $d_{yz}$-orbitals. This can be related to the transitions for $\hbar\omega_{\mathrm{L}}=1.00$~eV in Figure~2E, where angles of 0° and 90° excite both orbitals equally, while angles of 45° and 135° excite only one orbital.

\section{Non-magnetic RuO\(_2\)\label{ch:non-magnetic-ruo2}}

\begin{figure}[t]
    \centering
    \includegraphics[width=\linewidth]{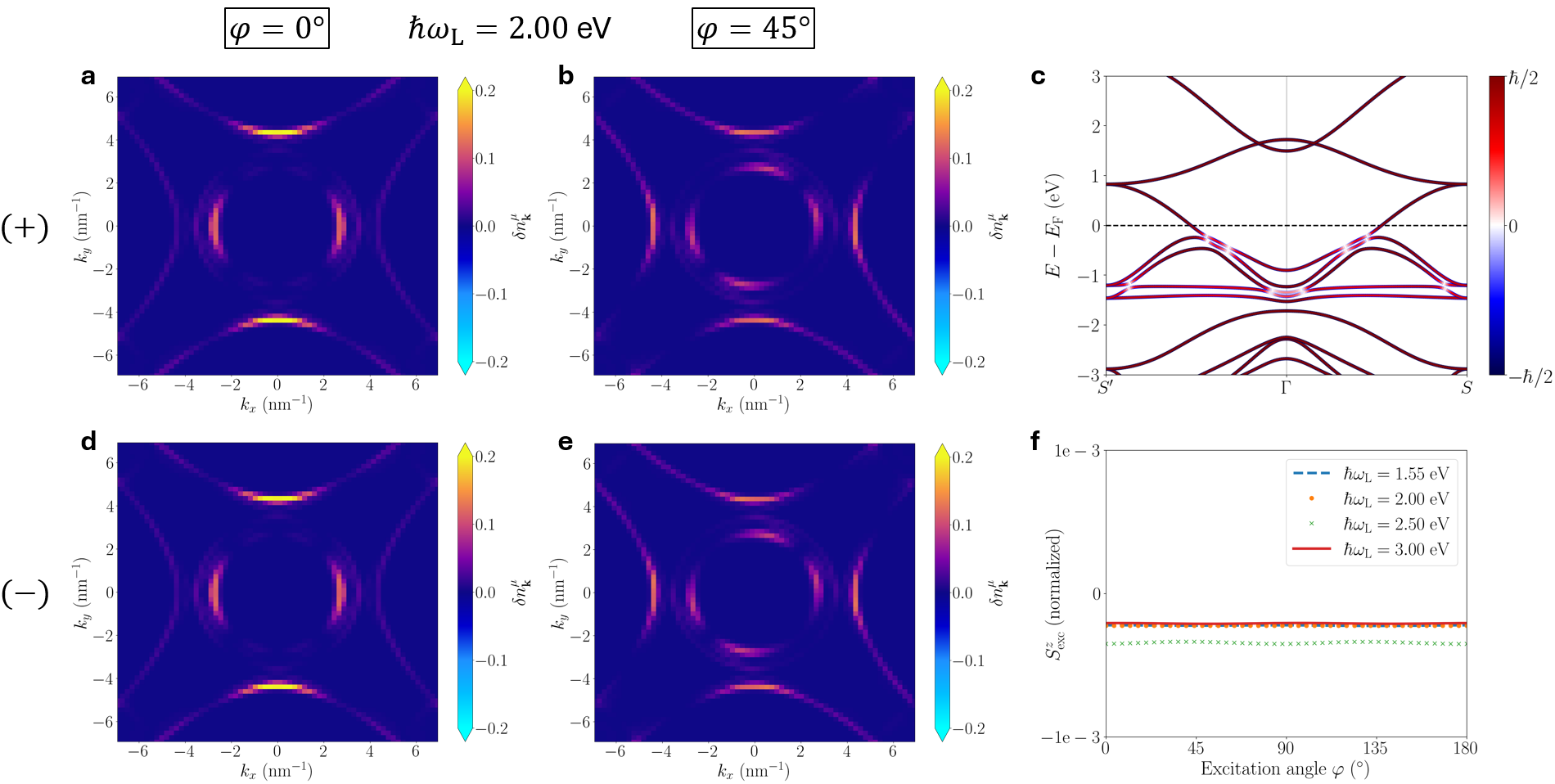}
    \caption{\textbf{Computed optical excitation in non-magnetic RuO$_2$} Change in occupation on the 2-dimensional Brillouin zone slice at $k_z=0$ for excitation with $\hbar\omega_{\mathrm{L}}=2.00$ eV in the spin-up band for a) $\varphi=0$° and b) $\varphi=45$° and for the spin-down band d) $\varphi=0$° and e) $\varphi=45$°. The resulting spin polarization can be seen in f). c) shows the band structure along the path $S^{\prime}-\Gamma-S$ in the non-magnetic phase including SOC.}
    \label{fig:bands_paramag}
\end{figure}

To ensure that the effects observed in the main manuscript are exclusive to the case of altermagnetic RuO$_2$, we performed further DFT calculations for RuO$_2$ in the non-magnetic phase using the ELK code within the PBE approximation of the exchange correlation functional. We omitted the Hubbard correction and calculated the ground state on a $22\times 22\times 32$ $k$-grid including spin-orbit coupling. The lattice parameters are the same as in the main manuscript. Figure~\ref{fig:bands_paramag}~c) shows the resulting band structure along the path $S^{\prime}-\Gamma-S$, where we now have only degenerate bands for spin-up and spin-down electrons. 
The influence of spin-orbit coupling in the bands below the Fermi level results in regions of high spin mixing, but no gap opening or splitting between the bands. Fig.~\ref{fig:bands_paramag}~a),b),d),e) show the occupation number changes in the first two degenerate bands above the Fermi edge in the 2-dimensional Brillouin zone slice for $k_z=0$ using the excitation angles of $\varphi=0$° and $\varphi=45$° respectively. The spin-up band is denoted by $(+)$ and the spin-down band by $(-)$. For illustration purposes, we choose a linearly polarized laser pulse with photon energy $\hbar\omega_{\mathrm{L}}=2.00$ eV. Note that the excitation is still highly anisotropic and changes for different excitation angles, but it has the same structure in both degenerate bands, resulting in a vanishingly small spin polarization when summed over $k$ points and bands, as shown in Fig.~\ref{fig:bands_paramag}~f), where the excited spin polarization is normalized to the same value as the spin polarization plot in Fig. 2F of the main paper. The maximum values are at least three orders of magnitude smaller than for altermagnetic RuO$_2$. Importantly, no change in sign is observed when the excitation angle is varied. The same behavior can be observed for other photon energies and excitations.
\begin{figure}
    \centering
    \includegraphics[width=\linewidth]{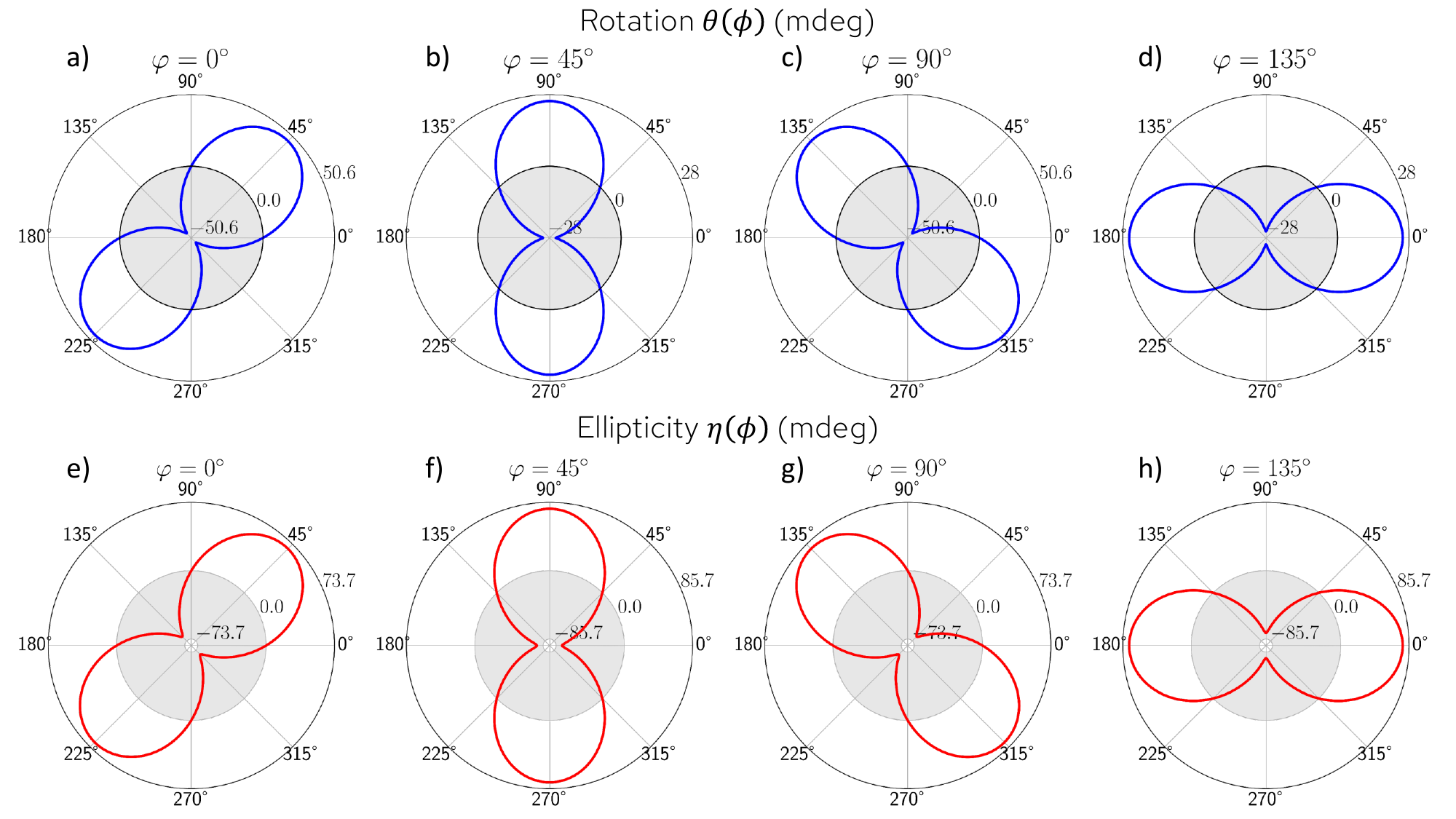}
    \caption{\textbf{Computed probe-angle dependence of Kerr response for non-magnetic RuO$_2$.}Angle dependent rotation (a)-(d) and ellipticity (e)-(h) of the probe beam in degrees for different pump excitation angles after pumping paramagnetic RuO$_2$ with a photon energy of $\hbar\omega_{\mathrm{L}}=1.55$ eV and probing at $\hbar\omega_{\mathrm{L}}=3.10$ eV.}
    \label{fig:birefringence_para}
\end{figure}
Finally, we also want to highlight the the probe angle dependent birefringence and magneto-optical signatures as we did for the altermagnetic band structure in section \ref{ch:birefringence_theory}. We apply the same model to the paramagnetic RuO$_2$. In figure~\ref{fig:birefringence_para} we show the rotation and ellipticity of the reflected probe beam for different pump excitation angles if we pump the system with 1.55 eV and probe with 3.10 eV. Like in the altermagnetic case one can observe the birefringent signatures in the rotation and ellipticity of the probe beam. This time however the ``peanut'' shaped rotation signals show a counter-clockwise rotation. Importantly, unlike in the altermagnetic case all angle integrated signals vanish, as can be seen in Fig. \ref{fig:kerr_para}, since there are no isotropic magnetic signatures present.
\begin{figure}
    \centering
    \includegraphics[width=\linewidth]{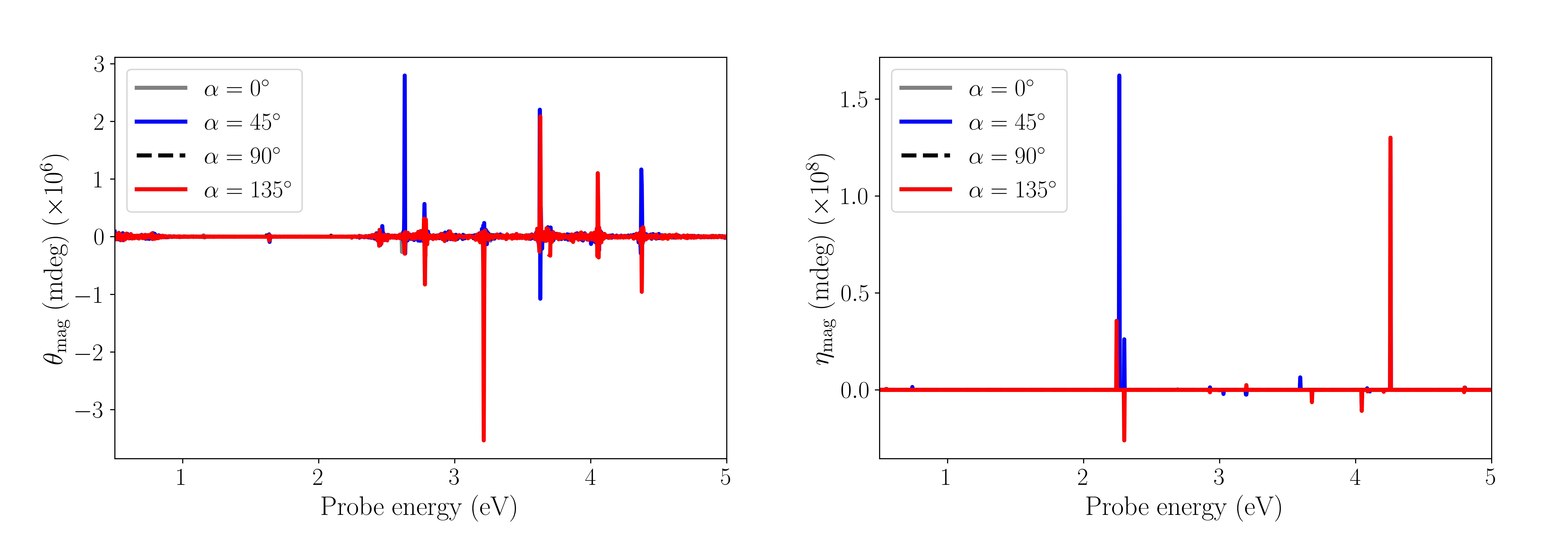}
    \caption{\textbf{Vanishing true magnetic Kerr response for non-magnetic RuO$_2$} Magnetic Kerr rotation $\theta_{\mathrm{mag}}$ (left side) and magnetic Kerr ellipticity $\eta_{\mathrm{mag}}$ (right side) obtained by the probe angle integrated signals for probe photon energies in the visible range. The system was pumped with an energy of $\hbar\omega_{\mathrm{L}}=1.55$ eV with the excitation angles $\varphi=0$° (gray line), 45° (blue line), 90° (black dashed line) and 135° (red line). Note that $\theta_{\mathrm{mag}}$ is scaled by $10^{6}$ and $\eta_{\mathrm{mag}}$ by $10^{8}$.}
    \label{fig:kerr_para}
\end{figure}

\newpage

%